\documentclass[aps,prc,superscriptaddress,showpacs,nofootinbib,twocolumn,floatfix]{revtex4}

\usepackage[dvipsnames]{xcolor}
\usepackage{amssymb,amsmath,mathtools}
\usepackage{hyperref}
\usepackage[normalem]{ulem}
\usepackage{afterpage}
\usepackage{float}

\usepackage{dcolumn}
\newcolumntype{d}[1]{D{.}{.}{#1}}

\newcommand{\unit}[1]{\ensuremath{\,\mathrm{#1}}}
\newcommand{\unitn}[1]{\ensuremath{\mathrm{#1}}}
\newcommand{\sym}{\ensuremath{\mathrm{sym}}}
\newcommand{\etal}{\textit{et al.} }
\newcommand{\etaln}{\textit{et al.}}
\newcommand{\LS}{L\&S }
\newcommand{\LSn}{L\&S}

\newcommand{\ie}{\textit{i.e.}, }

\newcommand{\figurewidth}{0.8}
\newcommand{\figurehalfwidth}{0.50}

\usepackage{libertineotf}
\usepackage[percent]{overpic}

\begin{document}

\title{A New Open-Source Nuclear Equation of State Framework based on the Liquid-Drop Model with Skyrme Interaction }
\author{A. S. Schneider}\email{andschn@caltech.edu}
\affiliation{TAPIR, Walter Burke Institute for Theoretical Physics, MC 350-17, California Institute of Technology, Pasadena, CA 91125, USA}
\author{L. F. Roberts}\email{robertsl@nscl.msu.edu}
\affiliation{National Superconducting Cyclotron Laboratory and Department of Physics and Astronomy, Michigan State University, East Lansing, MI 48824, USA}
\author{C. D. Ott}\email{cott@tapir.caltech.edu}
\affiliation{TAPIR, Walter Burke Institute for Theoretical Physics, MC 350-17, California Institute of Technology, Pasadena, CA 91125, USA}
\affiliation{Center for Gravitational Physics and International Research Unit of Advanced Future Studies, Yukawa Institute for Theoretical Physics, Kyoto University, Kyoto,  Kyoto Prefecture 606-8317, Japan}
\date{\today}
\begin{abstract}
The equation of state (EOS) of dense matter is an essential ingredient
for numerical simulations of core-collapse supernovae and neutron star
mergers.  The properties of matter near and above nuclear saturation
density are uncertain, which translates into uncertainties in
astrophysical simulations and their multi-messenger signatures.
Therefore, a wide range of EOSs spanning the allowed range of nuclear
interactions are necessary for determining the sensitivity of these
astrophysical phenomena and their signatures to variations in input
microphysics.  We present a new set of finite temperature EOSs based on
experimentally allowed Skyrme forces. We employ a liquid drop model
of nuclei to capture the non-uniform phase of nuclear matter at
sub-saturation density, which is blended into a nuclear statistical
equilibrium EOS at lower densities. We also provide a new, open-source
code for calculating EOSs for arbitrary Skyrme parametrizations. We
then study the effects of different Skyrme parametrizations on
thermodynamical properties of dense astrophysical matter, the neutron
star mass-radius relationship, and the core collapse of 15 and 40 solar
mass stars.

\end{abstract}



\maketitle

\section{Introduction}\label{sec:Intro}

Core-collapse supernovae (CCSNe) and neutron star (NS) mergers, the
birth places of neutron stars and black holes (BH), can only be
understood in the light of the microphysics that drives them.  A clear
picture of these astrophysical phenomena is directly tied to our
understanding of the properties of matter and radiation at high energy
densities.  Therefore, one of the essential microphysical ingredients
in computational simulations of these phenoma is the equation of state
(EOS) of dense matter (e.g., \cite{oertel:16,lattimer:16}).


An EOS for CCSNe and NS merger simulations must encompass a very large
range in density, temperature, and composition.  The temperatures
encountered in these events range from zero up to hundreds of MeV,
densities from $\lesssim10^4$ to $10^{15}\unit{g\,cm}^{-3}$, and
proton fractions $y$ may be close to zero or as high as 0.60.  Over
this wide parameter space, matter may be in a gas, liquid, or solid
phase, and in its ground state or in a highly-excited state
\cite{lattimer:91,lattimer:16,oertel:16}.


At low densities and temperatures, isospin symmetric matter with the
same number of protons and neutrons clusters into heavy nuclei.  By
making the system isospin asymmetric, \ie having an excess of neutrons
with respect to protons or vice-versa, nuclei become neutron or proton
rich.  If the isospin asymmetry is large enough, nucleons drip out of
nuclei to form a background gas.  Keeping proton fraction and density
constant, heavy nuclei split into lighter ones as the temperature is
increased.  At very high temperatures all nuclei dissociate and only a
gas of free nucleons immersed in the electron and photon gas exists.
If, instead, composition and temperature are kept constant as density
is increased, nuclei become more and more packed.  Just below nuclear
saturation density, a series of phase transitions in which nucleons
arrange themselves into complex shapes known as nuclear ``pasta''
occurs \cite{ravenhall:83b,hashimoto:84}.  At even higher densities,
nucleons form a free gas and the EOS stiffens due to short range
nuclear repulsive forces.  The EOS may soften at densities much higher
than nuclear saturation density due to the appearance of heavier
leptons, hyperons, kaon condensates, or a quark-gluon plasma
\cite{prakash:97, steiner:01, pons:00}.  We do not consider these
phases in the present work.

  %

The EOS is poorly constrained in regions of parameter space relevant
for CCSNe and NS mergers, as matter in these sites is under extreme
conditions that cannot be easily reproduced in laboratory experiments.
Hence, any EOS built for astrophysical applications depends on
extrapolations based on theoretical models of microscopic interactions
as well as astrophysical and experimental inputs.  Ideally, these
models should be supported by available nuclear experimental data
\cite{dutra:12,dutra:14} and make predictions that fulfill known
astrophysical \cite{demorest:10, nattila:16, lattimer:12,
  ozel:16, lattimer:01, lim:15} and theoretical constraints
  \cite{tews:16,hebeler:13,hebeler:10}. 

Broadly, there are two approaches for generating nuclear interactions
used in calculating the properties of dense matter. Phenomenological
interactions employ reasonable forms for the nuclear interaction and
fit the force parameters to the measured properties of nuclei and
other constraints from laboratory experiments \cite{jose:11,
  meisel:16, wiescher:12,schatz:16} and astrophysical observations of
NSs \cite{steiner:13}. Since they are constrained mainly by
observations of nearly isospin-symmetric systems accessible in the
laboratory, extrapolating to the highly isospin-asymmetric matter
encountered in NS mergers and CCSNe introduces significant uncertainties. 
More microscopic treatments, such as chiral effective field theory, use
interactions that obey symmetries of QCD and fit the small number
of free parameters in the interaction based on observed properties of
the nucleon-nucleon interaction \cite{machleidt:11}. These approaches are very
accurate at near nuclear saturation density and below  and should capture the
properties of highly asymmetric matter, but difficult to calculate higher-order
interactions become increasingly important with increasing density
\cite{rrapaj:16}.

Another source of uncertainty, in addition to the uncertainty in the form of the
effective nuclear interaction, comes from the many-body
techniques used to predict the thermodynamic properties of an ensemble
of nucleons. For a given 
effective nuclear interaction, the properties of a system of nucleons
can be calculated exactly using modern quantum many-body techniques
\cite{bogner:10, gandolfi:09}, but such calculations are too expensive
to cover the wide range of conditions required for an astrophysical
EOS \cite{sagert:16,gandolfi:09}. Because of their relative simplicity
and their ability to capture the properties of nuclear matter near
saturation density, phenomenological interactions combined with
mean-field techniques are often applied when calculating astrophysical
EOSs \cite{lattimer:91, shen:98, steiner:13}. For instance,
phenomenological Skyrme models assume a zero-range effective
interaction that, in the mean field
approximation\footnote{The mean field approximation
    assumes that nucleons only interact with other nucleons through
    the average field produced by all of the other nucleons, removing
    the possibility of any correlations from the system.}, results in
a parametrized energy functional that can be fit to measured
properties of nuclei \cite{stone:07}.

Non-uniform phases of matter appear at low temperatures and
sub-saturation densities, with a high-density phase near nuclear
saturation density and a low density nucleon gas phase. The treatment
of these non-uniform phases is another source of uncertainty in a 
high-density astrophysical EOS.  A number of different approaches for 
treating the non-uniform phases have been used in previous work (in 
addition to using different treatments of uniform nuclear matter).
For a review, see Oertel \etal \cite{oertel:16} and references
therein.

One often used approach is to treat the non-uniform nuclear matter 
using the single nucleus approximation (SNA)
\cite{baym:71, lamb:78, lamb:81, lattimer:85, lattimer:91, shen:98,
  shen:98b, shen:11b, lim:12}. The SNA assumes that there is one
representative nucleus (or, more generally, a high density structure
such as a pasta phase) and calculates its properties from equilibrium
conditions within a spherical Wigner-Seitz cell, possibly including
surface, Coulomb, and translational energy corrections using either a
liquid drop or a Thomas-Fermi model for the surface corrections. At very
low temperature this should be a good approximation, but at
intermediate temperatures an ensemble of nuclei is likely to be
present.

A second approach is to use a nuclear statistical equilibrium
(NSE)-like description of nuclei along with Coulomb corrections and
exclude the low density gas from regions inside the nuclei
\cite{blinnikov:11, hempel:10, hempel:12, steiner:13, furusawa:11,
  furusawa:13, furusawa:17}.  This gives a more reasonable
distribution of nuclei at finite temperature and the excluded volume
approximation makes nuclei naturally dissappear just below saturation
density. However, such approaches cannot easily incorporate the
presence of nuclear pasta and may have trouble including the exotic
nuclei formed at very high density.  Additionally, a number of works
have used a hybrid approach where NSE is used at low density and the
SNA is used closer to nuclear saturation density \cite{hillebrandt:84,
  arnett:85, shen:10, shen:10b, shen:11}.


Motivated by the need for a wide array of finite-temperature nuclear
EOSs consistent with experimental and observational constraints, we
build an open-source code to construct EOSs for wide range of Skyrme
interactions available in the literature and use the code to generate
EOS tables using broad range of Skyrme interactions.
For the inhomogeneous phase, we follow the open-source model of Lattimer
\& Swesty \cite{lattimer:91} (hereafter referred to as \LSn; available
at \url{http://www.astro.sunysb.edu/dswesty/lseos.html}) at high
density and transition to an NSE model at low-density. We extend the
\LS model to include non-local isospin asymmetric terms, treat the
size of heavy nuclei consistently, and include an improved method to
treat nuclear surfaces. Rather than use a Gibbs construction to go
from inhomogeneous to homogeneous nuclear matter, we simplify the
treatment and choose either the uniform or non-uniform phase based on 
which has a lower free energy, which sets the phase transition to be 
first order. Additionally, the algorithm used in the new EOS code
converges across a much wider range of temperature, density,
composition space than the original L\&S code.  At very high
densities, we allow for additional terms in the Skyrme parametrization
that can be used to stiffen the high-density EOS while leaving the
saturation-density EOS essentially unchanged.  This allows one to use
a specific Skyrme parametrization that agrees with well determined
nuclear matter constraints, but varies the maximum NS mass.

We thoroughly test our new EOSs to ensure thermodynamic consistency.
Using the LS220 parametrization, we find excellent agreement with the
original work of L\&S. We present zero-temperature NS mass-radius
relations for all considered Skyrme parametrizations and demonstrate
how the new high-density adjustments translate to NS structure.
Finally, we employ the open-source general-relativistic \texttt{GR1D}
code \cite{oconnor:10,oconnor:11,oconnor:15} to carry out
spherically-symmetric core collapse and postbounce supernova
simulations with our new EOSs. We consider $15$ solar-mass and $40$
solar-mass presupernova stars and follow the $40$ solar-mass
simulations to black hole formation. Burrows \& Lattimer 
\cite{burrows:84} argued that thermodynamic quantities
obtained in the SNA approximation differ little from the general
case. Our simulations show that the small differences between SNA
and NSE low-density treatments translate to mild variations in the
inner core's collapse time to core bounce and in the postbounce
accretion rate. There is, however, little impact on the overall
postbounce evolution and black hole formation.

The remainder of this paper is organized as follows.  In Sections 
\ref{sec:Formalism} and \ref{sec:Solutions} we review, respectively, the 
formalism to obtain the EOS in the SNA and the methodology to solve the 
system of equations that minimize the free energy of nuclear matter. 
We compare the results of our code to those of \LS in Section 
\ref{sec:Results}.  In the same section, we compare EOSs obtained from 
different Skyrme parametrizations, as well as nuclear matter
properties obtained for selected Skyrme parametrizations.  In Section
\ref{sec:NS}, we study properties of cold NSs obtained for different Skyrme
parametrizations and the effects to NS mass-radius curve obtained from
adding extra stiffening terms to the EOS. We then briefly study adiabatic
compression of nuclear matter in Section \ref{sec:Compression} and spherical
core collapse of $15$ and $40$ solar mass stars in Section
\ref{sec:Collapse}. Due to its relevance for core collapse, in Section 
\ref{sec:NSE}, we discuss an implementation of an NSE EOS and a method to 
transition from the SNA EOS at high densities to NSE at low densities. 
Finally, we conclude in Section \ref{sec:Conclusions}.
In the appendices, we provide details left out in the main text, including 
the lepton and photon contributions and details on our NSE treatment. The 
EOS source code and example EOS tables are available at 
\url{https://stellarcollapse.org/SROEOS}.

Throughout this paper, we use the convention of measuring temperature 
in $\mathrm{MeV}$, setting the Boltzmann constant $k_B = 1$ 
unless otherwise explicitly mentioned.
For thermodynamic quantities, we use upper case letters when refering 
to quantities per volume and lower case letters for specific (per 
baryon or per mass) quantities. Furthermore, we define the zero point 
of the specific internal energy based on the free neutron rest mass 
$m_n$, set the neutron and proton masses $m_n$ and $m_p$ 
to their experimental values unless otherwise noted, and explicitly 
include the neutron--proton mass difference where necessary. 
Finally, we use the neutron mass $m_n$ to convert from number density
(fm$^{-3}$) to rest-mass density (g\,cm$^{-3}$).

\section{Single Nucleus Approximation Formalism}\label{sec:Formalism}

Here, we describe the formalism we use for determining a
self-consistent EOS from a given Skyrme parametrization across a wide
range of density $n$, temperature $T$ and proton fraction $y$. At high
density, our model closely follows L\&S \cite{lattimer:91}, while at
lower densities we employ an NSE EOS, which we describe in Section
\ref{sec:NSE}. We assume that the medium contains neutrons, protons,
alpha particles, electrons, positrons, and photons.  The electrons,
positrons, and photons are treated as uniform free gases and charge
neutrality is assumed, so that the number of electrons per baryon is
equal to the number of protons per baryon.  Electron/positron and
photon contributions are discussed in detail in Appendix
\ref{app:ele}.  In what follows, \emph{nucleonic matter} refers to a
bulk system of protons and neutrons with uniform density.  We use
\emph{uniform matter} to refer to a free gas of nucleons and alpha
particles, while we use \emph{non-uniform matter} to describe matter
including heavy nuclei.
 
The possible presence of heavy nuclei or pasta-like phases at high
density is treated via the single nucleus approximation (SNA), which
is essentially a two-phase construction including surface effects.  In
this construction, each heavy nucleus occupies a volume $V_N$ inside a
Wigner-Seitz cell of volume $V_{\mathrm{cell}}$. We define the volume
fraction occupied by heavy nuclei as $u=V_N/V_{\mathrm{cell}}$.  In
the interior of the heavy nucleus, nucleonic matter is assumed to have
a constant density ($n_i$) and proton fraction ($y_i$), and have
thermodynamic properties determined from a Skyrme interaction in the
mean field approximation.  In each cell, nuclei are surrounded by a
free gas of nucleons and alpha particles that occupy a volume
$V_{\mathrm{cell}}-V_N$.  The alpha particles have density $n_\alpha$
and are assumed to be hard spheres of volume $v_\alpha=24 \,
\textrm{fm}^{-3}$ \citep{lattimer:85} that exclude nucleons, so that
they occupy a fraction $n_\alpha v_\alpha$ of the exterior
volume. This leaves a fraction of the total cell volume $u_o = (1 -
u)(1 - n_\alpha v_\alpha)$ for the exterior nucleons.  They have a
density $n_o$ and a proton fraction $y_o$ in this volume.  The
nucleons in the exterior portion of the cell are treated using the
same Skyrme interaction as the material inside the nucleus. With these
definitions, we can write the total baryon and proton number densities as
\begin{subequations}\label{eq:const5}
 \begin{align}
  n&=un_i+(1-u)[4n_\alpha + n_o (1-n_\alpha v_\alpha)]\,,\\
  ny&=un_iy_i+(1-u)[2n_\alpha + n_o y_o (1-n_\alpha v_\alpha)]\,.
 \end{align}
\end{subequations}
When $u \rightarrow 0$, uniform matter consisting of neutrons, protons, and
alpha particles is recovered. All of the material is assumed to be in thermal
equilibrium and it is therefore characterized by a single temperature $T$.

The Helmholtz free energy of the system, from which all other thermodynamical
quantities may be derived, is the sum of free-energies of the individual
components, that is,
\begin{equation}\label{eq:free}
F = F_o + F_\alpha + F_h + F_e + F_\gamma\,,
\end{equation}
where $F_o,\,F_\alpha,\,F_h,\,F_e$, and $F_\gamma$ are, respectively,
the free-energy densities of the nucleon gas outside the heavy nuclei,
alpha particles, nucleons clustered into heavy nuclei, electrons and
positrons, and photons. The free energies of the leptons and photons
are simply those of arbitrarily degenerate and relativistic free gases
(see Appendix~\ref{app:ele} for details).  The alpha particles are
treated as a free Boltzmann gas present only in the exterior volume,
so that their contribution to the free energy is given by
\begin{equation}\label{eq:falpha}
F_\alpha = (1-u) n_\alpha (\mu_\alpha - B_\alpha - T)\,,
\end{equation}
where $B_\alpha$ is the alpha particle binding energy\footnote{Unless otherwise
noted, we set the binding energy of the alpha particles to the experimentally
measured value, $B_\alpha=30.887\unit{MeV}$.  This follows the discussion of
Horowitz and Schwenk \cite{horowitz:06} that noticed that \LS set particle
energies with respect to the neutron vacuum mass, but did not include the
neutron-proton mass difference in their calculations for the alpha particle
binding energy.}. The alpha particle chemical potential is
\begin{equation}\label{eq:mu_alpha}
\mu_\alpha = T \ln \left( \frac{n_\alpha}{8 n_Q} \right)\,,
\end{equation}
where $n_Q = (m_n T/2 \pi \hbar^2)^{3/2}$.
The exterior nucleon contribution to the free energy is
\begin{equation}
F_o = u_o n_o f_B(n_o, y_o, T)\,,
\end{equation}
where $f_B$ is the specific free energy of a bulk nucleon gas
(nucleonic matter), which is assumed to come from a particular model for the
properties of bulk nuclear matter. In this work, we assume that bulk
nuclear matter is described by Skyrme interactions in the mean field
approximation as is discussed in the Section~\ref{sec:bulk}.

The free energy density of the heavy nuclei is further decomposed as
\begin{align}\label{eq:Fheavy}
 F_h&=F_i+F_S+F_C+F_T,
\end{align}
where $F_i,\,F_S,\,F_C$, and $F_T$ are, respectively, the free energy densities
due to the assumed interior bulk nucleon gas, surface effects, Coulomb forces, 
and bulk translational motion of the heavy nuclei.  The free energy density of 
bulk nucleons inside nuclei is $F_i=u_in_if_i$, where $f_i\equiv f_B(n_i,y_i,T)$
and $u_i$ is the total heavy nuclei volume. Ignoring the surface volume, $u_i=1-u_o$. 
If the surface, Coulomb, and translational contributions to the free energy are
neglected, we would arrive at a Gibbs two phase construction. These
finite size contributions are important for recovering a
semi-realistic description of nuclei and the pasta phases and we
describe the models we use for them below in
Sections~\ref{ssec:surface} -- \ref{ssec:translation} after discussing
bulk nuclear matter in the next Section~\ref{sec:bulk}.

\subsection{Bulk nuclear matter}
\label{sec:bulk}

Assuming a Skyrme type interaction in the mean field approximation,
the internal energy density $E_B$ of nucleonic matter with density $n$,
proton fraction $y$, and temperature $T$ can be written in the 
form\footnote{In principle, contributions of
spin-orbit and Coulomb interaction terms should also be included 
in the equation for the internal energy $E_B$. However, since they 
constitute only a small portion of the total energy, we neglect them.}
\begin{align}
\label{eq:Skyrme}
 E_B(n,y,T)=&\frac{\hbar^2\tau_n}{2m_n^*}+\frac{\hbar^2\tau_p}{2m_p^*}
 +\left(a+4by(1-y)\right)n^2\nonumber\\
 &+\sum_{i}\left(c_i+4d_iy(1-y)\right)n^{1+\delta_i}-yn\Delta\,,
\end{align}
where $a$, $b$, $c_i$, $d_i$, and $\delta_i$ are parameters of the
Skyrme force and $\tau_t$ ($t \in \{n,p\}$) are the kinetic energy
densities of neutrons and protons.  We include in Equation
\eqref{eq:Skyrme} a summation over index $i$ in the fourth term as
introduced by Agrawal~\emph{et al.}~\cite{agrawal:06}.  The first two
right-hand side terms represent the non-relativistic kinetic energy
density of neutrons $n$ and protons $p$, respectively.  The term
proportional to $n^2$ represents two-body nucleon interactions while
the terms proportional to $n^{1+\delta_i}$ approximate the effects of
many-body or density dependent interactions.  The last right-hand size
term includes the mass difference between neutrons and protons
$\Delta=m_n-m_p$, since we measure all energies relative to the free
neutron rest mass $m_n$.  The kinetic energy terms depend on the
density-dependent effective nucleon masses $m^*_t$ given by
\begin{equation}\label{eq:mstar}
 \frac{\hbar^2}{2m_t^*}=\frac{\hbar^2}{2m_t}+\alpha_1 n_{t}+\alpha_2n_{-t}\,.
\end{equation}
Here, $m_t$ is the vacuum nucleon mass and $-t$ denotes the opposite isospin of
$t$. The quantities $\alpha_1$ and $\alpha_2$ are also parameters of the
model. Additional terms that mix the neutron and proton densities in Equation
\eqref{eq:mstar}, as used by Chamel \etal \cite{chamel:09}, are omitted here.

The temperature dependence of the nuclear force is implicitly included in the
$\tau_t$ term,
\begin{equation}\label{eq:taut}
 \tau_t=\frac{1}{2\pi^2}\left(\frac{2m_t^*T}{\hbar^2}\right)^{\frac{5}{2}}\mathcal{F}_{3/2}(\eta_t)\,,
\end{equation}
where the Fermi integral $\mathcal{F}_k(\eta)$ is given by
\begin{equation}\label{eq:Fermi}
\mathcal{F}_k(\eta)=\int_0^{\infty}\frac{u^kdu}{1+\exp(u-\eta)}\,.
\end{equation}
The Fermi integral is a function of the degeneracy parameter
\begin{equation}\label{eq:eta}
 \eta_t=\frac{\mu_t-V_t}{T}\,.
\end{equation}
Here, $\mu_t$ is the nucleon chemical potential and $V_t$ is the single-particle potential,
\begin{equation}\label{eq:V}
 V_t\equiv\left.\frac{\delta E_B}{\delta n_t}\right|_{\tau_t,\tau_{-t},n_{-t}}\,.
\end{equation}
The degeneracy parameter $\eta_t$ can be obtained from the nucleon
density and temperature by inverting the relation
\begin{equation}\label{eq:nt}
 n_t=\frac{1}{2\pi^2}\left(\frac{2m_t^*T}{\hbar^2}\right)^{\frac{3}{2}}\mathcal{F}_{1/2}(\eta_t)\,.
\end{equation}
We obtain the Fermi integrals and their inverses using the routines provided by 
Fukushima~\cite{fukushima:15,fukushima:15b}.  These proved to be fast, accurate 
and  thermodynamic consistent. 

In Equations \eqref{eq:Skyrme} and \eqref{eq:mstar}, $a$, $b$, $c_i$, $d_i$,
$\delta_i$, $\alpha_1$, and $\alpha_2$ are parameters of the model that are
chosen to reproduce observables of infinite nuclear matter, an idealized system
of many nucleons interacting only through nuclear forces.  These parameters are
directly related to the more often used Skyrme parameters $x_j$ and $t_j$
($j \in \{0, 1, 2\}$), $t_{3i}$ and $\sigma_i$ through
\cite{dutra:12,steiner:05}
\begin{subequations}\label{eq:parameters}
\begin{align}
\label{eq:Skyrme_a}
a       &=\frac{t_0}{4}(1-x_0)\,,\\
\label{eq:Skyrme_b}
b       &=\frac{t_0}{8}(2x_0+1)\,,\\
\label{eq:Skyrme_c}
c_i     &=\frac{t_{3i}}{24}(1-x_{3i})\,,\\
\label{eq:Skyrme_d}
d_i     &=\frac{t_{3i}}{48}(2x_{3i}+1)\,,\\
\label{eq:Skyrme_delta}
\delta_i&=\sigma_i+1\,,\\
\label{eq:Skyrme_alpha1}
\alpha_1&=\frac{1}{8}\left[t_1(1-x_1)+3t_2(1+x_2)\right]\,,\\
\label{eq:Skyrme_alpha2}
\alpha_2&=\frac{1}{8}\left[t_1(2+x_1)+t_2(2+x_2)\right]\,.
\end{align}
\end{subequations}
For most Skyrme parametrizations, $t_{3i}$, $x_{3i}$, and $\sigma_{i}$
are only non-zero for a single value of $i$, which we set to $i=1$.
Even limiting $i$ to a single value $i=1$, we note that if we compare
Equation \eqref{eq:Skyrme} with Equation (2.8) of L\&S, we have an
extra term, the one proportional to the parameter $d_1$.  This term is
necessary to obtain the correct EOS whenever $x_{31}\ne-1/2$.  This is
the case for almost every Skyrme parametrization found in the
literature, albeit not for that of \LSn.

\LS obtain the parameters of the bulk internal energy density
(Equation~\ref{eq:Skyrme}) from experimentally determined values for
symmetric matter at saturation density $n_0$, the
binding energy $E_0$, the incompressibility $K_0$, and the symmetry
energy at saturation $J$.  We implement three different methods to
determine the parameters of Equation~\ref{eq:Skyrme}:
\begin{enumerate}
 \item Following \LSn, input experimental values for $n_0$, $E_0$,
   $K_0$, and $J$, which are then used to determine $a$, $b$, $c$ and
   $\delta$.  For consistency with L\&S, this assumes $d=0$ and
   $\alpha_1=\alpha_2=0$.
 \item Direct input of the parameters $a$, $b$, $c_i$, $d_i$, $\delta_i$,
   $\alpha_1$, and $\alpha_2$.
 \item Input Skyrme parameters $x_j$ and $t_j$ ($j \in \{0,1,2,3i\}$),
   and $\sigma_i$, which are used to determine $a$, $b$, $c_i$, $d_i$,
   $\delta_i$, $\alpha_1$, and $\alpha_2$ as shown in Equations
   \eqref{eq:parameters}.
\end{enumerate}
The last method has advantages over the first two. First, 
specifying only a few known nuclear experimental values to obtain the
Skyrme coefficients (as in \LSn) is unlikely to correctly predict
other well determined physical constraints. Also, direct input of the
parameters $a$, $b$, $c_i$, $d_i$, $\delta_i$, $\alpha_1$, and
$\alpha_2$ does not uniquely define the surface properties of nuclear
matter for a Skyrme parametrization, specifically the parameters
$\lambda$, $q$, and $\alpha$ discussed in Section~\ref{ssec:surface}.
On the other hand, input of the Skyrme parameters $x_j$, $t_j$, and
$\sigma_i$ makes it straightforward to determine the surface properties
of finite nuclei and to calculate nuclear matter properties that can
be directly compared with experiments. Finally, most studies on Skyrme
parametrizations in the literature explicitly give $x_j$, $t_j$, and
$\sigma_i$.

For completeness, we give the expressions for the bulk specific entropy
$s_B$, bulk specific free energy $f_B$, and bulk pressure
$P_B$ \cite{lattimer:78,lattimer:91}:
\begin{align}
\label{eq:sbulk} s_B&=\frac{1}{n}\sum_t\left(\frac{5\hbar^2\tau_t}{6m_t^*T}-n_t\eta_t\right),\\
\label{eq:fbulk} f_B&=E_B/n-Ts_B\,,\\
\label{eq:Pbulk} P_B&=\sum_tn_t\mu_t-nf_B\,.
\end{align}

\subsection{Nuclear surface}\label{ssec:surface}

For a given density $n$, proton fraction $y$, and temperature $T$,
nuclear matter may be uniform or phase separate into dense and dilute
phases that are in thermal equilibrium.  If the latter is the case,
there will be some energy stored in the surface between the two
phases.  \LS parametrize the nuclear surface free-energy density $F_S$
in terms of a surface shape function $s(u)$, a generalized nuclear
size $r$, and the surface tension per unit area $\sigma(y_i,T)$, which
is a function of the proton fraction $y_i$ in the dense phase and
temperature $T$.  The surface free energy density is written as
\cite{lattimer:91,lim:12}
\begin{equation}\label{eq:FS}
 F_S=\frac{3s(u)}{r}\sigma(y_i,T)\,.
\end{equation}
Both the generalized nuclear size $r$ and the surface shape function
$s(u)$ depend on the geometry of the heavy nuclei formed.  While at
low densities nuclei are spherical, as density increases and
approaches nuclear saturation density, nuclei may assume shapes such
as cylinders (think of pasta), slabs, cylindrical holes, and bubbles
(think of Swiss cheese) \cite{ravenhall:83b,hashimoto:84}, as well as
more exotic shapes \cite{nakazato:11,pais:12,schuetrumpf:15}.  In this
picture, the generalized nuclear size $r$ represents the radius of
spherical nuclei or bubbles, the radius of cylinders or cylindrical
holes, or the thickness of slabs. It is unclear what $r$ should be for
more exotic shapes.  Following \LSn, we do not consider specific
geometries for the heavy nuclei and simply determine $r$ by solving
the nuclear virial theorem, see Equation \eqref{eq:r} and the
discussion in Section \ref{ssec:Coulomb}.  The surface shape function
$s(u)$, meanwhile, is chosen as an interpolating function that
reproduces the low and high density limits for the shape of nuclei,
which are, respectively, spheres ($\lim_{u\rightarrow0}s(u)=u$) and
bubbles ($\lim_{u\rightarrow1}s(u)=1-u$).  The simplest choice for
this function is $s(u) = u ( 1 - u )$, which is what \LS
use\footnote{Note that this choice is not obvious in \LSn's paper.  It
  is, however, what is implemented in their source code.} and we adopt
here.

Following the prescription of \cite{lattimer:85,lim:12}, the surface
tension per unit area, $\sigma(y_i,T)$, is fitted by
\begin{equation}\label{eq:sigma}
 \sigma(y_i,T)=\sigma_s h\left(y_i,T\right)
 \frac{2\cdot2^{\lambda}+q}{y_i^{-\lambda}+q+(1-y_i)^{-\lambda}},
\end{equation}
where
$\sigma_s\equiv\sigma(0.5,0)$.
The function $h(y_i,T)$ contains the temperature dependence in the form
\begin{equation}\label{eq:h}
 h\left(y_i,T\right)=
 \begin{dcases}
    [1-({T}/{T_c(y_i)})^2]^p\,,& \mathrm{if\, } T\leq T_c(y_i)\,;\\
    0\,,              & \mathrm{otherwise}\quad.
\end{dcases}
\end{equation}
In Equations \eqref{eq:sigma} and \eqref{eq:h}, $\lambda$, $q$, and
$p$ are parameters to be determined (see below), while $T_c(y_i)$ is the
critical temperature for which the dense  and the dilute phases coexist.
The dense phase is assumed to have density $n_i$ and proton fraction $y_i$
while the dilute phase has density $n_o\leq n_i$ and proton fraction $y_o$.

To obtain the parameters $\lambda$, $q$, and $p$ and a functional form
for $T_c(y_i)$, we follow \cite{lim:12,lattimer:85,steiner:05} and study
the two phase equilibrium of bulk nucleonic matter.
For a given proton fraction $y$, there exists a critical temperature
$T_c$ and a critical density $n_c$ in which both the dense and dilute
phases have the same density $n_i=n_o$ and same proton fraction $y_i=y_o$
(cf.\ Figures (2.3) and (2.4) of \cite{lim:12}).  The quantities
$n_c$ and $T_c$ are obtained by simultaneously solving \cite{lim:12}
\begin{equation}\label{eq:coexistence}
 \left.\frac{\partial P_B}{\partial n}\right|_T=0
 \qquad \mathrm{and} \qquad
 \left.\frac{\partial^2P_B}{\partial n^2}\right|_T=0\,,
\end{equation}
for proton fractions $y\leq0.50$. Here, $P_B$ is the bulk pressure
given by Equation \eqref{eq:Pbulk}. Because we ignore Coulomb
contributions to the surface tension, the formalism presented in this
Section is almost symmetric under a $y\rightarrow1-y$ transformation.
The symmetry is only slightly broken by the small difference $\Delta$
in the neutron and proton rest masses, $m_n=m_p+\Delta$, which we
ignore here when considering $y > 0.5$.  Once the critical temperature
$T_c$ has been determined for a range of proton fractions $y$, we fit
it using the function
\begin{align}\label{eq:Tcrit}
 T_c(y)=T_{c0}\left[a_c+b_c\delta(y)^2+c_c\delta(y)^4+d_c\delta(y)^6\right],
\end{align}
where $T_{c0}\equiv T_{c}(y=0.5)$ is the critical temperature for
symmetric nuclear matter and $\delta(y)=1-2y$ is the neutron excess.

After determining $T_c(y)$, we compute the properties of semi-infinite
nucleonic matter, that is, matter for which the density varies along
one direction (the $z$ axis) and is constant in the remaining two.
Ignoring Coulomb effects, we assume that in the limits
$z\rightarrow\pm\infty$ matter saturates at densities $n_i$ and $n_o$
and proton fractions $y_i$ and $y_o$.
These two phases are in equilibrium if their pressures as well as their neutron
and proton chemical potentials are the same, \ie
\begin{align}\label{eq:phaseequilibrium}
 P_i=P_o\,, \quad \mu_{ni}=\mu_{no}\,, \quad\mathrm{and}\quad \mu_{pi}=\mu_{po}\,.
\end{align}
Equations \eqref{eq:phaseequilibrium} are solved simultaneously with
\begin{equation}\label{eq:yi}
 y_i=\frac{n_{pi}}{n_{ni}+n_{pi}}
\end{equation}
to obtain the neutron and proton densities of the high and low density phases
$n_{ni}$, $n_{pi}$, $n_{no}$, and $n_{po}$, respectively.

Once the neutron and proton densities of the two coexisting phases have been
calculated, we determine the surface shape that minimizes $\sigma(y_i,T)$.
Since we assume the system to be homogeneous across two dimensions, the surface
tension per unit area is given by
\cite{ravenhall:83,steiner:05}
\begin{align}\label{eq:tension}
 \sigma(y_i,T)=\int_{-\infty}^{+\infty}\bigg[&F_B(z)+E_S(z)+P_o\nonumber\\
                                      &-\mu_{no}n_n(z)-\mu_{po}n_p(z)\bigg]dz\,.
\end{align}
Here, $P_o$, $\mu_{no}$, and $\mu_{po}$ or, alternatively, $P_i$,
$\mu_{ni}$, and $\mu_{pi}$ are solutions to Equations
\eqref{eq:phaseequilibrium}.  Meanwhile, $F_B(z)=n(z)f_B(n(z),y(z),T)$
is the bulk free energy density across the $z$ axis, while $E_S(z)$ is
the spatially-varying contribution to the energy density of a
Skyrme-like Hamiltonian (see Equations 1--4 of Steiner \emph{et
  al.}~\cite{steiner:05}). It has the form
\cite{steiner:05,lattimer:85,ravenhall:83}
\begin{align}\label{eq:HS}
 E_S(z)=\frac{1}{2}\bigg[&q_{nn}\left(\boldsymbol{\nabla}n_n\right)^2
 +q_{np}\boldsymbol{\nabla}n_n\cdot\boldsymbol{\nabla}n_p\nonumber\\
 &+q_{pn}\boldsymbol{\nabla}n_p\cdot\boldsymbol{\nabla}n_n
 +q_{pp}\left(\boldsymbol{\nabla}n_p\right)^2\bigg]\,,
\end{align}
where $n_t\equiv n_t(z)$ ($t \in \{n,p\}$).
The parameters $q_{tt'}$ are related to the Skyrme coefficients by
\begin{subequations}
\begin{align}\label{eq:qtt}
 q_{nn}=q_{pp}&=\frac{3}{16}\left[t_1(1-x_1)-t_2(1+x_2)\right]\,,\\
 q_{np}=q_{pn}&=\frac{1}{16}\left[3t_1(2+x_1)-t_2(2+x_2)\right]\,.
\end{align}
\end{subequations}
As Steiner \etal point out \cite{steiner:05}, for Skyrme-type forces,
the $q_{tt'}$ are constants and the relations $q_{nn}=q_{pp}$ and
$q_{np}=q_{pn}$ are always true.  In the
general case, however, $q_{tt'}$ may be density dependent and $q_{nn}$
may be different from $q_{pp}$, though $q_{np}=q_{pn}$ is still
expected to hold.

To minimize Equation \eqref{eq:tension}, we assume that the neutron and
proton densities have a Woods-Saxon form, \ie
\begin{equation}\label{eq:WS}
 n_t(z)=n_{to}+\frac{n_{ti}-n_{to}}{1+\exp\left((z-z_t)/a_t\right)}\,,
\end{equation}
where $z_n$ and $a_n$ ($z_p$ and $a_p$) are, respectively, the neutron
(proton) half-density radius and its diffuseness \cite{woods:54}.
This form has the expected limits
$\lim_{z\rightarrow-\infty}n_t(z)=n_{ti}$ and
$\lim_{z\rightarrow+\infty}n_t(z)=n_{to}$.  Following References
\cite{steiner:05,lattimer:85,ravenhall:83}, we set the proton
half-density radius $z_p$ at $z=0$ and minimize the surface tension
per unit area with respect to the three other variables $z_n$, $a_n$,
and $a_p$.  This allows us to tabulate values of the surface tension
per unit area $\sigma(y_i,T)$ as a function of the proton fraction
$y_i$ of the dense phase and the temperature $T$ of the semi-infinite
system.  This is used to determine the parameters $\alpha$ and $q$ in
Equation \eqref{eq:sigma} and $p$ in Equation \eqref{eq:h} performing
a least squares fit.

It is worth mentioning that the surface free energy density should, in
general, include a contribution from the neutron skin
$\sigma\rightarrow\sigma+\mu_n\nu_n$, where $\nu_n$ is the neutron
excess \cite{ravenhall:83,lim:12}.  However, we follow \LSn, and
neglect this term.  In future work, this term should be included since
its effects are important for very neutron rich matter \cite{lim:12}.

\subsection{Coulomb energy}\label{ssec:Coulomb}

Following \LSn, we approximate the Coulomb free energy density using
the static Wigner-Seitz approximation,
\begin{equation}\label{eq:FC}
 F_C=\frac{4\pi\alpha_C}{5} (y_i n_i r)^2 c(u)\,.
\end{equation}
Here $\alpha_C$ is the fine structure constant, $y_i$ is the proton
fraction inside heavy nuclei, $n_i$ the nuclear density also inside
heavy nuclei, $r$ is the generalized nuclear size, and $c(u)$ is the
Coulomb shape function, discussed below.

In this model, only the surface and Coulomb energy densities depend on
the generalized nuclear size $r$.  Thus, minimizing the total energy
density with respect to the nuclear size $r$ implies that $F_S=2F_C$,
known as the nuclear virial theorem. With this, the generalized nuclear 
size becomes
\begin{equation}\label{eq:r}
 r=\frac{9\sigma}{2\beta}\left[\frac{s(u)}{c(u)}\right]^{1/3}\,,
\end{equation}
where
\begin{equation}\label{eq:beta}
 \beta=9\left[\frac{\pi\alpha_C}{15}\right]^{1/3}\left(y_in_i\sigma\right)^{2/3}\,,
\end{equation}
and $\sigma\equiv\sigma(y_i,T)$ is the surface tension per unit area discussed
in Section \ref{ssec:surface}.  Using the results of this section,
the surface and Coulomb energy densities may be combined in the form
\begin{equation}\label{eq:FSFC}
 F_S+F_C=\beta\left[c(u)s(u)^2\right]^{1/3}\equiv\beta\mathcal{D}(u)\,.
\end{equation}
This defines $\mathcal{D}(u)$ in terms of the surface
and Coulomb shape functions, $s(u)$ and $c(u)$, respectively.

As is the case for the surface shape function $s(u)$ discussed in
Section \ref{ssec:surface}, the function $c(u)$ is also chosen to
reproduce known physical limits
\cite{ravenhall:83,lattimer:91,lim:12}.  At low densities, nuclei are
spherical and the generalized nuclear size $r$ is the nuclear radius.
Considering the nuclei to occupy a small volume fraction of the
Wigner-Seitz cell, $u\simeq0$, the Coulomb shape function is given by
$\lim_{u\rightarrow0}c(u)=uD(u)$, where
$D(u)=1-\tfrac{3}{2}u^{1/3}+\tfrac{1}{2}u$ \cite{baym:71}.  Just below
nuclear saturation density, $u\simeq1$ and nuclei turn ``inside out''
and low-density spherical bubbles form inside an otherwise dense
nucleonic phase.  Here, the generalized nuclear size $r$ is the bubble
radius and $\lim_{u\rightarrow1}c(u)=(1-u)D(1-u)$ \cite{baym:71}.
Between these two limits matter may be more stable assuming
non-spherical shapes, such as cylindrical and planar geometries
\cite{ravenhall:83b,hashimoto:84}.  Using the results of Ravenhall
\etal \cite{ravenhall:83b} for the structures that minimize the energy
density of nucleonic matter with non-spherical geometries at zero
temperatures, \LS showed that the function $\mathcal{D}(u)$ is well
approximated by 
\begin{equation}
 \mathcal{D}(u)=u(1-u)\frac{(1-u)D(u)^{1/3}+uD(1-u)^{1/3}}{u^2+(1-u)^2+0.6u^2(1-u)^2}\,\,,
\end{equation}
again with $D(u)=1-\tfrac{3}{2}u^{1/3}+\tfrac{1}{2}u$.
For simplicity, we make the same choice in our implementation.

\subsection{Translational energy}\label{ssec:translation}

Assuming that the heavy nuclei form a non-degenerate and
non-relativistic Boltzmann gas with no internal degrees of freedom
that is free to move within a Wigner-Seitz cell, we have \cite{lim:12}
\begin{equation}\label{eq:FT}
 F_T=\frac{u(1-u)n_i}{\bar{A}}h(y_i,T)\left(\mu_T-T\right)\,,
\end{equation}
where
\begin{equation}\label{eq:muT}
 \mu_T=T\log\left(\frac{u(1-u)n_i}{n_Q\bar{A}^{5/2}}\right)
\end{equation}
is the chemical potential of heavy nuclei with $n_Q=(m_nT/2\pi\hbar^2)^{3/2}$.
Here
\begin{equation}\label{eq:barA}
 \bar{A}=\frac{4\pi n_i r^3}{3}
\end{equation}
is the mass number of the representative heavy nucleus.  One
difference between our treatment and \LS is that they choose to set a
fixed value for $\bar{A}=60$ in the translational energy calculation.
We, on the other hand, compute the value of the heavy nucleus mass
number $\bar{A}$ and the translational energy $F_T$ self-consistently.
In order to guarantee that the translational free energy $F_T$ also
vanishes at the critical temperature $T_c(y_i)$, as is the case for
the surface tension, we set $F_T$ to be proportional to the
function $h(y_i,T)$ \cite{lattimer:85} (see Equation~\ref{eq:h}). 

Also, note that the heavy nuclei of course have internal degrees of 
freedom. These are accounted for in $F_i$ (see Section~\ref{sec:bulk}).

\section{Solving the EOS}\label{sec:Solutions}

The model free energy described in Section \ref{sec:Formalism} depends
upon the variables $u$, $r$, $n_{i}$, $y_{i}$, $n_{no}$, $n_{po}$,
$n_\alpha$, and $T$.  In thermodynamic equilibrium, the system will
assume a state in which the free energy is minimized with respect to
these variables, subject to the constraints of fixed baryon density,
proton fraction, and temperature.

Our procedure is to search for extrema in the free energy surface, which is done
by setting the derivatives of the free energy to zero and using standard root
finding algorithms to find solutions to the resulting system of equations.
First, we reduce the number of variables by using Equations \eqref{eq:const5} to
express $n_{no}$ and $n_{po}$ in terms of the other variables and automatically
obey baryon number and charge conservation.  We then carry
out minimization with respect to five independent variables: $r$, $n_i$, $y_i$,
$u$, and $n_\alpha$. Minimization with respect to $r$ results in the constraint
given by Equation \eqref{eq:r}.  Setting the derivative of $F$ with respect to
$n_\alpha$ equal to zero gives
\begin{equation}
\label{eq:alpha_constraint}
\mu_\alpha = 2(\mu_{no} + \mu_{po}) + B_\alpha - P_o v_\alpha\,,
\end{equation}
which is just a condition for alpha particles in chemical equilibrium with the
exterior protons and neutrons with an excluded volume correction. The
derivatives with respect to the interior densities and the volume fraction give
the constraints
\begin{subequations}\label{eq:solve3}
 \begin{align}
 A_1 &= P_i-B_1-P_o-P_\alpha =0\,,\\
 A_2 &= \mu_{ni}-B_2-\mu_{no}=0\,,\\
 A_3 &= \mu_{pi}-B_3-\mu_{po}=0\,.
 \end{align}
\end{subequations}
In Equations \eqref{eq:solve3}, we use the quantities
\begin{subequations}\label{eq:B}
 \begin{align}
 B_1 &=\frac{\partial\hat{F}}{\partial u}-\frac{n_i}{u}\frac{\hat{F}}{\partial n_i}\,,\\
 B_2 &=\frac{1}{u}\left[\frac{y_i}{n_i}\frac{\partial\hat{F}}{\partial y_i}
       -\frac{\partial\hat{F}}{\partial n_i}\right]\,,\\
 B_3 &=-\frac{1}{u}\left[\frac{1-y_i}{n_i}\frac{\partial\hat{F}}{\partial y_i}
       +\frac{\partial\hat{F}}{\partial n_i}\right]\,,
 \end{align}
\end{subequations}
where $\hat{F}=F_S+F_C+F_T$. The derivatives of $F_S$, $F_C$, and $F_T$
with respect to the variables $u$, $n_i$, and $y_i$ are readily obtained
from Equations \eqref{eq:FS}, \eqref{eq:FC}, and \eqref{eq:FT}, respectively.
This system of equations can then be solved to find the equilibrium values of
the independent variables for fixed $n$, $y$, and $T$. These, in turn, can be
used to calculate the pressure, entropy, and other thermodynamic quanties
required by simulation codes.

We solve this system of non-linear equations by first using Equations
\eqref{eq:r} and \eqref{eq:alpha_constraint} to explicitly find
$n_\alpha$ and $r$. We then search for solutions to the three
remaining contstraint equations using the independent variables
$\zeta=[\log_{10}(u), \log_{10}(n_{no}), \log_{10}(n_{po})]$ and the
root finding routines provided by \cite{hasselman:16}. Solving the
system of equations requires initial guesses for the independent
variables $\zeta$. Often, an initial choice of $\zeta$ may not result
in convergence of the root finding algorithm. Therefore, we perform an
extensive search of possible initial guesses when the root finding
algorithm fails, which allows us to gain convergence over a wider
range of thermodynamic conditions than the original implementation of
\LSn. Since we are building tables, rather than using the EOS code
directly in simulations, the increased computational expense is not
burdensome.

In some regions of parameter space, uniform matter has a lower free
energy than the non-uniform phase and is therefore the favored
state. In uniform matter, $u=0$ and the free energy has to be
minimized with respect to $n_{no}$, $n_{po}$, and $n_\alpha$, since
the portion of the free energy that depends on $r$, $n_i$, and $y_i$
is multiplied by $u$.  Therefore, the properties of uniform matter can
be found by solving Equation~\eqref{eq:alpha_constraint} subject to
the neutron and proton number conservation constraints.

A significant difference between our EOS and that of
\LS is our treatment of the transition between uniform and
non-uniform mattter. \LS assume a continuous transition between
uniform and non-uniform matter that is obtained using a Maxwell
construction.  In this picture, two phases with densities $n_h$ and
$n_l$, where $n_h>n_l$, are in thermal and chemical equilibrium with
each other.  The uniform higher density phase occupies a volume
fraction $v=(n-n_l)/(n_h-n_l)$ of the system, while the non-uniform
lower density phase occupies a volume fraction $(1-v)$.  Hence, the
free energy density in the boundary between both phases is
\begin{equation}
 F(n,n_l,n_h,y,T)=vF_h(n_h,y,T)+(1-v)F_l(n_l,y,T)\,,
\end{equation}
and the equilibrium conditions used to obtain $n_h$ and $n_l$ are
\begin{equation}
 \left.\frac{\partial F}{\partial n_l}\right|_{n_h,n,y,T}=
 \left.\frac{\partial F}{\partial n_h}\right|_{n_l,n,y,T}=0\,.
\end{equation}

Instead of using the \LS procedure,
we determine what type of solutions may exist (uniform, non-uniform, or
both) and solve the necessary system of equations.  If only one of the
systems has a physical solution then that is assumed to be the most
stable configuration of nuclear matter.  If both systems have
solutions we choose the one with the lowest free energy as the
favorable solution.  This assumes that the transition from uniform to
non-uniform matter is first order and, therefore, there is no
coexistent phase as assumed by \LS and no need for a Maxwell
construction.

We note that there are rare cases where non-uniform matter has lower
free energy density than uniform matter, but we still set the latter
as the favorable configuration.  We make this choice whenever the
adiabatic index
\begin{equation}\label{eq:gamma}
 \Gamma=\left.\frac{\mathrm{d}\log P}{\mathrm{d}\log n}\right\vert_s
\end{equation}
of non-uniform matter is negative, implying an unphysical imaginary
speed of sound.  This occurs rarely and typically at intermediate
proton fraction $y\sim0.20$ to $0.35$, high density $n\sim0.08$ to
$0.11\unit{fm}^{-3}$, and low temperatures $T\lesssim0.5\unit{MeV}$.
In these cases, uniform and non-uniform matter have very similar
free-energy densities and, therefore, we do not expect that choosing
the phase with slightly higher free energy density will affect the EOS
significantly.

\begin{table*}[htbp]
\caption{\label{Tab:parametrizations} Parameters of the considered
  Skyrme interactions with the exception of L\&S, for which there are
  multiple ways to set $t_1, t_2, x_1$, and $x_2$ to achieve $\alpha_1
  = \alpha_2 = 0$ in Equations \eqref{eq:Skyrme_alpha1} and
  \eqref{eq:Skyrme_alpha2}.  Here, $t_0$ is in \unit{MeV\,fm^{3}},
  $t_1$ and $t_2$ are in \unit{MeV\,fm^{5}}, $t_{31}$ is in
  \unit{MeV\,fm}$^{3+3\sigma_1}$, and $x_0$, $x_1$, $x_2$, $x_{31}$, and
  $\sigma_1$ are dimensionless. See References
  \cite{stone:07,dutra:12,stone:03} for general discussions of these
  parameters.}
\begin{ruledtabular}
\begin{tabular}{l  D{.}{.}{5.2} D{.}{.}{4.2} D{.}{.}{4.2} D{.}{.}{5.1} D{.}{.}{1.5} D{.}{.}{1.5} D{.}{.}{1.5} D{.}{.}{1.5} D{.}{.}{1.5}}
\multicolumn{1}{c}{Parametrization}&
\multicolumn{1}{c}{$t_0$}&
\multicolumn{1}{c}{$t_1$}&
\multicolumn{1}{c}{$t_2$}&
\multicolumn{1}{c}{$t_{31}$}&
\multicolumn{1}{c}{$x_0$}&
\multicolumn{1}{c}{$x_1$}&
\multicolumn{1}{c}{$x_2$}&
\multicolumn{1}{c}{$x_{31}$}&
\multicolumn{1}{c}{$\sigma_1$}\\
\hline
KDE0v1 \cite{agrawal:05}
& -2553.08 & 411.69  & -419.87  & 14063.61  &  0.6483  & -0.3472   & -0.9268  & 0.9475    & 0.1673 \\
LNS \cite{cao:06}
& -2484.97 & 266.735 & -337.135 & 14588.2   &  0.06277 &  0.65845  & -0.95382 & -0.03413  & 0.16667 \\
NRAPR \cite{steiner:05}
& -2719.70 & 417.64 & -66.687   & 15042.0   & 0.16154  & -0.047986 &  0.02717 & 0.13611   & 0.14416 \\
SKRA \cite{rashdan:00}
& -2895.4  & 405.5  & -89.1     & 16660.0   & 0.08     &  0.0      &  0.2     & 0.0       & 0.1422 \\
SkT1 \cite{tondeur:84}
& -1794.0  & 298.0  & -298.0    & 12812.0   & 0.154    & -0.5      & -0.5     & 0.089     & 0.33333 \\
Skxs20 \cite{brown:07}
& -2885.24 & 302.73 & -323.42   & 18237.49  & 0.13746  & -0.25548  & -0.60744 & 0.05428   & 0.16667 \\
SLy4 \cite{chabanat:98}
& -2488.91 & 486.82 & -546.39   & 13777.0   & 0.834    & -0.344    & -1.0     & 1.354      & 0.16667 \\
SQMC700 \cite{guichon:06}
& -2429.10 & 370.97 & -96.67    & 13773.42  & 0.10     &  0.0      &  0.0     & 0.0        & 0.16667 \\
\end{tabular}
\end{ruledtabular}
\end{table*}

\begin{table*}[t]
\caption{\label{Tab:bulk} Properties of nuclear matter calculated for
  the considered Skyrme interactions. $n_0$ (in fm$^{-3}$) is the
  saturation density of symmetric nuclear matter (SNM) and
  $\epsilon_0$ (in $\mathrm{MeV}\,\mathrm{baryon}^{-1}$) is the
  binding energy of SNM at $n_0$. Given in
  $\mathrm{MeV}\,\mathrm{baryon}^{-1}$ are the incompressibility
  $K_0$, the skewness $K'$, the symmetry energy parameters $J$, $L$,
  $K_\mathrm{sym}$, and $Q_\mathrm{sym}$, and the volume part of the
  isospin incompressibility $K_{\tau,v}$.  $M^*_n / m_n$ is the
  dimensionless ratio of the neutron effective mass to the neutron
  rest mass in SNM at $n_0$ and $\Delta M^*$ (in MeV) is the
  proton--neutron effective mass difference in SNM at $n_0$. Also
  given in MeV is the critical temperature $T_c$ for two-phase
  coexistence.  Small deviations between the LS220 results listed here
  and the original results of \LS are due to differences in the
  employed proton masses, the inclusion of the neutron proton mass
  difference (see the discussion in Section \ref{ssec:comparison}) and
  from calculating the symmetry energy expansion parameters explicitly
  from the derivatives of $\epsilon_B$ and not from the 
  difference in $\epsilon_B$ between SNM and pure neutron matter.
  }
\begin{ruledtabular}
\begin{tabular}{l  D{.}{.}{1.4} D{.}{.}{3.2} D{.}{.}{3.2} D{.}{.}{3.2} D{.}{.}{2.2} D{.}{.}{2.2} D{.}{.}{3.2} D{.}{.}{3.2} D{.}{.}{3.2} D{.}{.}{1.3} D{.}{.}{1.4} D{.}{.}{2.2} }
\multicolumn{1}{c}{Parametrization}&
\multicolumn{1}{c}{$n_0$}&
\multicolumn{1}{c}{$\epsilon_0$}&
\multicolumn{1}{c}{$K_0$}&
\multicolumn{1}{c}{$K'$}&
\multicolumn{1}{c}{$J$}&
\multicolumn{1}{c}{$L$}&
\multicolumn{1}{c}{$K_{\mathrm{sym}}$}&
\multicolumn{1}{c}{$Q_{\mathrm{sym}}$}&
\multicolumn{1}{c}{$K_{\tau,\nu}$}&
\multicolumn{1}{c}{$M_n^*/m_n$}&
\multicolumn{1}{c}{$\Delta M^*$}&
\multicolumn{1}{c}{$T_c$}\\
\hline
LS220 \cite{lattimer:91}        & 0.1549 & -16.64 & 219.85 & 410.80 & 28.61 & 73.81 & - 24.04 &  96.17 & -328.97 & 1.000 & 1.2933 & 16.80 \\
KDE0v1 \cite{agrawal:05}        & 0.1646 & -16.88 & 227.53 & 384.83 & 34.58 & 54.70 & -127.12 & 484.44 & -362.79 & 0.744 & 0.7166 & 14.85 \\
LNS \cite{cao:06}               & 0.1746 & -15.96 & 210.76 & 382.50 & 33.43 & 61.45 & -127.35 & 302.52 & -384.45 & 0.826 & 0.8821 & 14.92 \\
NRAPR \cite{steiner:05}         & 0.1606 & -16.50 & 225.64 & 362.51 & 32.78 & 59.64 & -123.32 & 311.60 & -385.32 & 0.694 & 0.6224 & 14.39 \\
SKRA \cite{rashdan:00}          & 0.1594 & -16.43 & 216.97 & 378.73 & 31.32 & 53.04 & -139.28 & 310.83 & -364.92 & 0.748 & 0.7243 & 14.35 \\
SkT1 \cite{tondeur:84}          & 0.1610 & -16.63 & 236.14 & 383.49 & 32.02 & 56.18 & -134.83 & 318.99 & -380.68 & 1.000 & 1.2933 & 17.05 \\
Skxs20 \cite{brown:07}          & 0.1617 & -16.46 & 201.94 & 425.53 & 35.50 & 67.06 & -122.31 & 328.52 & -383.37 & 0.964 & 1.2015 & 15.37 \\
SLy4 \cite{chabanat:98}         & 0.1595 & -16.62 & 229.90 & 363.07 & 32.00 & 45.96 & -119.70 & 521.48 & -322.84 & 0.695 & 0.6241 & 14.52 \\
SQMC700 \cite{guichon:06}       & 0.1704 & -16.14 & 219.59 & 367.98 & 33.40 & 59.14 & -140.23 & 312.66 & -395.42 & 0.755 & 0.7385 & 14.72 \\
\end{tabular}
\end{ruledtabular}
\end{table*}

\section{The equations of state}\label{sec:Results}

The model and approach described in the previous sections can be used
to compute thermodynamically consistent EOSs for a wide range of
Skyrme parametrizations.  There are over 200 Skyrme parametrizations
in the literature. We focus on eight parametrizations that are able to
reproduce most or all known experimental nuclear matter constraints
according to Dutra \etal \cite{dutra:12}. Since it has seen such wide
use, we are also including the \LS EOS with $K_0 = 220 \unit{MeV}$
although it does not fulfill many current nuclear physics constraints.

Specificially, we consider the following parametrizations (and provide 
EOS tables at \url{https://stellarcollapse.org/SROEOS}): 
NRAPR \cite{steiner:05},
SLy4 \cite{chabanat:98},
SkT1 \cite{tondeur:84},
SKRA \cite{rashdan:00},
LNS \cite{cao:06},
SQMC700 \cite{guichon:06},
Skxs20 \cite{brown:07}, 
KDE0v1 \cite{agrawal:05},
and \LS with $K_0=220\unit{MeV}$ (LS220 hereafter). 
Note that SLy4 does not fulfill one out of the eleven experimental
constraints studied by Dutra \etaln: its isospin incompressibility
(Equation~\ref{eq:Ktau}) is slightly below the experimentally
allowed range. We include it since its zero-temperature variant has
seen use in NS merger simulations (e.g.,
\cite{bauswein:13,taniguchi:10}).

We summarize the Skyrme parameters $t_i$, $x_i$, and $\sigma$ in Table
\ref{Tab:parametrizations} for the considered parametrizations.  
Note, however, that we exclude the \LS parametrization since there 
are multiple ways to set $t_1$, $t_2$, $x_1$ and $x_2$ that reproduce 
$\alpha_1=\alpha_2=0$, see Equations \eqref{eq:Skyrme_alpha1} and \eqref{eq:Skyrme_alpha2}.  Furthermore, it is not straightforward to 
chose a combination of these four parameters that also reproduces the 
fit parameters for the surface tension per unit area $\sigma(y,T)$ 
used by \LSn, see Equation \eqref{eq:sigma} and Table \ref{Tab:surf}.

\begin{figure}[t]
\centering
\includegraphics[trim = 0.5in 0.4in 0.1in 0.in, clip, width=\figurehalfwidth\textwidth]{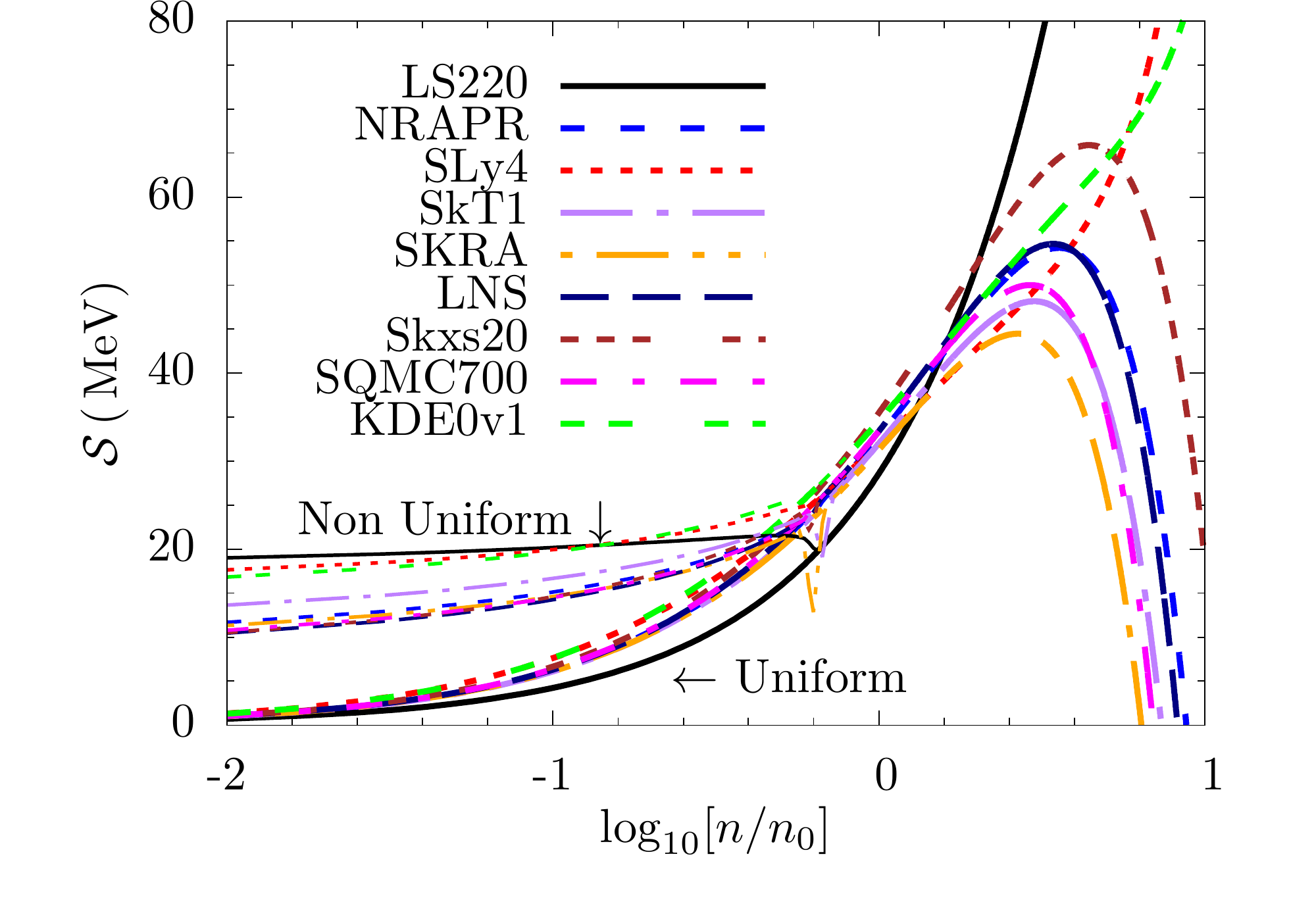}
\caption{(Color online) Density dependence of the symmetry energy
  $\mathcal{S}(n)$ for all considered Skyrme parametrizations. The
  thick curves show $\mathcal{S}(n)$ for uniform nuclear matter
  (neutrons and protons only) obtained from Equation  
  \eqref{eq:symmetry} with $\epsilon_B(n,y)=E_B(n,y)/n$ given by 
  Equation \eqref{eq:Skyrme}.  The thin curves correspond to 
  $\mathcal{S}(n)$ for the full high-density EOS at zero temperature, 
  allowing for non-uniform and uniform nuclear matter.  At densities 
  below the transition to uniform matter, $\mathcal{S}(n)$ is 
  obtained from Equation \eqref{eq:symmetry} with $\epsilon_B(n,y)$ 
  replaced by $F/n$ with $F$ from Equation~\eqref{eq:free}.  Note that 
  the high density ($n \gg n_0$) behavior of $\mathcal{S}(n)$ is highly
  uncertain. The high-density shape of $\mathcal{S}(n)$ for the Skyrme
  parametrizations shown here is a mere artifact of the expansion
  about $n_0$ and is not necessarily physical.  At low density, the
  binding energy of nearly symmetric nuclei increases the value of the
  symmetry energy for non-uniform matter.}
\label{fig:symmetry}
\end{figure}

For completeness, we list the zero-temperature properties of uniform
nuclear matter for all parametrizations in Table \ref{Tab:bulk}.  The
included properties of symmetric nuclear matter (SNM) are the nuclear
saturation density $n_0$, defined by
\begin{equation}
 P=\left.n^2\frac{\partial \epsilon_B(n,y)}{\partial n}\right|_{n=n_0,y=1/2}=0\,,
\end{equation}
the binding energy of SNM 
\begin{equation}
\label{eq:specific_energy}
\epsilon_0=\epsilon_B(n_0, y=1/2)\,,
\end{equation} 
the effective mass of nucleons at saturation density for SNM, 
$M_t^* = m_t^*(n_0,y=1/2)$, the incompressibility 
\begin{equation}
\label{eq:K} 
K_0 = 9n_0^2(\partial^2\epsilon_B/\partial n^2)\vert_{n=n_0,y=1/2}\,,
\end{equation}
and the skewness 
\begin{equation} 
K' = -27 n_0^3(\partial^3\epsilon_B/\partial n^3)\vert_{n=n_0,y=1/2}\,.
\end{equation}
Here, $\epsilon_B = E_B/n$ is the specific energy per baryon of uniform matter.
These quantities define the expansion of the specific energy of 
SNM around saturation density,
\begin{equation}
\epsilon_B(n,y=1/2) = \epsilon_0 + \frac{1}{2} K_0 x^2 - \frac{1}{6} K' x^3 +
\mathcal{O}(x^4)\,,
\end{equation} 
where $x = (n-n_0)/3 n_0$ and, as in Equation \eqref{eq:Skyrme},
$\Delta=m_n-m_p$.

The specific energy per baryon of asymmetric nuclear 
matter may be expanded around its value for symmetric matter,
\begin{equation}
\epsilon_B(n,y) = \epsilon_B(n,y=1/2) + \mathcal{S}(n)\delta(y)^2 + 
\mathcal{O}(\delta(y)^4)\,,
\end{equation} 
where $\delta(y)=1-2y$ is the isospin asymmetry and $S(n)$ is the
density-dependent symmetry energy, which is defined as
\begin{equation}\label{eq:symmetry}
 \mathcal{S}(n)=\frac{1}{8}\left.
 \frac{\partial^2\epsilon_B(n,y)}{\partial y^2}\right|_{n,y=1/2}\,.
\end{equation}
Sometimes the symmetry energy is defined as $S(n) = \epsilon_B(n,0) - 
\epsilon_B(n,1/2)$. The two definitions agree up to their quadratic terms. 
We plot $\mathcal{S}(n)$ for the considered Skyrme parametrizations in
Figure \ref{fig:symmetry}. We show curves for both uniform matter 
obtained from Equation \eqref{eq:symmetry} with $\epsilon_B(n,y)$
given by Equation \eqref{eq:Skyrme}, and for non-uniform matter, \ie 
accounting for the clustering of nucleons into heavy nuclei, which occurs 
for densities $n\lesssim0.10\unit{fm}^{-3}$.  For zero-temperature, 
non-uniform symmetric matter, the only non-negligible density-dependent
contribution comes from the term $F_i=u_in_if_B(n,y)$, which to first 
order is approximated by the binding energy of symmetric nuclear matter 
$n_0f_B(n_0,1/2)$. At high densities, $n\gtrsim3n_0$, nuclear physics 
observables are poorly constrained and the behavior of $\mathcal{S}(n)$, 
obtained as an expansion about $n_0$, is highly uncertain.

Expanding $\mathcal{S}(n)$ as a function of $x$, one obtains
\begin{equation}\label{eq:symmetry_expansion}
 \mathcal{S}(n)=J+Lx+\frac{1}{2}K_{\sym}x^2+\frac{1}{6}Q_{\sym}x^3+\mathcal{O}(x^4)\,,
\end{equation}
where
\begin{subequations}
\begin{align}
J&=\mathcal{S}(n_0)\,,\\
L&=3n_0(\partial\mathcal{S}/\partial n)\vert_{n=n_0}\,,\\
K_\sym&=9n_0^2(\partial^2\mathcal{S}/\partial n^2)\vert_{n=n_0}\,,\\
Q_\sym&=27n_0^3(\partial^3\mathcal{S}/\partial n^3)\vert_{n=n_0}\,.
\end{align}
\end{subequations}
These expansion parameters are listed in Table \ref{Tab:bulk}.
We note that the symmetry energy parameters $J$ and $L$ for all of the
Skyrme parametrizations, except for LS220, are consistent with recently
conjectured unitary gas constraints \cite{tews:16}.

We also show in Table \ref{Tab:bulk} the volume part of the isospin
incompressibility $K_{\tau,v}$ (e.g., \cite{dutra:12}), given by
\begin{equation}\label{eq:Ktau}
 K_{\tau,v}=\left(K_\sym-6L-\frac{Q_0}{K_0}L\right),
\end{equation}
the effective mass of neutrons in SNM $M_n^*$, the neutron proton
effective mass difference in SNM, $\Delta M^*=M_n^*-M_p^*$, and the
critical temperature $T_0$ discussed in Section \ref{ssec:surface}.
Note that most parametrizations have $T_c\simeq15\unit{MeV}$, the
exceptions being the SkT1 and LS220 parametrizations that have slightly
higher critical temperatures,
$T_c\simeq17\unit{MeV}$, which is due to their high effective masses.
For completeness, we provide the coefficients obtained for the critical 
temperature expansion (Equation \ref{eq:Tcrit}) in Appendix \ref{app:Tcrit}.

In Table \ref{Tab:surf}, we list the parameters $\sigma_s$, $q$, $\lambda$,
and $p$ that determine the surface tension per unit area, and which we 
obtain as described in Section \ref{ssec:surface}.
We also provide the values of the surface symmetry energy parameter $S_S$ 
and the surface level density parameter $A_S$ given by
\begin{subequations}\label{eq:SSAS}
 \begin{align}\label{eq:SS}
  S_S&=-\frac{A^{1/3}}{8}\left.\left(\frac{\partial^2 f_S(y,T)}{\partial y^2}\right)\right|_{y=1/2,T=0}\,,\\\label{eq:AS}
  A_S&=-\frac{A^{1/3}}{2}\left.\left(\frac{\partial^2 f_S(y,T)}{\partial T^2}\right)\right|_{y=1/2,T=0}\,\,,
 \end{align}
\end{subequations}
where $f_S$ is calculated for a spherical nucleus with mass number $A$ and density $n_0$, \ie
\begin{equation}
 f_S(y,T)=\frac{4\pi r_N^2\sigma(y,T)}{A}\,,
\end{equation}
with $r_N=(3/4\pi n_0 A)^{1/3}$.  Compared with the LS220
parametrization, all other Skyrme parametrizations have a much higher
surface symmetry energy parameter $S_S$, lower values for the
parameters $q$ and $p$, and a higher value for $\lambda$.  In
Reference \cite{lim:12}, Lim and Lattimer argue that the exponent
$\lambda$ is expected to be between 2 and 4.  This result agrees with
our results, though we find the range of $\lambda$ to be smaller for
all considered parametrizations, namely $3\lesssim\lambda\lesssim3.5$.
Finally, there are significant differences between the surface
properties we derive here for the SLy4 and those provided by Lim and
Lattimer \cite{lim:12}.  The differences reside in Lim and Lattimer
having an extra parameter that accounts for the surface tension of the
neutron skin of nuclei, $\sigma\rightarrow\sigma+\mu_n\nu_n$, as
discussed in Section \ref{ssec:surface}.

\begin{table}[t]
  \caption{\label{Tab:surf} Summary of the
      surface properties of nuclear matter obtained for the considered
      Skyrme parameterizations.  $S_S$ is the surface symmetry energy
      (in MeV; Equation~\ref{eq:SS}), $A_S$ is the surface level
      density (in MeV$^{-1}$; Equation~\ref{eq:AS}), $\sigma_s$ is the
      surface tension of symmetric nuclear matter at zero temperature
      (in MeV\,fm$^{-2}$; Equation~\ref{eq:tension}). $q$, $\lambda$,
      and $p$ are the dimensionless surface tension parameters in
      Equations~\eqref{eq:sigma} and \eqref{eq:h}.}
  \begin{ruledtabular}
\begin{tabular}{l  D{.}{.}{2.2} D{.}{.}{1.4} D{.}{.}{1.4} D{.}{.}{2.2} D{.}{.}{1.3} D{.}{.}{1.3}}
\multicolumn{1}{c}{Parametrization}&
\multicolumn{1}{c}{$S_S$}&
\multicolumn{1}{c}{$A_S$}&
\multicolumn{1}{c}{$\sigma_s$}&
\multicolumn{1}{c}{$q$}&
\multicolumn{1}{c}{$\lambda$}&
\multicolumn{1}{c}{$p$}\\
\hline
LS220 \cite{lattimer:91}        & 45.81 & 0.1365 & 1.150 & 24.40 & 3.000 & 2.000  \\
KDE0v1 \cite{agrawal:05}        & 78.63 & 0.1315 & 1.215 & 13.54 & 3.245 & 1.493  \\
LNS \cite{cao:06}               & 95.17 & 0.1089 & 1.044 &  7.78 & 3.507 & 1.506  \\
NRAPR \cite{steiner:05}         & 92.44 & 0.1316 & 1.140 & 13.96 & 3.522 & 1.467  \\
SKRA \cite{rashdan:00}          & 86.99 & 0.1332 & 1.125 & 14.26 & 3.464 & 1.492  \\
SkT1 \cite{tondeur:84}          & 78.71 & 0.0979 & 1.090 & 16.06 & 3.449 & 1.606  \\
Skxs20 \cite{brown:07}          &106.94 & 0.1117 & 1.045 &  6.48 & 3.540 & 1.555  \\
SLy4 \cite{chabanat:98}         & 64.31 & 0.1423 & 1.247 & 18.51 & 3.128 & 1.474  \\
SQMC700 \cite{guichon:06}       & 98.48 & 0.1280 & 1.191 &  9.90 & 3.442 & 1.486  \\
\end{tabular}
\end{ruledtabular}
\end{table}

\subsection{Comparison with \LS results}\label{ssec:comparison}

\begin{figure*}[t]
\centering
\includegraphics[trim = 0.4in 0.4in 0.2in 0.1in, clip, width=\figurewidth\textwidth]{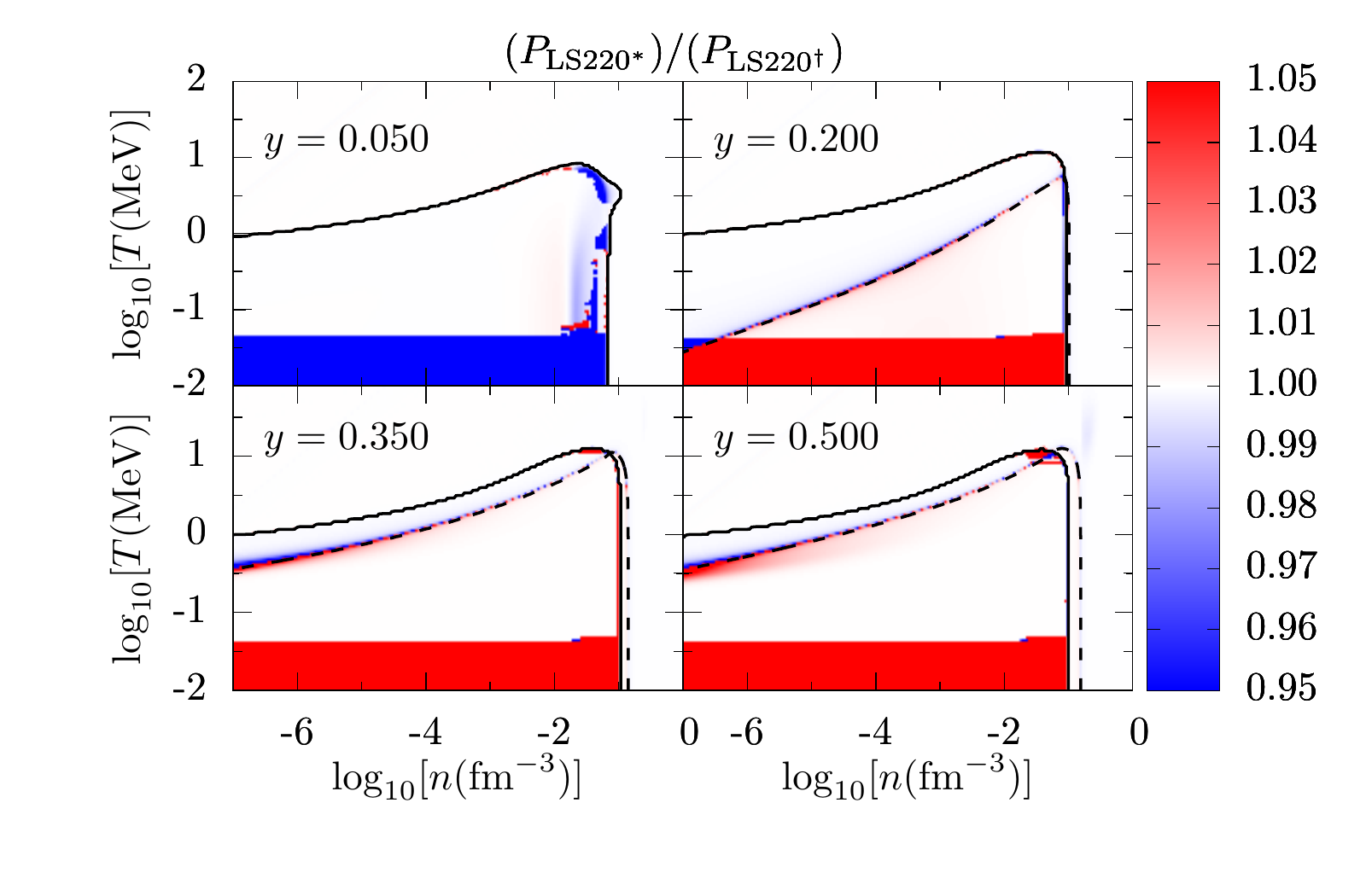}
\caption{(Color online) Comparison with \LS. We show the ratio of 
  nuclear pressure for a range of proton fractions obtained with 
  our LS220$^\dagger$ implementation and with the original \LS 
  implementation LS220$^*$. The solid black curve deliniates where the
  heavy-nuclei number fraction $X_i = u n_i / n$ changes from zero to a 
  non-zero value. Below and to the left of the curve, matter is non-uniform, 
  while above and to the right of the line it is uniform. The dashed line 
  shows where the nuclear pressure is zero. Differences between LS220$^\dagger$
  and LS220$^*$ are largest near this line. The wide horizontal
  band at the bottom of the panels marks the region where the
  original \LS implementation does not converge for non-uniform
  nuclear matter and assumes that the system is uniform.}
\label{fig:P_ratio}
\end{figure*}

\begin{figure*}[t]
\centering
\includegraphics[trim = 0.4in 0.4in 0.0in 0.1in, clip, width=\figurewidth\textwidth]{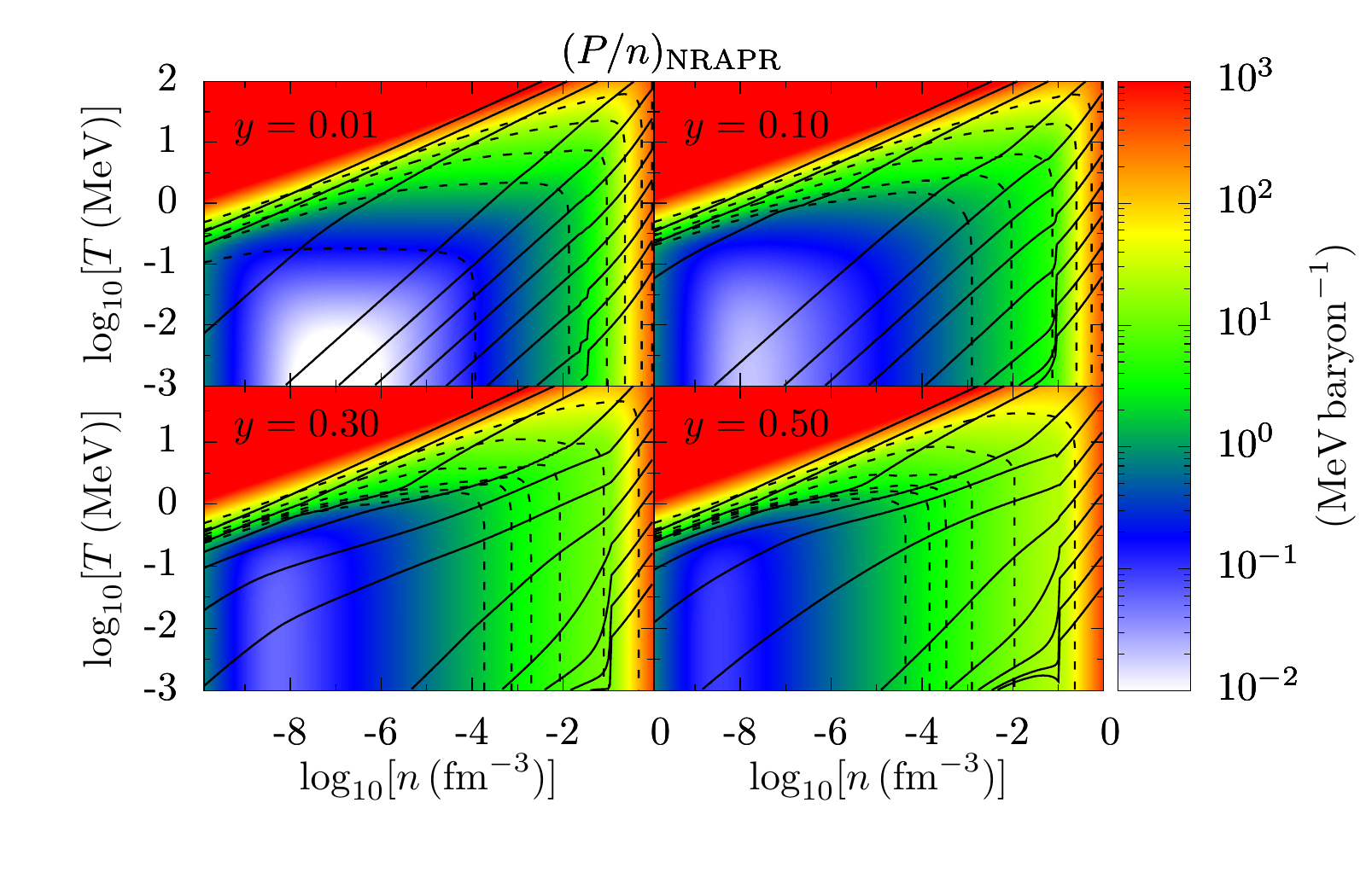}
\caption{(Color online) Total pressure per nucleon $P/n$ in 
  $\unit{MeV}\,\unit{baryon}^{-1}$ for the NRAPR Skyrme parametrization \cite{steiner:05} and proton fractions $y=0.01$, $0.10$, $0.30$, and $0.50$.  
  The solid black curves denote, from top to bottom, the adiabats at entropies
  $s=10^{n}k_B\,\mathrm{baryon}^{-1}$ for $n=2$ to $n=-3$ in $-0.5$
  increments.  The dashed black curves correspond, from right to left, to the
  isoergs for specific energies $\epsilon=100$, $30$, $10$, $3$, $0$,
  and $-3\unit{MeV}\,\mathrm{baryon}^{-1}$.  Note that only the
  $y=0.30$ and $y=0.50$ panels contain the 
  $\epsilon =-3\unit{MeV}\,\mathrm{baryon}^{-1}$ isoerg. 
  The pressure per nucleon is dominated by electrons, positrons, and photons 
  in large portions of the density--temperature space. Only at the highest 
  densities, at and above saturation density, is the pressure dominated 
  by the nucleon contributions and the impact of strong interactions.}
\label{fig:P_NRAPR}
\end{figure*}

Using the same \LS Skyrme parametrization which predicts a nuclear
incompressibility $K_0=220\unit{MeV}$, we compare the results from our
code, labeled here as LS220$^\dagger$, with the results of the original 
\LS implementation available at 
\url{http://www.astro.sunysb.edu/dswesty/lseos.html}, labeled here as
LS220$^*$.  In order to be consistent in our comparison with the original 
\LS implementation, for the cases discussed in this subsection only, we 
make the following choices:
(1) We set the alpha-particle binding energy to $B_\alpha = 28.3\,\mathrm{MeV}$. 
(2) We set $m_p = m_n = 939.5654\unit{MeV}$.  
(3) We set the proton-neutron mass difference to $\Delta=1.29\unit{MeV}$ 
and carry it explicitly. 
(4) We fix $\bar{A} = 60$ in Equation~\eqref{eq:barA}.


In Figure \ref{fig:P_ratio}, we plot the ratio of the total nuclear pressures (excluding electrons and photons) returned by the two LS220 implementations. 
We choose proton fractions of $y=0.05$, $0.20$, $0.35$, and $0.50$, densities 
in the range $10^{-7}\unit{fm}^{-3}\leq n\leq1\unit{fm}^{-3}$, and temperatures $0.01\unit{MeV}\leq T\leq100\unit{MeV}$.  We choose these ranges since the 
original \LS implementation only converges consistently for proton fractions 
in the range $0.03\leq{y}\leq0.51$, densities higher than $10^{-7}\unit{fm}^{-3}$, 
and temperatures higher than $10^{-1.5}\unit{MeV}$.  
In our implementation, however, we are able to compute the EOS for proton 
fractions $0.001\lesssim y\lesssim0.7$, and for temperatures and densities as 
low as $10^{-4}\unit{MeV}$ and $10^{-13}\unit{fm}^{-3}$, respectively.

Figure~\ref{fig:P_ratio} demonstrates that in uniform matter, with the 
exception of regions very close to $P\simeq0$, our results and those of 
\LS agree within 0.5\% or better. For non-uniform matter and very low 
temperatures, $T\lesssim0.04\unit{MeV}$, the \LS implementation is unable 
to find a non-uniform solution and assumes the system is uniform. This 
gives rise to the large ratio between the pressures in that region.  
In most of the non-uniform regions with temperatures above
$T\gtrsim0.04\unit{MeV}$, the agreement is, again, within 0.5\% or
better.  Exceptions occur near the transition from uniform to
non-uniform matter and regions where the nuclear pressure is close to
zero.  Even though the ratios are large in these regions, the absolute
pressure differences are relatively small.  Differences between the two
implementations also appear in regions of parameter space with very low 
proton fraction, represented in Figure~\ref{fig:P_ratio} by $y=0.05$, 
and densities $0.006\unit{fm}^{-3}\lesssim n\lesssim 0.03\unit{fm}^{-3}$.
Discrepancies are also visible in regions of non-uniform symmetric
nuclear matter at temperatures $T\simeq10\unit{MeV}$. In these
regions, the original \LS implementation has convergence issues for some 
values of density $n$ and temperature $T$. At very low proton
fraction, even in regions where both implementations converge, we
observe differences in the calculated pressures as large as 2\%.

We carry out similar comparison studies for   other thermodynamic quantities, 
including the specific energy, specific entropy, proton and neutron chemical 
potentials, average nuclear charge and mass, and the mass fractions of protons,
neutrons, alpha particles, and heavy nuclei. 
In all these comparisons we find differences that are qualitatively and 
quantitatively very similar to what is shown for the pressure in 
Figure~\ref{fig:P_ratio}.

\subsection{Comparing Equations of State}\label{ssec:EOSs}

\begin{figure*}[tbh]
\centering
\includegraphics[trim = 0.3in 0.4in 0.2in 0.1in, clip, width=0.8\textwidth]{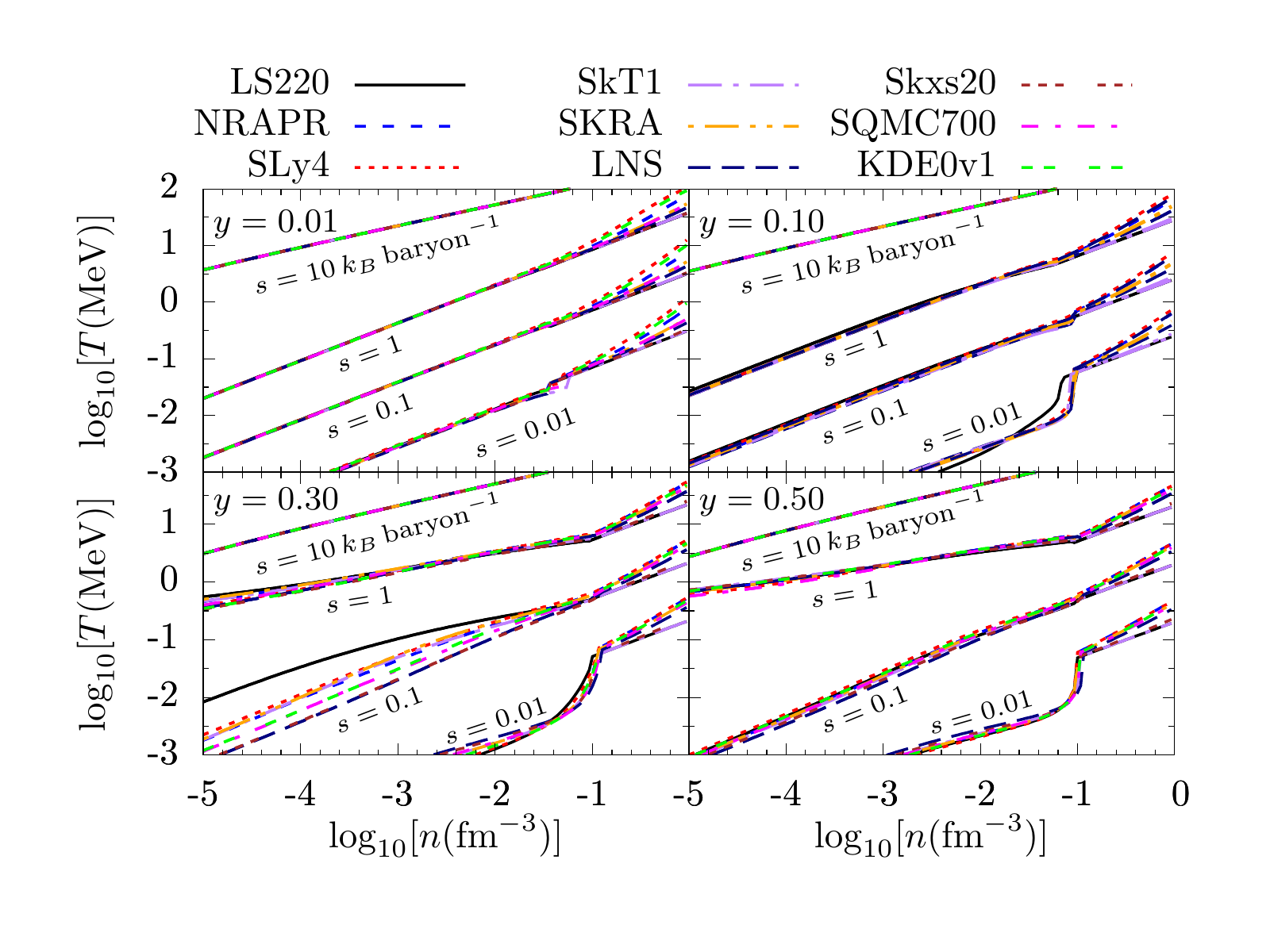}
\vspace*{-1em}
\caption{(Color online) Temperature along adiabats with specific entropy 
  $s=10$, $1.0$, $0.1$, and $0.01\,k_B\,\mathrm{baryon}^{-1}$ in the 
  single-nucleus approximation and for all considered Skyrme
  parametrizations. Note that electrons, positrons, and photons are 
  included. The adiabats differ mostly in regions dominated by nucleonic 
  pressure around and above saturation density $n_0\sim 0.16\,\unit{fm}^{-3}$
  and for $y\simeq0.30$ and $T\lesssim1\unit{MeV}$, where the number of
  free neutrons varies significantly between parametrizations.}
\label{fig:adiabats_T}

\vspace*{-1em}
\includegraphics[trim = 0.3in 0.4in 0.2in 0.1in, clip, width=0.8\textwidth]{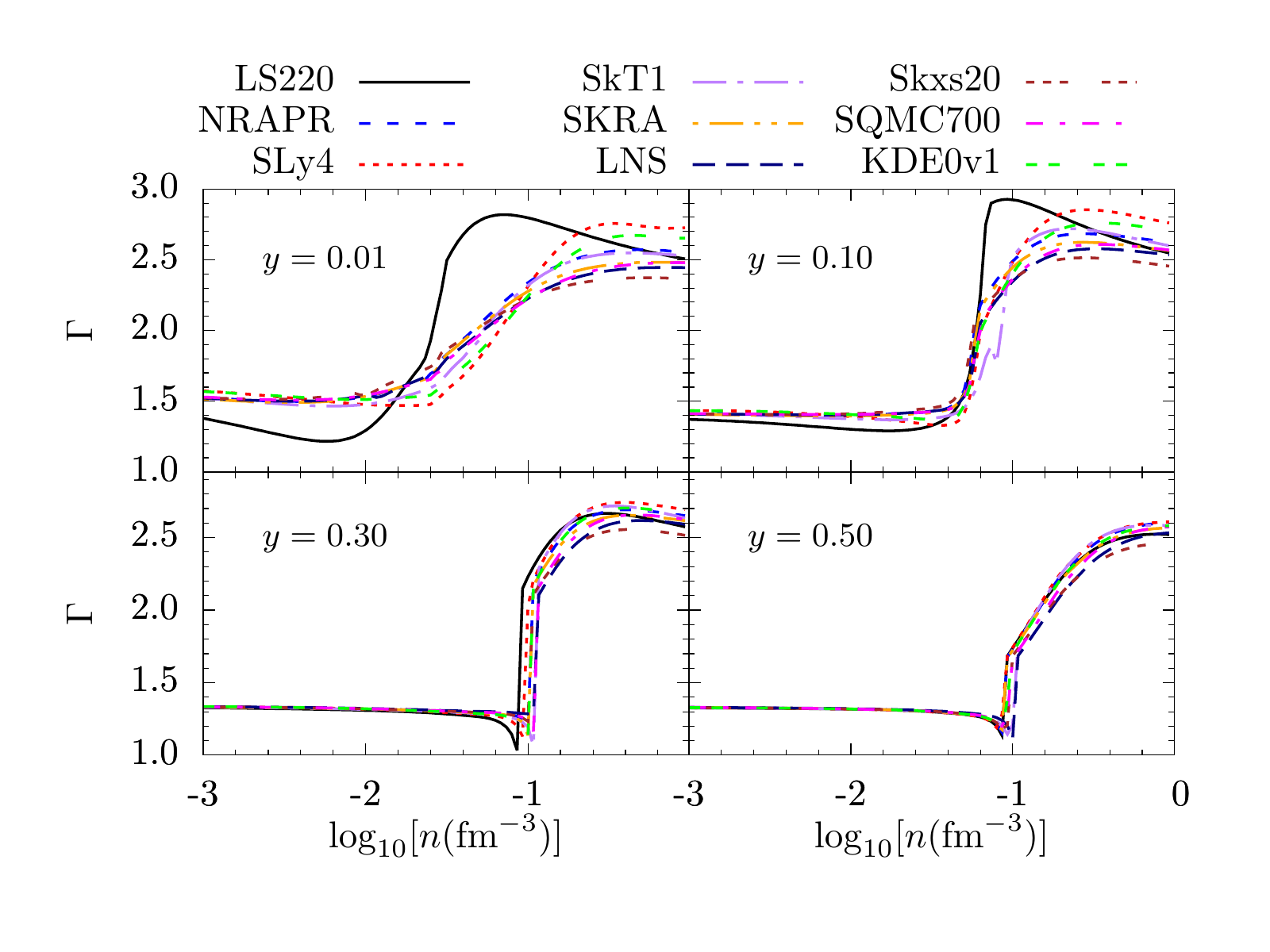}
\vspace*{-1em}
\caption{(Color online) Adiabatic index $\Gamma$ along the
  $s=1\,k_B\,\mathrm{baryon}^{-1}$ adiabat for full EOSs in the
  single-nucleus approximations. At low densities and proton fractions
  $y \gtrsim 0.1$, electrons dominate and $\Gamma \sim 4/3$.  At high
  densities and $y=0.50$, $\Gamma$ is roughly the same for all EOSs
  reflecting the well constrained properties of symmetric nuclear
  matter.  The sharp discontinuity is due to the transition between
  non-uniform and uniform nuclear matter at $n\simeq0.1\unit{fm}^{-3}$. It
  becomes smoother at lower proton fraction due to large free neutron
  contributions. Also, as the proton fraction decreases, differences
  between parametrizations increase due largely to variations in the
  density-dependent symmetry energy (cf.~Figure~\ref{fig:symmetry}.)}
\label{fig:adiabats_G}
\end{figure*}

We compare full EOSs obtained with the set of
  considered Skyrme parametrizations. We focus on SNA EOSs and defer a
  detailed discussion of our approach for matching to NSE at low
  densities to Section~\ref{sec:NSE}. In contrast to the
  previous section on the LS220 parametrization, we go back to an
  alpha particle binding energy of $B_\alpha=30.887\unit{MeV}$ since
  all free energies are computed  with respect to the free energy of 
  a gas of unbound neutrons. We set $m_p=938.2721\unit{MeV}$ and
  $m_n=939.5654\unit{MeV}$ \cite{mohr:16}.  The
  proton-neutron mass difference $\Delta=m_n-m_p$ is obtained
  self-consistently. Despite changing the proton mass, our LS220 
  implementation uses the same Skyrme parameters obtained by \LS and 
  used in Section~\ref{ssec:comparison}. This results in small 
  differences between the LS220 EOS and the LS220$^\dagger$ and 
  LS220$^*$ EOSs. The differences come from small changes in 
  the proton effective mass term, Equation~\ref{eq:mstar}. 
  Finally, we let $\bar{A}$ vary in the translational free energy 
  density (Equation~\ref{eq:barA}).

\begin{figure*}[t]
\centering
\includegraphics[trim =  0.3in 0.4in 0.2in 0.1in, clip, width=\figurewidth\textwidth]{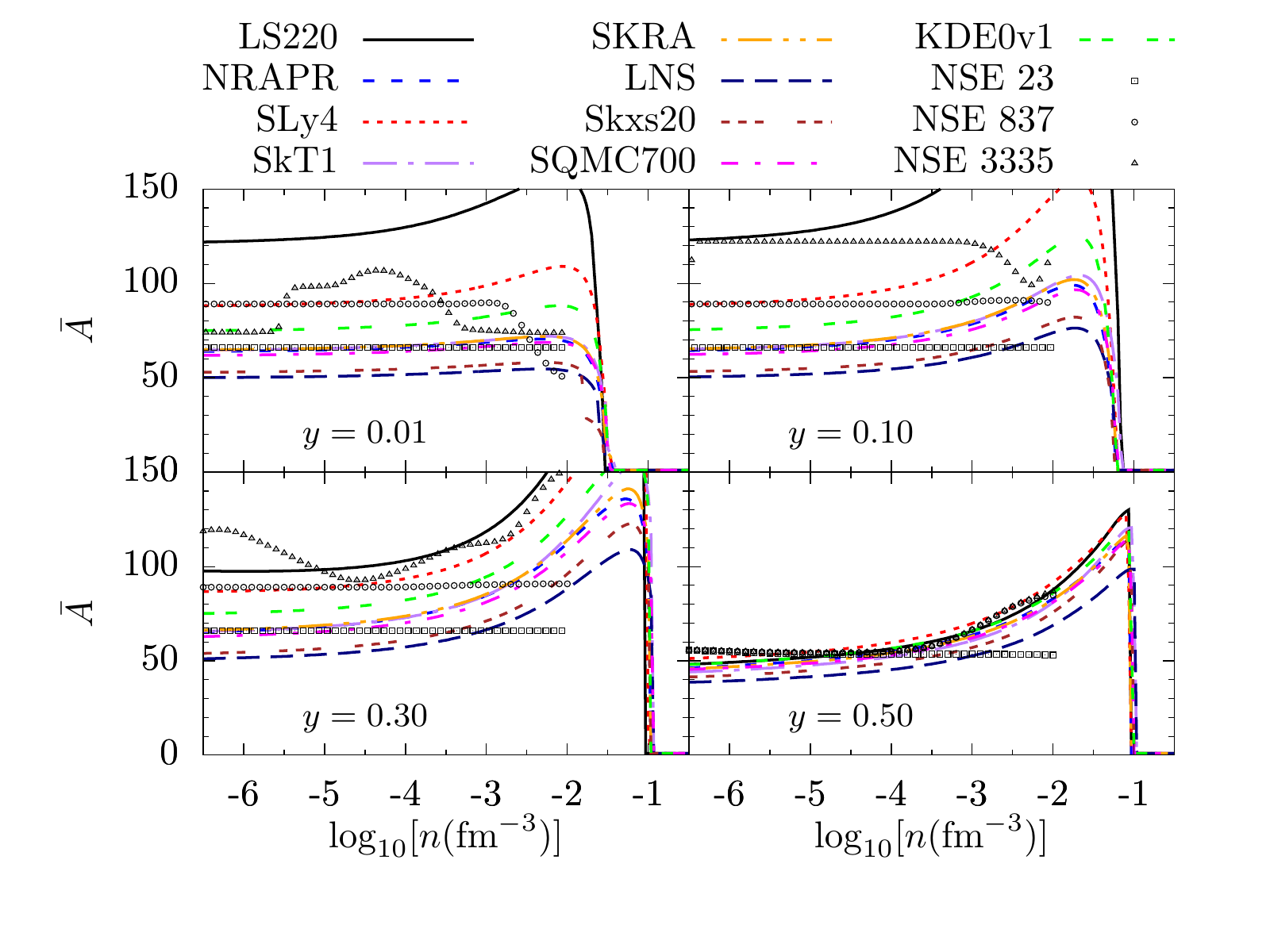}
\caption{(Color online) Average heavy nucleus mass number $\bar A$
  along the $s=1\,k_B\,\mathrm{baryon}^{-1}$ adiabat as a function of
  density for the considered Skyrme parametrizations in the
  single-nucleus approximation (SNA) and, at low densities, for
  nuclear-statistical equilibrium (NSE) with 23, 807, and 3\,335 nuclides 
  (we discuss our NSE treatment and matching to SNA in Section~\ref{sec:NSE}).
  For the SNA curves, differences in nuclear sizes result from
  differences in symmetry energy and the surface properties obtained
  for each parametrization. The 3\,335 nuclide NSE network exhibits large 
  oscillations in $\bar{A}$. These are due to nuclear shell effects included 
  implicitly in the nuclear masses.}
\label{fig:adiabats_A}
\end{figure*}

In Figure \ref{fig:P_NRAPR}, we plot the pressure per nucleon using
the NRAPR parametrization for proton fractions $y=0.01$, $0.10$
$0.30$, and $0.50$.  We also include in the plots eleven adiabats,
$s=10^{n}k_B\,\mathrm{baryon}^{-1}$ for $n=-3$ to $n=2$ in $0.5$
increments, and six isoergs at $\epsilon =-3$, $0$, $3$, $10$, $30$,
and $100\unit{MeV}\,\mathrm{baryon}^{-1}$.  The pressure per baryon is
dominated by the electron and photon contributions in large portions
of density--temperature space. At the highest temperatures, the
electrons, positrons, and photons behave as an ultra-relativistic gas
and drive the strong temperature dependence of the pressure seen
there. At lower temperatures ($T \lesssim 1 \, \textrm{MeV}$) and for
densities below saturation density, degenerate electrons give a large
contribution to the pressure and the pressure is relatively
insensitive to the temperature. Nevertheless, throughout the phase
diagram, the nuclear contribution to the pressure is often
significant, although subdominant. At the highest densities (i.e., at
and above saturation density), the pressure is dominated by the
nucleon contributions and the impact of strong interactions.  The EOSs
obtained from the other Skyrme parametrizations considered in this
study are qualitatively similar to the EOS resulting from the NPAPR
parametrization and shown in Figure \ref{fig:P_NRAPR}.

In Figure \ref{fig:adiabats_T}, we plot the temperatures along four
adiabats ($s=0.01$, $0.1$, $1$, and $10\,k_B\,\mathrm{baryon}^{-1}$) at
a range of proton fractions.  Except for very low entropies,
$s\lesssim0.1\,k_B\,\mathrm{baryon}^{-1}$, or very high densities,
$n\gtrsim0.1\unit{fm}^{-3}$, the entropy does not significantly depend
on the Skyrme parametrization. For uniform matter, the entropy depends 
only on the temperature, density, proton fraction, and nucleon effective 
masses. Therefore, we see systematically higher entropies for 
parametrizations with smaller effective masses at high density. 
At lower densities, variations between EOSs are caused by the different 
properties of the single nucleus predicted by the different Skyrme 
parametrizations.

In Figure \ref{fig:adiabats_G}, we compare the adiabatic index
$\Gamma$, Equation \eqref{eq:gamma}, along the
$s=1\,k_b\,\mathrm{baryon}^{-1}$ adiabats.  The largest differences
between the adiabatic indexes occur for very low proton fractions, $y
\lesssim 0.10$.  This follows from the Skyrme parameters being chosen
to fit properties of isospin symmetric matter and, therefore,
predicting significantly different properties of matter when
extrapolated to large isospin asymmetries.  In the very neutron rich
regime, $y \lesssim 0.10$, the LS220 parametrization shows results
that differ from the others not only quantitatively, but also
qualitatively.  Unlike the other parametrizations, at low proton
fractions, $\Gamma_{\mathrm{LS220}}$ exhibits a peak close to the
phase transition between non-uniform and uniform matter. The change of
$\Gamma_{\mathrm{LS220}}$ across the transition is overall much
smoother and occurs at lower densities than for the other
parametrizations.

\begin{figure*}[htbp]
\centering
\includegraphics[trim =  0.3in 0.4in 0.2in 0.1in, clip, width=\figurewidth\textwidth]{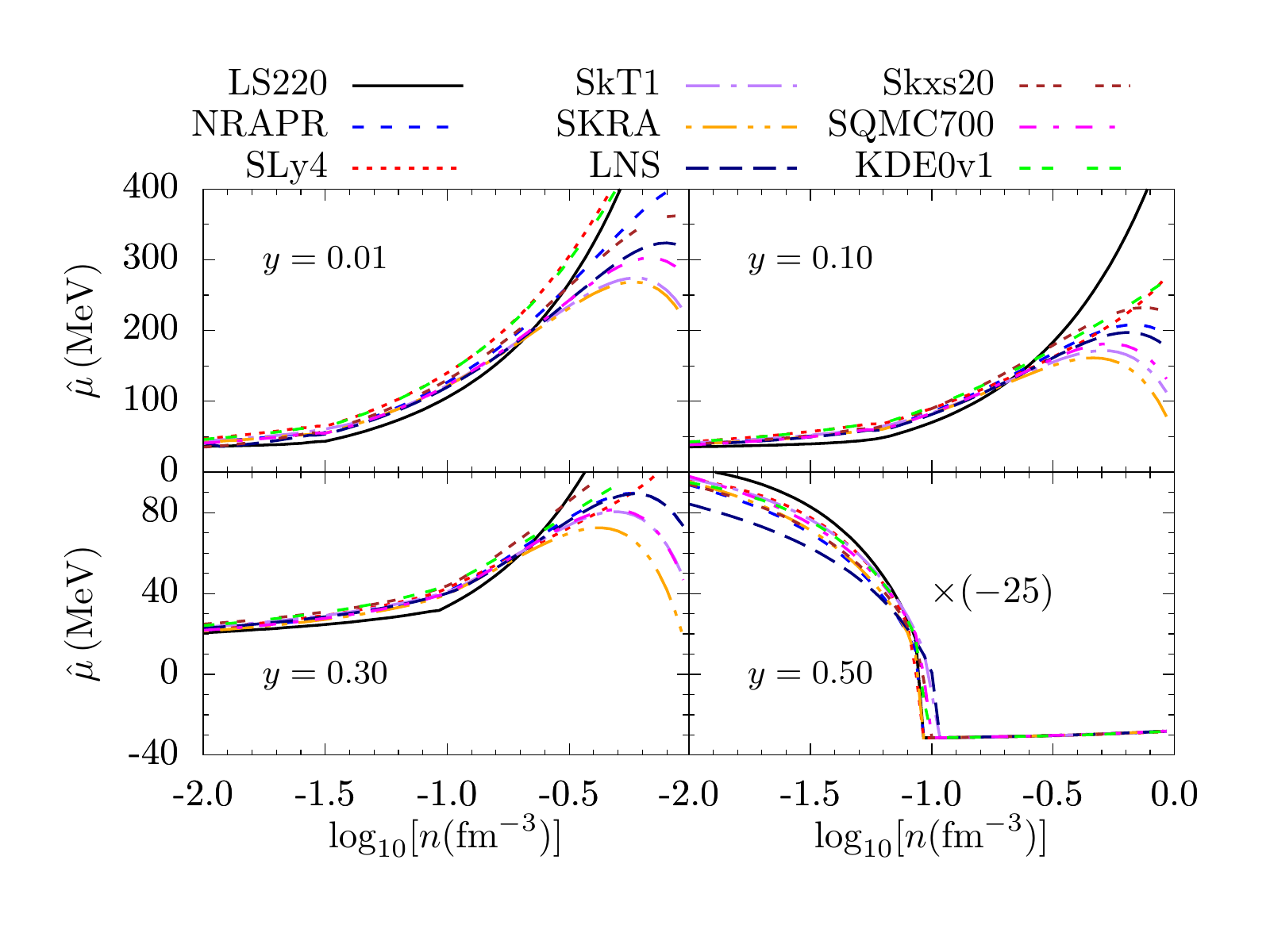}
\caption{(Color online) Neutron-proton chemical potential difference
  $\hat{\mu}=\mu_n-\mu_p$ along the $s=1.0\,k_B\,\mathrm{baryon}^{-1}$
  adiabat as a function of density for the considered Skyrme 
  parametrizations at proton fractions $y=0.01$, $10$, $0.30$, and
  $0.50$. Note that we multiply $\hat{\mu}$ by
    a factor of -25 in the bottom right panel showing the $y = 0.5$
    case. $\hat{\mu}$ is sensitive to the density dependence of the
    symmetry energy and the differences between parametrizations seen
    here correlate with those in Figure~\ref{fig:symmetry}.}
\label{fig:adiabats_x}
\end{figure*}

The composition of non-uniform matter influences the EOS and can
impact neutrino transport in CCSNe.  Each Skyrme parametrization
predicts different properties for the equilibrium nucleus in the SNA.
In Figure \ref{fig:adiabats_A}, we show the masses $\bar{A}$ of the
SNA nuclei formed along the $s=1\,k_B\,\mathrm{baryon}^{-1}$ adiabat
for different Skyrme parametrizations. We compare them with $\bar{A}$
obtained for ensembles of nuclei in NSE (see Section~\ref{sec:NSE}).
The LS220 parametrization produces much heavier nuclei at low $y$ than
any of the other parametrizations. By Equation \ref{eq:r}, the nuclear
size $r$ increases with the surface tension. Therefore, increasing
$\sigma$ increases $\bar{A}$, all other things being equal. LS220 has
the weakest $y$ dependence of the surface tension (which results in a
relatively larger surface tension at low $y$) and the smallest
symmetry energy of any of the Skyrme parameterizations. This explains
the large nuclei predicted by LS220.  The Skxs20 and LNS
parametrizations predict the lightest nuclei and have the smallest
surface tensions at low $y$.  Except for some parametrizations at very
low proton fractions, the SNA EOSs produce heavy nuclei that increase
with density for $n\gtrsim10^{-4}\unit{fm}^{-3}$ until close to the
phase transition to uniform nuclear matter.  This is the region where
the nuclear ``pasta'' phase is expected to appear.  The different
masses of nuclei may significantly alter neutrino cross sections and
CCSNe neutrino spectra as well as the cooling rates of NSs.  Since
Skyrme parametrizations are fitted to properties of SNM, all
parametrizations yield similar predictions for $\bar{A}$ at $y=0.5$.


Figure~\ref{fig:adiabats_A} includes NSE results for $\bar{A}$ that were 
obtained with ensembles of 23, 837, and 3\,335 nuclides. We see that $\bar{A}$ 
predicted by NSE for the $s=1\,k_B\,\mathrm{baryon}^{-1}$ adiabat is rather 
sensitive to the number of nuclides included. In the ensemble containing 
23 nuclei, which includes nuclides with $Z\leq26$, the only heavy and neutron 
rich nuclide included, $^{66}$Fe, dominates the composition for neutron 
rich matter.  The 837-nuclide ensemble includes nuclides with $Z\leq50$ and 
the dominant nucleus for neutron rich matter is $^{89}$Ge.

The 3\,335-nuclide NSE network includes nuclides up to $Z=85$ and sufficiently 
many neutron-rich heavy nuclides that there is no single nuclide that 
dominates in neutron rich matter. For SNM, on the other hand, all nuclide 
ensembles predict very similar compositions at low densities,
$n\lesssim10^{-4}\unit{fm}^{-3}$.

Finally, we present in Figure \ref{fig:adiabats_x} the difference
between the neutron and proton chemical potentials,
$\hat{\mu}=\mu_n-\mu_p$, along the $s=1\,k_B\,\mathrm{baryon}^{-1}$
adiabat.  The quantity $\hat\mu$ is relevant for charged current
neutrino interactions as it enters into the equilibrium neutrino
chemical potential, $\mu_{\nu}=\mu_e - \hat \mu$, which determines
detailed balance for charged current interactions and influences how
hard it is to turn neutrons into protons (or vice versa) in the
medium. Furthermore, $\hat\mu$ is correlated with the symmetry energy
$\mathcal{S}$, which gives a large contribution to the pressure at
high densities.  First, we note that for SNM, all Skyrme
parametrizations produce similar curves for $\hat{\mu}$, especially
for densities $n\gtrsim0.1\unit{fm}^{-3}$.  This is expected, since
the coefficients of each parametrization are chosen to reproduce
properties of uniform SNM where experimental constraints are abundant.

It is apparent from Figure~\ref{fig:adiabats_x} that for most proton 
fractions the LS220 parametrization predicts the lowest values for 
$\hat{\mu}$ in the range $0.01\unit{fm}^{-3}\lesssim n \lesssim n_0$ 
and the highest for densities above nuclear saturation density. 
In the neutron rich regime, the LS220, SLy4, and KDE0v1 parametrizations 
all predict $\hat{\mu}$ that increases monotonically with density.  
The other parametrizations, on the other hand, have a global maximum
above nuclear saturation density, which occurs in the range 
$2n_0\lesssim{n}\lesssim4n_0$ and is higher (lower) for Skxs20
(SKRA) than for the other parametrizations.  In the next Section~\ref{sec:NS}, 
we discuss the effects of this behavior on the radial profile and maximum 
mass of cold nonrotating NSs.

\section{Neutron star mass-radius relationship}\label{sec:NS}

We construct the mass-radius relationship of cold neutron stars (NS) 
by solving the Tolman-Oppenheimer-Volkoff (TOV) equations \cite{tolman:39} 
for neutrino-less beta equilibrated matter (BEM) near zero temperature.
We choose a low temperature of $T = 0.1\unit{MeV}$, and determine for 
each density $n$ the proton fraction $y$ where the neutrino chemical 
potential is zero, \ie $\mu_\nu=\mu_e - \mu_n + \mu_p =0$.  If no such solution 
can be found, or the solution implies a large discontinuity from $y\simeq0$
to $y>0.50$, we set the proton fraction to the minimum value available for 
a given combination of $n$ and $T$ in the EOS table.

\begin{figure}[t]
\centering
\includegraphics[trim = 0.3in 0.25in 0.2in 0.2in, clip, width=\figurehalfwidth\textwidth]{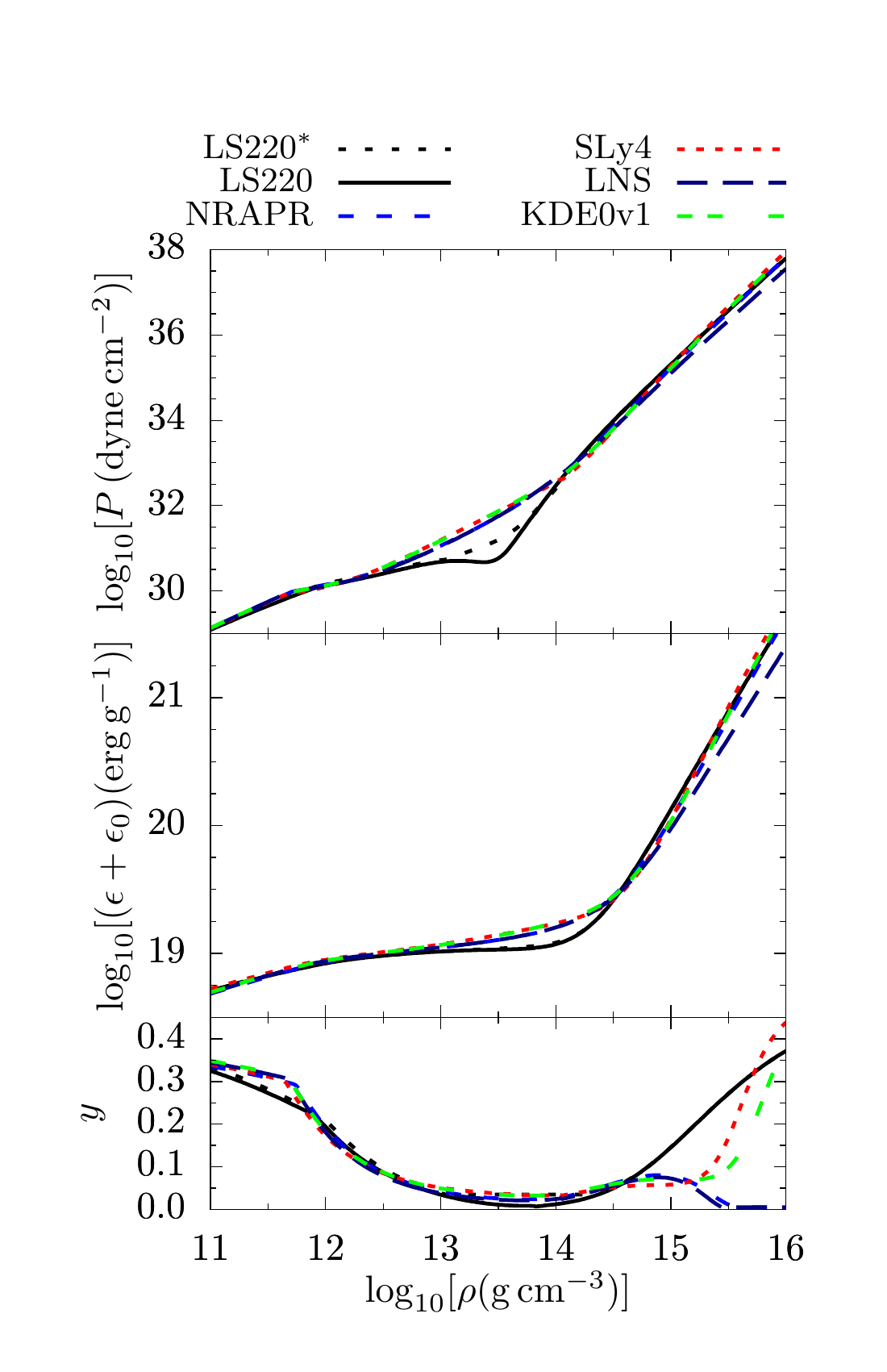}
\caption{(Color online) Pressure $P$ (top panel), specific internal energy
  $\epsilon$ (plus an additive constant $\epsilon_0 = 2 \times
  10^{19}\unit{erg\,g^{-1}}$; center panel), and proton fraction $y$ 
  for low temperature neutrino-less beta equilibrated matter (bottom panel). 
  We show results for five select Skyrme parametrizations that span
  the range of maximum neutron star masses shown in Figure \ref{fig:TOV}. 
  The LS220$^*$ (LS220) curve uses \LSn's (our) implementation of the \LS 
  $K_0=220\unit{MeV}$ parametrization. Differences between LS220
    and LS220$^*$ are due to the \LS implementation limit of proton
    fractions $y \ge 0.035$ and a small difference in the proton masses
    used (cf.~Section~\ref{ssec:comparison}).  The proton fraction $y$ at high 
    densities, $\rho\gtrsim10^{14.5}\unit{g\,cm^{-3}}$, mirrors the high 
    density behavior of the symmetry energy $\mathcal{S}(n)$
  (cf.~Figure~\ref{fig:symmetry}). }
\label{fig:EOS}
\end{figure}

In Figure \ref{fig:EOS}, we present density-dependent graphs of
pressure, specific internal energy, and proton fractions for the
LS220, NRAPR, SLy4, LNS, and KDE0v1 parametrizations.  For comparison,
we also show results obtained with the original \LS implementation
(LS220$^*$), which converges reliably only for proton fractions
$y\gtrsim0.035$.


\begin{figure*}[t]

\centering
\includegraphics[trim =  0.1in 0.1in 0.2in 0.05in, clip, width=\figurewidth\textwidth]{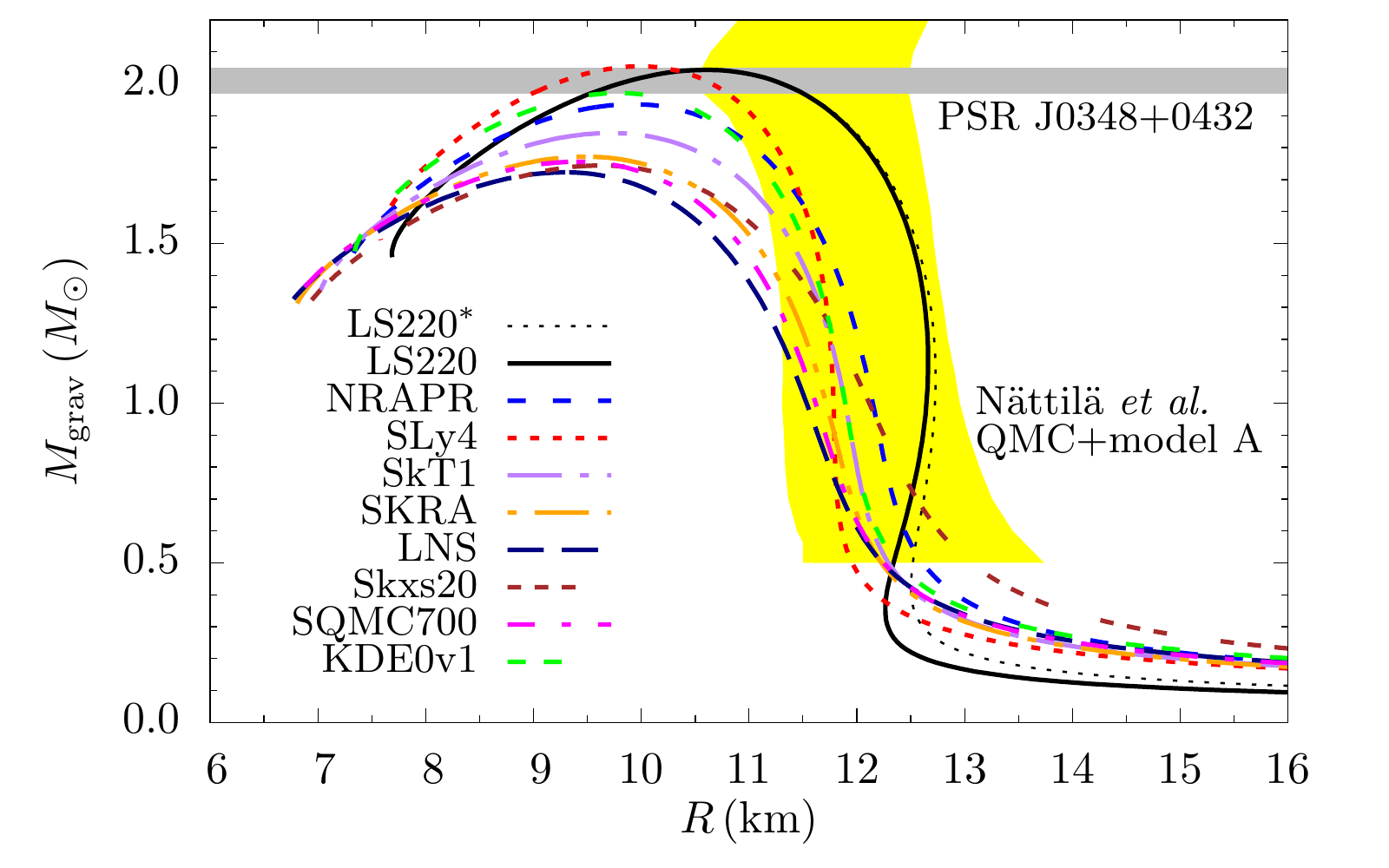}
\caption{(Color online) Mass-radius curves for cold,
  beta-equilibrated neutron stars (NSs) obtained by solving the
  Tolman-Oppenheimer-Volkoff equations for the considered Skyrme
  parametrizations. We summarize NS properties in Table 
  \ref{Tab:TOV}.  The gray strip represents the mass of the NS 
  $\mathrm{PSR\,J0348+0432}$, $M_\mathrm{J0348+0432}=2.01\pm0.04 M_\odot$
  \cite{antoniadis:13}. The yellow region indicates the NS mass-radius 
  constraints from model A of N\"attil\"a \etal \cite{nattila:16}.
  Besides the LS220 parameterization, only SLy4 and (barely) KDE0v1 and 
  NRAPR satisfy the $M_\mathrm{max} \gtrsim 2\,M_\odot$ constraint. 
  Differences between our implementation of LS220 and the original \LS
  implementation (LS220$^*$) are due to the lower limit of $y
  \ge 0.035$ in the latter (cf.~Figure~\ref{fig:EOS}).}
\label{fig:TOV}
\end{figure*}

Figure~\ref{fig:EOS} reveals some differences between the LS220 and
the LS220$^*$ curves.  These are due to small differences in chemical
potentials between the EOSs owing to the different treatments of
proton masses (cf.~Section~\ref{ssec:comparison}) and to the \LS
implementation limit of proton fractions $y\geq0.035$.  The four other
parametrizations shown in Figure \ref{fig:EOS}, SLy4, KDE0v1, NRAPR,
and LNS, have very similar qualitative and quantitative behavior below
nuclear saturation density in the three quantities plotted. For
densities above nuclear saturation density, on the other hand, the
EOSs can be separated into two groups according to their prediction
for the BEM proton fraction.  Group I EOSs, which includes the LS220,
LS220$^*$, SLy4, and KDE0v1 parametrizations, have proton fractions
that increase monotonically above nuclear saturation density.
Meanwhile, Group II EOSs, which include the NRAPR and LNS
parametrizations, have BEM proton fractions with a maximum near
nuclear saturation density and that decrease to zero at higher
densities. Group II also includes the other four parametrizations that
we consider in this study (SKRA, SkT1, Skxs20, and SQMC700), but do
not show in Figure~\ref{fig:EOS}.

The two different behaviors in the proton fraction above nuclear
saturation density can be traced back to the symmetry energy
$\mathcal{S}$ (shown in Figure \ref{fig:symmetry}) and the related
neutron-proton chemical potential difference $\hat\mu$ (see Figure
\ref{fig:adiabats_x}).  The EOSs in Group I have $\mathcal{S}$ and
$\hat\mu$ for neutron rich matter that increase monotonically with
density.  Therefore, above nuclear saturation density, their proton
fraction $y$ for BEM also increases monotonically with density.  In
Group II, meanwhile, both $\mathcal{S}$ and $\hat\mu$ have a maximum
at a density above nuclear saturation and then decrease for higher
densities. Figure \ref{fig:adiabats_x} shows the density dependence 
of $\hat\mu$ for the $s=1\,k_B\,\mathrm{baryon}^{-1}$ adiabat, which
is qualitatively similar to the density dependence near zero 
temperature and entropy. In Reference \cite{stone:03}, Stone \etal 
argued that a key quantity for distinguishing between these two 
groups of Skyrme parametrizations is the density dependence of the 
symmetry energy, expressed by the asymmetry parameter
\begin{equation}
\label{eq:as} 
a_s(n)=\epsilon_B(n,y=1/2)-\epsilon_B(n,y=0)\,.
\end{equation} 
Stone \etal argued that parametrizations for which $a_s$ and, thus, 
$\hat\mu$ increases monotonically with density above saturation density, 
are more realistic since their behavior matches that observed for 
realistic nuclear potentials.
Realistic nuclear potentials, such as the Argonne 
$v_{18}$ \cite{wiringa:95}, CD-Bonn \cite{machleidt:01} and Nijmegen II 
\cite{stoks:94} are obtained by fitting 40 to 60 adjustable parameters to 
thousands of experimental data points of free nucleon-nucleon scattering 
and properties of the deuteron.

In Figure~\ref{fig:TOV}, we show the NS mass-radius curves that we
obtain by solving the TOV equations with our EOSs. We also indicate
the mass of the currently most massive known NS (PSR J0348+0432
\cite{antoniadis:13}) and the $2\sigma$ confidence region for the NS
mass-radius relationship given by ``model A'' of N\"{a}till\"{a} \etal
\cite{nattila:16}.  They obtained these constraints via a Bayesian
analysis of Type-I X-ray burst observations. For completeness, we
summarize in Table~\ref{Tab:TOV} key properties of the TOV NS
sequences obtained with all considered Skyrme parametrizations.

We note from Figure~\ref{fig:TOV} that there is a small difference between 
the mass-radius relation curves for NSs obtained with the LS220 and LS220$^*$ 
EOSs for low-mass NSs.  Recall that LS220 represents results from our full
SNA implementation of the LS220 parametrization and LS220$^*$ represents 
results obtained with an EOS table generated with the original code by \LS.  
The differences in the $M-R$ curves come from pressure differences in the 
range $10^{13}\unit{g\,cm^{-3}}\lesssim\rho\lesssim10^{14}\unit{g\,cm^{-3}}$
that are a result of the lower proton fraction limit of $y=0.035$ for 
LS220$^*$. 

Most of the considered Skyrme parametrizations are unable to support a
$2\,M_\odot$ NS. To date, the most massive observed NS have masses
$M_{J1614-2230}=1.97\pm0.04 M_\odot$ \cite{demorest:10} (recently
revised to $1.928\pm0.017$ by Fonseca \etal \cite{fonseca:16}) and
$M_{J0348+0432}=2.01\pm0.04 M_\odot$ \cite{antoniadis:13}.  The latter
is shown as a gray strip in Figure \ref{fig:TOV}.  Besides the LS220
parametrizations, only the NRAPR, SLy4, and KDE0v1 EOSs can account
for the existence of $2M_\odot$ NSs.

The radius $R_{1.4}$ of a canonical $1.4\,M_\odot$ NS was constrained
by Lattimer \etal to be in the range $10.5$ to $12.5\unit{km}$
\cite{lattimer:13}, by Guillot \etal to be in the $10$ to
$11.5\unit{km}$ range (\cite{guillot:13} as updated by
\cite{ozel:16}), and by N\"{a}till\"{a} \etal to be
$R_{1.4}=12.0\pm0.7\unit{km}$ \cite{nattila:16}.  As shown in Table
\ref{Tab:TOV}, the results for $R_{1.4}$ from all considered Skyrme
parametrizations are in agreement with these constraints.  Combining
the results for $R_{1.4}$ with the lower limit of the maximum NS mass
from observations, we see that LS220, NRAPR, SLy4, and KDE0v1
parametrizations are the ones which more closely fulfill current
astrophysical constraints.  Note, however, that the LS220
parametrization is an outlier and predicts $R_{1.4}$ about 1\,km
larger than the upper limit obtained by Guillot \etaln

\begin{table*}[htbp]
  \caption{\label{Tab:TOV} Summary of neutron star (NS) properties for 
  the considered Skyrme parametrizations. $M_{\mathrm{max}}$ is the 
  maximum NS mass, $R_{\mathrm{max}}$ is the radius of the maximum-mass 
  NS, ${(c_s/c)}_{\mathrm{max}}$ is its central speed of sound relative 
  to the speed of light $c$ and ${(n_c/n_0)}_{\mathrm{max}}$ is its central
  density relative to saturation density $n_0$.  $R_{1.4}$,      ${(c_s/c)}_{\mathrm{1.4}}$, 
  and ${(n_c/n_0)}_{\mathrm{1.4}}$ are the radius of a $1.4\,M_\odot$ NS, 
  its central speed of sound, and its central density, respectively.}
\begin{ruledtabular}
\begin{tabular}{l  D{.}{.}{1.2} D{.}{.}{1.2} D{.}{.}{1.3} D{.}{.}{1.3} D{.}{.}{2.2} D{.}{.}{1.3} D{.}{.}{1.3}}
\multicolumn{1}{c}{Parametrization}               &
\multicolumn{1}{c}{$M_{\mathrm{max}}$ $(M_\odot)$}       &
\multicolumn{1}{c}{$R_{\mathrm{max}}$ (km)}  &
\multicolumn{1}{c}{${(c_s/c)}_{\mathrm{max}}$}  &
\multicolumn{1}{c}{${(n_c/n_0)}_{\mathrm{max}}$}  &
\multicolumn{1}{c}{$R_{1.4}$ (km)} &
\multicolumn{1}{c}{${(c_s/c)}_{\mathrm{1.4}}$}  &
\multicolumn{1}{c}{${(n_c/n_0)}_{\mathrm{1.4}}$}  \\[0.2em]
\hline
LS220~\cite{lattimer:91}         &  2.04 & 10.61 & 0.880 &  7.18 & 12.66 & 0.556 & 2.84 \\
KDE0v1~\cite{agrawal:05}         &  1.97 &  9.80 & 0.966 &  7.75 & 11.67 & 0.617 & 3.46 \\
LNS~\cite{cao:06}                &  1.72 &  9.29 & 0.839 &  8.58 & 11.02 & 0.612 & 4.18 \\
NRAPR~\cite{steiner:05}          &  1.94 &  9.94 & 0.913 &  7.94 & 11.87 & 0.594 & 3.46 \\
SKRA~\cite{rashdan:00}           &  1.77 &  9.48 & 0.852 &  8.98 & 11.31 & 0.600 & 4.15 \\
SkT1~\cite{tondeur:84}           &  1.85 &  9.74 & 0.868 &  8.30 & 11.55 & 0.595 & 3.73 \\
Skxs20~\cite{brown:07}           &  1.74 &  9.63 & 0.811 &  8.86 & 11.52 & 0.587 & 4.12 \\
SLy4~\cite{chabanat:98}          &  2.05 &  9.99 & 0.990 &  7.47 & 11.72 & 0.624 & 3.35 \\
SQMC700~\cite{guichon:06}        &  1.76 &  9.40 & 0.853 &  8.60 & 11.16 & 0.609 & 4.06 \\
\end{tabular}
\end{ruledtabular}
\end{table*}

\begin{figure}[htbp]
\centering
\includegraphics[trim = 0.15in 0.3in 0.2in 1.0in, clip, width=\figurehalfwidth\textwidth]{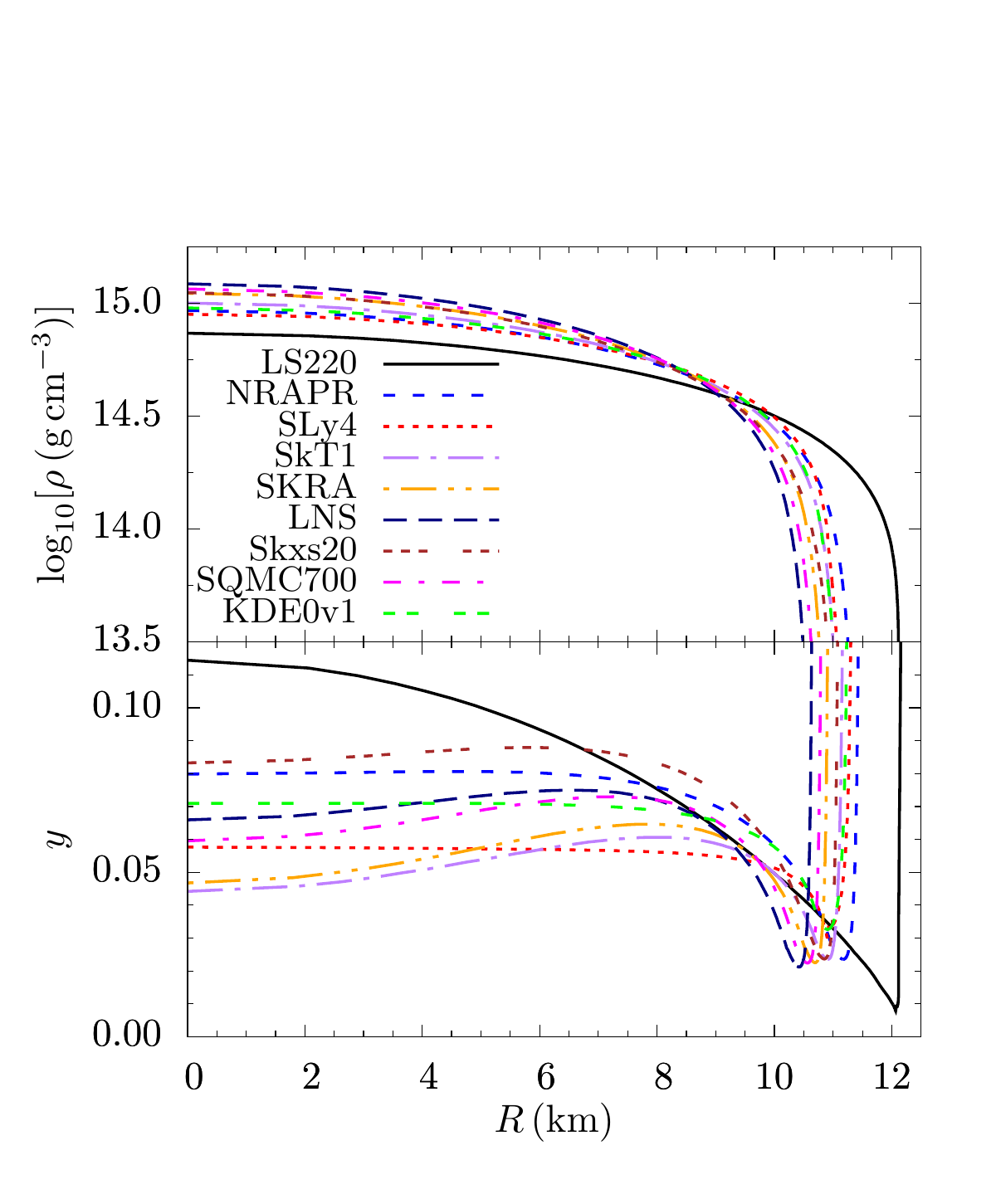}
\caption{(Color online) Radial rest-mass density (top panel) and proton 
  fraction profiles (bottom panel) of cold, beta-equilibrated $1.4\,M_\odot$
  neutron stars (NSs) obtained with the considered Skyrme parametrizations. 
  Note that LS220 is an outlier, yielding the lowest central density, the 
  largest radius, and the highest central proton fraction. This is due 
  primarily to its large $L$ parameter and the linear behavior of its 
  density-dependent symmetry energy, which results in the smallest symmetry 
  energies below saturation density and the largest symmetry energies above
  saturation density of all the EOS considered here (cf.~Figure~\ref{fig:symmetry}).
  We summarize key NS quantities in Table~\ref{Tab:TOV}. }
\label{fig:1.4}
\end{figure}

\begin{figure}[htbp]
\centering
\includegraphics[trim = 0.15in 0.3in 0.2in 1.0in, clip, width=\figurehalfwidth\textwidth]{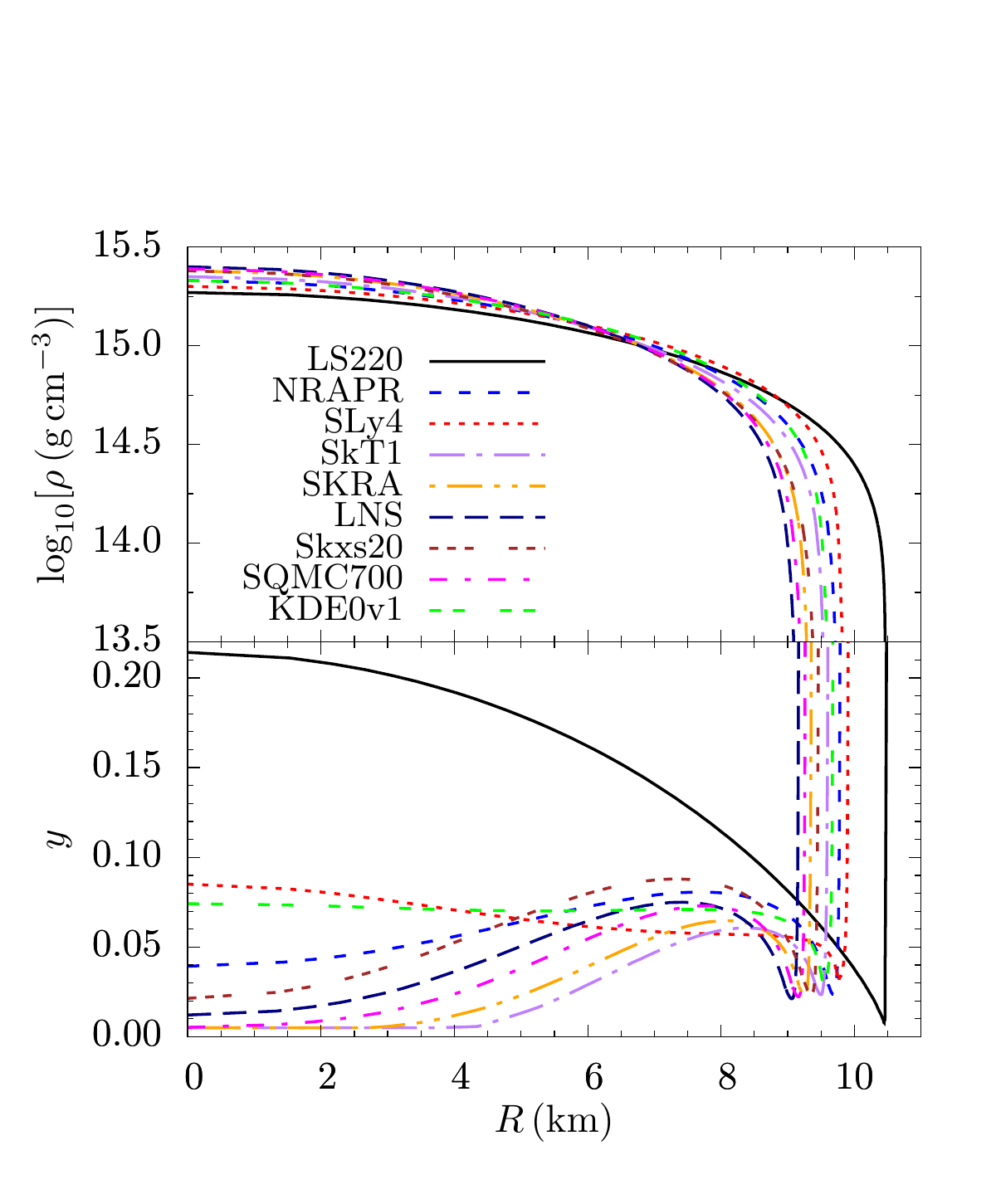}
\caption{(Color online) Radial rest-mass density (top panel) and 
  proton fraction (bottom panel) profiles of cold, beta-equilibrated 
  maximum-mass neutron stars (NSs) as predicted by the considered Skyrme
  parametrizations. Note that the maximum mass varies between 
  parametrizations. Table~\ref{Tab:TOV} summarizes key NS quantities for 
  all parametrizations. As in the $1.4\,M_\odot$ NS case shown in
    Figure~\ref{fig:1.4}, the LS220 parametrization is an outlier and
    yields the lowest central density, the highest central proton
    fraction, and the largest radius for its maximum-mass NS
    configuration.}
\label{fig:max}
\end{figure}

We plot density and proton fraction profiles for $1.4\,M_\odot$ NSs in
Figure~\ref{fig:1.4} and for maximum mass NS configurations in
Figure~\ref{fig:max}. We note that the LS220 parametrization predicts
lower densities and higher central proton fractions than the other
parametrizations.  This results from the LS220 EOS being stiffer than
all other considered EOSs and having significantly different
predictions for the density-dependent symmetry energy $\mathcal{S}(n)$
(cf.\ Equation~\ref{eq:symmetry_expansion} and
Table~\ref{Tab:bulk}.)  In the maximum-mass all NSs have central
densities far above $n_0$ and we can again separate the EOSs into two
groups.  In Group I, which encompasses the LS220, SLy4,and KDE0v1
parametrizations, the proton fraction increases toward the center of
the NS. In contrast, for Group II, which includes the other six
parametrizations, the proton fraction decreases toward the center of
the NS, even reaching $y=0$ for SKRA, SkT1, and SQMC700.

\begin{figure*}[htbp]
\centering
\includegraphics[trim = 0.1in 0.1in 0.2in 0.05in, clip, width=\figurewidth\textwidth]{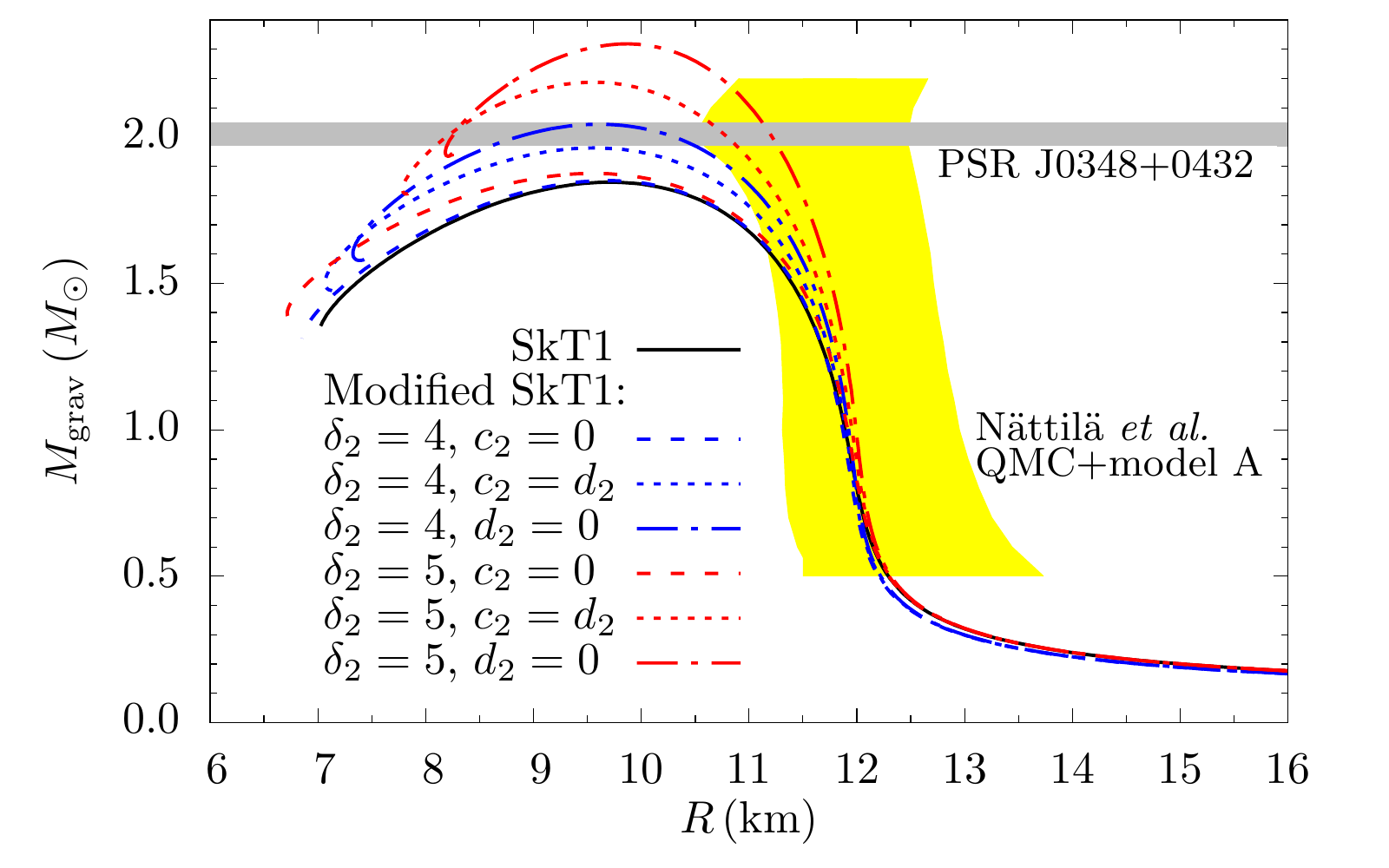}
\caption{(Color online) Mass-radius curves for cold, beta-equilibrated
  neutron stars (NS) obtained with the SkT1 parametrizations and its
  various high-density modifications.  The gray strip represents the
  mass $\mathrm{PSR\,J}0348+0432$ $M_{\mathrm{J}0348+0432}=2.01\pm0.04
  M_\odot$ \cite{antoniadis:13} and the yellow region indicates the NS
  mass-radius constraints from model A of N\"attil\"a \etal
  \cite{nattila:16}.}
\label{fig:TOV_SkT1}
\end{figure*}

\subsection{High density EOS modifications}\label{ssec:highNS}

Most Skyrme parametrizations fail to produce $2M_\odot$ NSs (see,
e.g., \cite{dutra:12} and Table~\ref{Tab:TOV}).  Since $2M_\odot$ NSs
have been observed in nature
\cite{antoniadis:13,demorest:10,fonseca:16}, a Skyrme parametrization
intended for astrophysical simulations should satisfy this lower limit
on the maximum NS mass.  However, Skyrme parameters are often chosen
to produce properties of nearly symmetric nuclear matter in the range
$\sim n_0/2-3n_0$ while densities in the center of a NS near maximum
mass may reach $\sim10n_0$ and matter may be very neutron rich.  Under
these conditions, the properties of matter are still fairly
unconstrained. Therefore, Skyrme interactions are not expected to be
valid beyond a density $n\sim3n_0\sim0.5\unit{fm}^{-3}$
\cite{stone:03,dutra:12}.  Thus, the maximum NS mass should not be
necessarily used to invalidate a Skyrme parametrization.  Ideally, a
model of high density matter should be matched to the Skyrme model at
high densities.  Dutra \etal use the Skyrme interaction up to about $3
n_0$ and match it to a different high-density EOS at higher densities
\cite{dutra:12}.  For $n \gtrsim 3 n_0$, they chose a zero-temperature
full quark-meson-coupling (FQMC) model \cite{stone:07b}, which
includes a full baryon octet in the high-density matter and predicts
$2M_\odot$ NSs, in agreement with observations.  Since we are
interested in finite-temperature EOSs, we instead propose a direct
modification of the Skyrme parametrization that affects its behavior
at high densities, but leaves the EOS properties at and below
saturation density unchanged.

For most Skyrme parametrizations the terms $\{c_i, d_i, \delta_i\}$ in
Equation \eqref{eq:Skyrme} are only non-zero for a single value of
$i$, \ie $i=1$.  The generalization to include extra non-zero terms is
straightforward.  We proceed as follows: we add an extra set of terms
$\{c_2, d_2, \delta_2\}$ to the sum in Equation \eqref{eq:Skyrme} with
$\delta_2 > 3$.  We adjust the values of $c_2$ and $d_2$ to minimally
impact the properties of saturation-density matter. As an example of
this high density EOS modification, we consider the SkT1
parametrization, which predicts a maximum NS mass of
$M_{\mathrm{max}}=1.85M_\odot$. We add extra terms to it so that the
contribution to the nuclear incompressibility $K_0=K(n_0,0.5)$ (see
Equation \ref{eq:K}) from the $i=2$ term is 1\% of the $i=1$ term
contribution, \ie
\begin{equation}\label{eq:ratio}
 \frac{\delta_2}{\delta_1}\frac{\delta_2-1}{\delta_1-1}
 \frac{c_2+d_2}{c_1+d_1}\frac{n_0^{\delta_2}}{n_0^{\delta_1}}=0.01.
\end{equation}
This choice, along with $\delta_1<\delta_2\lesssim10$, leaves all nuclear matter
properties at saturation density $n_0$ well within current known experimental
constraints, but significantly increases the pressure at high densities.

\begin{table*}[htbp]
  \caption{\label{Tab:TOV_SkT1} Summary of
      neutron star (NS) properties for the SkT1 parametrization and
      its high-density modifications. $M_{\mathrm{max}}$ is the
      maximum NS mass, $R_{\mathrm{max}}$ is the radius of the
      maximum-mass NS, ${(c_s/c)}_{\mathrm{max}}$ is its central speed
      of sound relative to the speed of light $c$ and
      ${(n_c/n_0)}_{\mathrm{max}}$ is its central density relative to
      saturation density $n_0$.  $R_{1.4}$,
      ${(c_s/c)}_{\mathrm{1.4}}$, and ${(n_c/n_0)}_{\mathrm{1.4}}$ are
      the radius of a $1.4\,M_\odot$ NS, its central speed of sound,
      and its central density, respectively. Note that the central
      speed of sound in the maximum-mass NS is superluminal for most
      of the modified EOSs.}
\begin{ruledtabular}
\begin{tabular}{l  D{.}{.}{1.2} D{.}{.}{1.2} D{.}{.}{1.3} D{.}{.}{1.3} D{.}{.}{2.2} D{.}{.}{1.3} D{.}{.}{1.3}}
\multicolumn{1}{c}{Parametrization}               &
\multicolumn{1}{c}{$M_{\mathrm{max}}$ $(M_\odot)$}       &
\multicolumn{1}{c}{$R_{\mathrm{max}}$ (km)}  &
\multicolumn{1}{c}{${(c_s/c)}_{\mathrm{max}}$}  &
\multicolumn{1}{c}{${(n_c/n_0)}_{\mathrm{max}}$}  &
\multicolumn{1}{c}{$R_{1.4}$ (km)} &
\multicolumn{1}{c}{${(c_s/c)}_{\mathrm{1.4}}$}  &
\multicolumn{1}{c}{${(n_c/n_0)}_{\mathrm{1.4}}$}  \\[0.2em]
\hline
SkT1~\cite{tondeur:84}          &  1.85 &  9.74 & 0.868 &  8.30 & 11.55 & 0.595 & 3.73 \\
$\delta_2=4,\,\,c_2=0\,$        &  1.85 &  9.65 & 0.891 &  8.49 & 11.55 & 0.597 & 3.69 \\
$\delta_2=4,\,c_2=d_2$          &  1.96 &  9.59 & 1.091 &  8.11 & 11.63 & 0.608 & 3.55 \\
$\delta_2=4,\,\,d_2=0\,$        &  2.04 &  9.58 & 1.209 &  7.92 & 11.68 & 0.616 & 3.45 \\
$\delta_2=5,\,\,c_2=0\,$        &  1.88 &  9.56 & 0.989 &  7.85 & 11.60 & 0.602 & 3.37 \\
$\delta_2=5,\,c_2=d_2$          &  2.19 &  9.56 & 1.533 &  6.84 & 11.73 & 0.638 & 3.12 \\
$\delta_2=5,\,\,d_2=0\,$        &  2.32 &  9.88 & 1.615 &  6.23 & 11.83 & 0.655 & 2.98 \\
\end{tabular}
\end{ruledtabular}
\end{table*}

We study here six modified SkT1 parametrizations.  Besides the choice
defined by Equation \eqref{eq:ratio}, we chose the exponent values
$\delta_2=4$ and $5$, and set the constants $c_2$ and $d_2$ such that
$c_2=0$, or $d_2=0$, or $c_2=d_2$.

In Figure~\ref{fig:TOV_SkT1}, we plot NS mass-radius curves for these
modified parametrizations. We summarize key properties of the TOV NS
sequences in Table \ref{Tab:TOV_SkT1}. Both figure and table show, as
expected, that the higher the exponent $\delta_2$ and the larger $c_2$
is with respect to $d_2$, the stiffer the EOS for cold BEM becomes.
This results in a higher maximum NS mass and a larger radius for the
$1.4M_\odot$ NS.  The main drawback of the proposed modifications is
that the speed of sound increases significantly for densities above
$3n_0$. It becomes superluminal at densities lower than those at the
center of maximum-mass NSs
(cf.~Table~\ref{Tab:TOV_SkT1}). Nevertheless, the modifications can be
useful for studying the impact of a higher maximum NS mass on
astrophysical simulations while keeping the properties of
saturation-density nuclear matter fixed.

\section{Adiabatic Compression}\label{sec:Compression}

To check the thermodynamic consistency of our code and of the EOS
tables it generates, we perform adiabatic compression tests.  An
isolated system that is slowly compressed from a lower to a higher
density should retain its initial entropy.  To test this, we generate
EOS tables for different Skyrme parametrizations in the ranges of
density $n$, temperature $T$, and proton fraction $y$, given in
Table~\ref{Tab:range}.  We set the table resolution to 30 points per
decade in temperature and density and 1 point every 0.01 in proton
fraction. We also consider tables with double the resolution across
each EOS dimension. The lower resolution is similar to that of the
tables available at
\url{https://stellarcollapse.org/equationofstate}. These older tables
are described by O'Connor \& Ott in \cite{oconnor:10} and have been
used frequently in astrophysical simulations. Following
\cite{oconnor:10}, we interpolate tri-linearly in $n$, $T$, and
$y$. We find $T$ for a given $n$, $y$, and specific internal energy
$\epsilon$ or specific entropy $s$ via Newton-Raphson root finding.

\begin{table}[tbp]
  \caption{\label{Tab:range} Ranges in density
      $n$, temperature $T$, and proton fraction $y$, and the number of
      EOS table points in each dimension for our standard-resolution
      EOS tables.  The high-resolution tables have the same range, but
      contain twice the number of points in each dimension.}
\begin{ruledtabular}
\begin{tabular}{c | D{.}{.}{3.3} D{.}{.}{1.3} D{.}{ }{3.1} }
\multicolumn{1}{c}{Parameter}&
\multicolumn{1}{c}{minimum}&
\multicolumn{1}{c}{maximum}&
\multicolumn{1}{c}{points}\\
\hline
$\log_{10}[n(\unitn{fm}^{-3})]$  & -12.2   &  0.8    & 391  \\
$\log_{10}[T(\unitn{MeV})]$      &  -3.0   &  2.4    & 163  \\
$y$                              &   0.005 &  0.655  &  66  \\
\end{tabular}
\end{ruledtabular}
\end{table}

\begin{figure*}[htbp]
\centering
\includegraphics[trim = 0.15in 0.35in 0.1in 0.3in, clip, width=\figurewidth\textwidth]{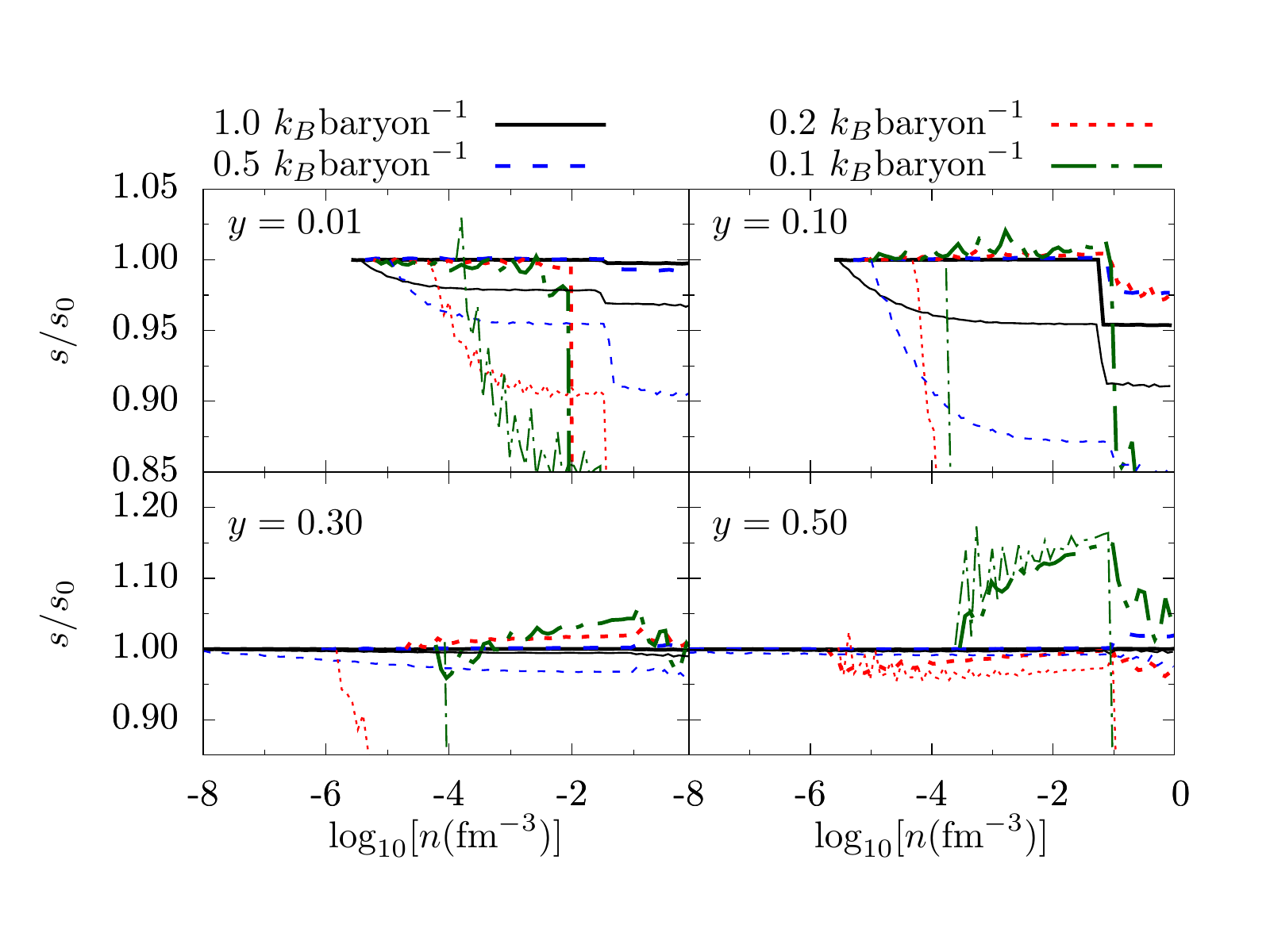}
\caption{(Color online) Relative error in the
    specific entropy as a function of density during adiabatic
    compression tests with the full EOS based on the SLy4
    parametrization and the single-nucleus approximation at all
    densities. The different panels show results for proton fractions
    $y = 0.01, 0.10, 0.3$, and $0.5$. All curves start at
    $T=0.01\,\unit{MeV}$ and we choose the initial densities to obtain
    starting values of the specific entropy of $s_0=0.1$, $0.2$,
    $0.5$, and $1.0\,k_B\,\mathrm{baryon}^{-1}$. Thick (thin) curves
    correspond to tests with the high (standard) resolution tables
    (cf.~Table~\ref{Tab:range}).  For $s_0 \gtrsim
    0.5\,\mathrm{k_B\,baryon}^{-1}$ and proton fractions $y \gtrsim
    0.3$, the standard-resolution tables perform very well. In stellar
    collapse, the specific entropy always stays higher than
    $0.5\,\mathrm{k_B\,baryon}^{-1}$ and proton fractions below
    $\sim$$0.3$ are not reached until the final phase of
    collapse. Errors at lower $s_0$ and $y$ are largely numerical and
    are reduced by employing the high resolution table. 
    However, large changes in entropy can still occur near the first-order
    phase transition between non-uniform and uniform matter at $n
    \simeq 0.1\,\mathrm{fm}^{-3}$.}
\label{fig:compression}
\end{figure*}

In our adiabatic compression tests, for a given proton fraction, we
set the system to an initial temperature $T=10^{-2}\unit{MeV}$ and
determine the initial densities for which the entropy has values of
$0.1$, $0.2$, $0.5$, and $1.0\,k_B\,\mathrm{baryon}^{-1}$. Every step,
the density is increased by $\delta n = 10^{-3}n$ until the system
reaches a density of $1\unit{fm}^{-3}$ or its temperature excedes the
maximum of our tables ($T_{\mathrm{max}} = 250\unit{MeV}$).  As the
system is compressed, we integrate the first law of thermodynamics
using a fourth-order Runge-Kutta integrator and determine the ratio of
the entropy $s(n)$ to the initial entropy $s_0$ as a function of
density. As a representative example result, we show in
Figure~\ref{fig:compression} the fractional changes in entropy during
the compression for both the high-resolution and standard-resolution
SLy4 tables.

We see from Figure~\ref{fig:compression} that for specific entropies
of $s \gtrsim$\,$0.5\,k_B\,\mathrm{baryon}^{-1}$ and proton fractions
$y \gtrsim$\,$0.3$, even the standard-resolution tables yield nearly
perfectly adiabatic compression. This bodes very well for stellar
collapse and CCSN simulations, since entropies always stay higher than
$0.5\,k_B\,\mathrm{baryon}^{-1}$ and proton fractions below
$\sim$\,$0.3$ are not reached until the final phase of collapse.

At lower entropies and proton fractions, we observe substantial
deviations from adiabatic compression with entropy errors of order
$10\%$ or greater with the standard-resolution tables. This issue is
largely numerical and due to interpolation and root-finding errors,
since the high-resolution tables yield much better results. However,
large changes in entropy can still occur near the first-order phase
transition between non-uniform and uniform nuclear matter near
$n\simeq10^{-1}\unit{fm}^{-3}$.
  
For comparison, we carry out adiabatic compression tests also for the
tables of \cite{oconnor:10}. We find that even our standard-resolution
tables yield smaller entropy errors than any of the EOS tables of
\cite{oconnor:10} available at
\url{https://stellarcollapse.org/equationofstate}.

Finally, adding a transition from SNA to NSE at low densities, as we
discuss below in Section \ref{sec:NSE}, only leads to small
quantitative changes compared to the results presented in this
section.

\section{Application to Stellar Core Collapse}
\label{sec:Collapse}

We carry out a set of example core collapse and postbounce CCSN
simulations to investigate how our new EOSs perform in this important
astrophysical scenario and how they influence core collapse,
postbounce evolution, and black hole formation. 
Before discussing the CCSN simulations, we describe how we
modify our EOSs at low density to include an ensemble of nuclei in NSE.

\subsection{Nuclear statistical equilibrium (NSE)}\label{sec:NSE}

NSE holds for temperatures $T \gtrsim 0.5\,\mathrm{MeV}$ at which
forward and backward nuclear reaction rates are so high that
equilibrium is obtained faster than any other timescale in the
system. At low density and moderate temperatures, the NSE equilibrium
state of matter includes an ensemble of nuclear species (nuclides) and
SNA is not a good approximation for describing the thermodynamics. SNA
predicts different thermodynamic quantities, average nuclear binding
energies, and neutrino opacities than a model assuming NSE
\cite{hix:03, souza:09}.  Furthermore, SNA predicts a single average
nucleus whose properties can differ significantly from the observed
properties of nuclei due to shell closures, pairing, and many
body-body effects missing from the simple liquid-drop SNA. Conversely,
NSE breaks down at high densities when interactions between the
nuclear interior and the surrounding medium become important.  This
can be partially overcome by including excluded volume corrections in
the NSE formulation \cite{hempel:10}. However, such an approach does
not account for changes in nuclear shapes and requires a very large,
neutron rich ensemble of nuclei to reasonably reproduce the high
density, low proton fraction composition. To alleviate the
aforementioned issues with the SNA while still retaining its
advantages at high density, we transition from SNA to an NSE EOS at
densities where nuclear interactions are small and SNA and NSE can be
smoothly matched.

Another reason for transitioning from the SNA to NSE is that at low
density and temperature, the abundances of nuclei can fall out of
equilibrium, which requires smoothly transitioning from material in
NSE to following a network of reactions between separate nuclides. To
perform such a transition in a thermodynamically consistent manner,
the same set of nuclei and nuclear partition functions must be used to
calculate both the equilibrium number densities and for the
non-equilibrium evolution. The SNA will not satisfy this consistency
condition, but it can be easily enforced by using an NSE EOS at
moderate temperature and density.

For completeness, we provide a full discussion of our standard
treatment for an ensemble of nuclei in NSE in Appendix~\ref{app:NSE}.
Once the free energy densities for the SNA and NSE phases,
$F_{\mathrm{SNA}}$ and $F_{\mathrm{NSE}}$, respectively, have been
determined, we combine the two using a density dependent function
$\chi(n)$, \ie
\begin{equation}
F_{\mathrm{MIX}} = \chi(n) F_{\mathrm{SNA}} 
                  + [1-\chi(n)] F_{\mathrm{NSE}}\,.
\end{equation}
Here, the $\mathrm{SNA}$ ($\mathrm{NSE}$) subscripts denote the
contribution to the thermodynamical quantities from the high (low)
density parts of the EOS. $F_{\mathrm{SNA}}$ is given by Equation
\eqref{eq:free} while $F_{\mathrm{NSE}}$ is given by Equation
\eqref{eq:FNSE}.  The limits of the function $\chi(n)$ are chosen so
that it goes to zero at low densities and to one at high densities.
We mix the two using the smooth choice for $\chi(n)$,
\begin{equation}\label{eq:merge}
\chi(n) = \frac{1}{2}\left[1 + \tanh
\left(\frac{\log_{10}(n)-\log_{10}(n_t)}{n_\delta}\right)\right],
\end{equation}
where $n$ is the density of the system, $n_t$ the center of the 
transition, and $n_\delta$ its width.  We set the center of the 
transition density $n_t=10^{-4}\unit{fm}^{-3}$ $(\simeq 1.7\times
10^{11}\,\mathrm{g\,cm}^{-3})$ and its dimensionless width 
$n_\delta=0.33$. This choice guarantees that the transition happens 
in a region where differences in the nuclear contributions to the 
total pressure, entropy, and energy density in the NSE and SNA 
treatments are relatively small, at least for matter with small 
isospin asymmetry, where EOS constraints are more accurately known. 
Furthermore, this transition is at sufficiently low densities that the 
EOS is dominated by the electron (photon) contribution at low (high) 
temperatures. At the same time, the transition density is high enough
that above $n_t$ we expect large deformed nuclei and the pasta phases 
to dominate, which are well described in the SNA approximation.

Because $\chi(n)$ is density dependent, the transition procedure
introduces corrections to the pressure and other derivatives with
respect to density in the transition region that are of order
$F_{\mathrm{SNA}}-F_{\mathrm{NSE}}$. For example, in the mixing
region, the pressure is given by
\begin{align}
 P_{\mathrm{MIX}}
    &=n^2\left.\frac{\partial (F_B/n)}{\partial n}\right|_{T,y}\nonumber\\
    &=\chi(n) P_{\mathrm{SNA}} + [1-\chi(n)] P_{\mathrm{NSE}} \nonumber\\
    &\qquad + n^2 \frac{\partial \chi(n)}{\partial n}\left(F_{\mathrm{SNA}}-F_{\mathrm{NSE}}\right)\,.
\end{align}
Other quantities are readily computed.  In practice, we find that the
corrections due to $\chi(n)$ are small compared to the other
contributions to the free energy.  Although this procedure is ad-hoc,
it results in a thermodynamically consistent EOS and does not require
the calculation of a more complicated phase transition.

\subsection{Stellar Collapse}

To study the impact of our new EOSs on stellar collapse, we employ the 
open-source spherically-symmetric (1D) general-relativistic hydrodynamics 
code \texttt{GR1D} \cite{oconnor:10,oconnor:11,oconnor:15}.  For simplicity 
and efficiency, we employ its neutrino leakage/heating scheme described in
\cite{oconnor:10} and postpone detailed radiation-hydrodynamics studies 
using \texttt{GR1D}'s two-moment transport solver to future work. 
Deleptonization during the collapse phase is handled via a parametrization 
of  the proton fraction $y$ as a function of rest-mass density $\rho$ as 
proposed by Liebend\"orfer \cite{liebendoerfer:05fakenu} with the parameters 
given in \cite{oconnor:10}. \texttt{GR1D}'s EOS routines interpolate tabulated
thermodynamic variables such as pressure, specific internal energy, specific 
entropy, etc. linearly in $\log_{10} \rho$, $\log_{10} T$, and $y$, and do not 
obtain them via the interpolated free energy (and its derivatives). 
This means that thermodynamic consistency is not guaranteed, is subject to 
interpolation errors and EOS table resolution, and must be checked 
\cite{swesty:96}.

We study core collapse and postbounce evolution in two progenitor
stars: (1) In the $15$-$M_\odot$ progenitor of Woosley and Weaver
(W\&W hereafter) \cite{woosley:95}, which has been used widely in
the literature. (2) In the $40$-$M_\odot$ progenitor of Woosley and
Heger (W\&H hereafter) \cite{woosley:07}, which has a very massive,
high-compactness core and is expected to form a black hole (BH)
\cite{oconnor:11}. For the $15$-$M_\odot$ progenitor, we use a
computational grid with $1000$ grid cells, constant cell size of
$100\,\mathrm{m}$ out to a radius of $20\,\mathrm{km}$, and then
geometrically increasing cell size to an outer radius of
$10\,000\,\mathrm{km}$. For the $40$-$M_\odot$ progenitor, whose
collapse we evolve until BH formation, we use $1500$ grid
cells, a constant cell size of $75\,\mathrm{m}$ out to
$25\,\mathrm{km}$, and geometrically increasing cell size to an outer
radius of $10\,000\,\mathrm{km}$.

Stellar evolution codes use EOSs (e.g., \cite{timmes:99}) that can
differ substantially from the EOSs presented in this paper. On the one
hand, in the NSE region, the predicted pressure, entropy, etc.\ 
depend on the number of nuclides tracked in the stellar model. On the
other hand, in the non-NSE region, compositional details will depend
on the employed nuclear reaction network and, again, composition will
affect the thermodynamical variables. These differences between EOSs are
not negligible for core collapse simulations: at the onset of
collapse, small variations in the pressure profile between stellar and
core collapse EOSs can alter the hydrodynamics of the core, and may
accelerate or delay collapse.

In order to start our simulations in a way that is as consistent as
possible with the hydrodynamical structure of our progenitor models,
we map the stellar rest-mass density $\rho$, proton fraction $y$, and
pressure $P$ to \texttt{GR1D}, and then find temperature $T$ (and
specific internal energy, entropy, etc.) using the EOS table. We
stress that our approach for setting up the initial conditions results
in differences between the original stellar profile and the
\texttt{GR1D} initial conditions in all quantities except $\rho$, $y$,
and $P$.  Also note that for the purpose of this study, we assume NSE
throughout the part of the star mapped to \texttt{GR1D}'s grid. This
is an approximation that will need to be relaxed in the future, since
the outer regions of the core and the silicon-rich and oxygen-rich
layers are not in NSE.

In most of our core collapse simulations, we use our
standard-resolution EOS tables described in Table~\ref{Tab:range}. Our
adiabatic compression tests in Section~\ref{sec:Compression} suggest
that higher resolution tables lead to more accurate adiabatic collapse
results, in particular for low entropies. However, in our collapse
simulations, entropies are always sufficiently high that using our
standard-resolution tables yields excellent results. Tests with the
high-resolution tables show only negligible differences in the
simulation results. Only in the case of very stiff EOS, such as
SkT1$^*$ (see below), we find it necessary to use higher-resolution
tables to accurately track simulations on the route to BH formation at
central proto-NS densities above $\sim$\,$10^{15}\unit{g\,cm}^{-3}$.

\subsubsection{$15$-$M_\odot$ Progenitor}
\label{ssec:15M}

\begin{figure}[t]
\centering
\includegraphics[trim = 0.in 0.25in 0.in 0.in, clip, width=\figurehalfwidth\textwidth]{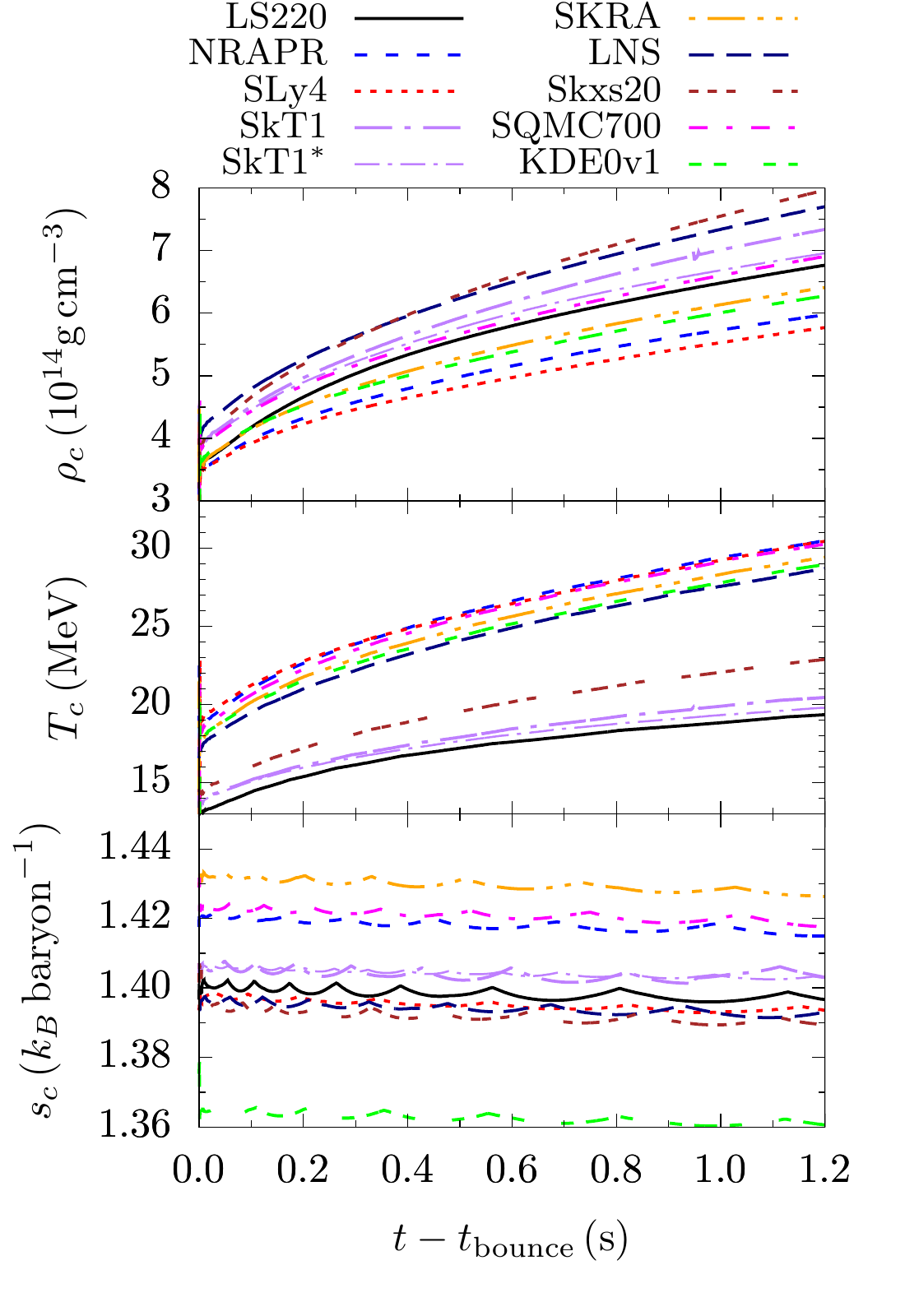}
\caption{(Color online) Results from core collapse simulations with
  the $15$-$M_\odot$ progenitor and EOSs generated with various Skyrme
  parametrizations and a $3\,335$-nuclide NSE EOS at low
  densities. From top to bottom, we show $1.2\,\mathrm{s}$ of
  postbounce evolution of the central density $\rho_c$, central
  temperature $T_c$, and central specific entrop $s_c$. Note that the
  entropy stays roughly constant (modulo mild numerical oscillations)
  throughout the postbounce evolution, as it should for
  thermodynamically consistent EOSs. As postbounce accretion adds mass
  to the proto-NS, it contracts, which is marked by an increase in
  $\rho_c$ and softer EOSs result in a steeper increase. The splitting
  of the $T_c$ evolutions into two groups of parametrizations can be
  understood by considering that those resulting in lower temperatures
  have a larger effective nucleon mass (see~Section~\ref{sec:bulk} and
  Table~\ref{Tab:bulk}).}
\label{fig:15M_FULLNSE}
\end{figure}

\begin{figure}[t]
\centering
\includegraphics[trim = 0.in 0.25in 0.in 0.in, clip, width=\figurehalfwidth\textwidth]{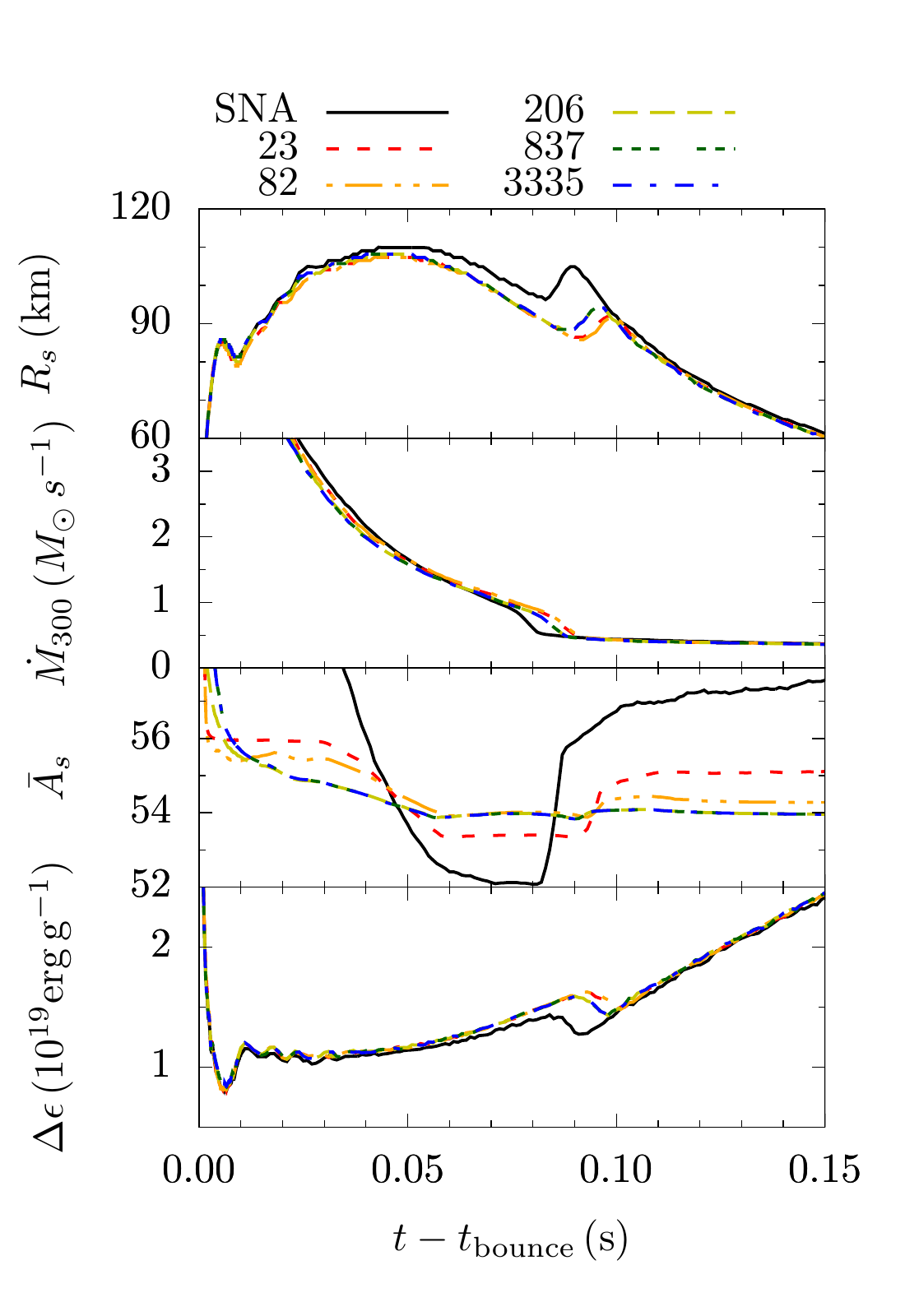}
\caption{(Color online) Results of core collapse simulations with the
  $15$-$M_\odot$ progenitor and the LS220 Skyrme parametrization. We
  compare results obtained for pure SNA (at all densities) with
  results from simulations that use SNA at high densities, smoothly
  matched to an NSE EOS with varying number of nuclides at low
  densities (see Section~\ref{sec:NSE} for details). From top to
  bottom, we plot the postbounce evolution of shock radius $R_s$,
  accretion rate $\dot{M}_{300}$ at a radius of $300\unit{km}$,
  average nuclear mass number $5\unit{km}$ above the shock, and
  difference in specific internal energy, $\Delta\epsilon$
  $5\unit{km}$ above and $5\unit{km}$ below the
  shock. Note that the pure-SNA simulation predicts a
    slightly larger shock radius than the SNA+NSE simulations between
    $\sim$$50\,\mathrm{ms}$ and $\sim$$80\,\mathrm{ms}$ after bounce.
    This is a consequence of the SNA simulation having a slightly
    lower accretion rate and slightly less bound nuclei crossing the
    shock in that interval.}
\label{fig:15M_shock}
\end{figure}

We follow core collapse and postbounce evolution up to
$1.2\,\mathrm{s}$ after bounce in the $15$-$M_\odot$ progenitor. While
this star is expected to explode in nature (e.g., \cite{whw:02}), we
use the default scaling factor $f_\mathrm{heat} = 1$ for neutrino
heating in \texttt{GR1D} and do not obtain an explosion in our
\texttt{GR1D} simulations.  This is consistent with more elaborate 1D
radiation-hydrodynamic simulations (e.g., \cite{buras:06a}).

In a first set of simulations, we focus on the effects of different
Skyrme parametrizations. We employ ten different EOSs -- the nine
Skyrme parametrizations discussed in Section~\ref{sec:Results} and one
of the modified versions of the SkT1 parametrizations stiffened at
high density studied in Section \ref{ssec:highNS}. We call this
parametrization SkT1$^*$ and use $\delta_2=5$, $d_2=0$, which produces
the highest cold NS mass for SkT1. We merge the SNA Skyrme EOSs with
an NSE EOS containing 3\,335 nuclides following the prescription
detailed in Section \ref{sec:NSE}. We employ a transition density
$n_t=10^{-4}\unit{fm}^{-3}$ ($\rho_t \simeq 1.67 \times
10^{11}\,\mathrm{g\,cm}^{-3}$) and dimensionless width of $n_\delta =
0.33$ (cf.~Equation~\ref{eq:merge}).

The time from the onset of collapse to core bounce is approximately
the same for all simulations, $t_{\mathrm{bounce}}=0.331\pm0.008\unit{s}$, 
since it is mostly a function of the low density part of the EOS, which is 
the same for all tables which include NSE at low densities.  In
Figure~\ref{fig:15M_FULLNSE}, we plot the postbounce evolution of the
central density, central temperature, and central specific entropy
resulting from the 10 different Skyrme parametrizations.  As the
proto-NS's mass increases due to the settling of material that
accretes through the stalled supernova shock, its core is
adiabatically compressed since the time scale for neutrino diffusion
is much longer than the accretion time scale. Core density and
temperature increase, while the central entropy stays nearly constant
over the $1.2\,\unit{s}$ of postbounce time we simulate.  The latter
is a further demonstration of the thermodynamic consistency of our
EOSs. We attribute the small wiggles and the small secular drift in
the central entropy to interpolation errors and the finite resolution
of our EOS tables.

The postbounce central density and temperature evolutions shown in
Figure~\ref{fig:15M_FULLNSE} exhibit significant dependence on Skyrme
parametrization. The ordering of the central density evolution and its
slope roughly follows the stiffness of the EOS. Softer EOSs (lower
maximum NS mass) have higher densities at bounce and a steeper
postbounce slope in $\rho_c$ than stiffer EOSs. The two bracketing
cases are SLy4 ($M_\mathrm{max} \sim 2.05 M_\odot$) and Skxs20
($M_\mathrm{max} \sim 1.74\,M_\odot$). Note that the SkT1 and the
SkT1$^*$ parametrizations start out at the same $\rho_c$ at bounce,
but that the slope of $\rho_c$ in the SkT1$^*$ simulation becomes
gradually shallower as the proto-NS contracts. This is a direct
consequency of the stiffened high-density part of SkT1$^*$.

The $T_c$ evolution in Figure~\ref{fig:15M_FULLNSE} is divided into
two groups. In the first group, containing LS220, Skxs20, SkT1, and
SkT1$^*$, $T_c$ right after bounce is $\sim 13 - 15\unit{MeV}$ and
rises to $T_c\sim 19 - 22\unit{MeV}$ within the first second after
bounce.  For the second group, containing all other parametrizations,
we find $T_c \sim 17 - 19\unit{MeV}$ right after bounce, rising to
$\sim 28 - 30\unit{MeV}$ a second after bounce. These pronounced
differences in core temperatures result from different treatments of
the nucleon effective masses in Equations~\eqref{eq:Skyrme} and 
\eqref{eq:mstar} with the parameters in Table~\ref{Tab:bulk}.  
At a fixed density and proton fraction, the thermal contribution to the 
free energy of uniform matter only depends on the chosen Skyrme 
parametrization through the effective masses, at least in the mean field
approximation. For non-relativistic particles, temperature enters the baryon
entropy for fixed neutron and proton densities only through the
combinations $m_t^* T$. Therefore, if $m_t^*\eqsim m_t$,
then temperatures at similar density and entropy will be smaller than
in cases where $m_t^*<m_t$. This explains the $T_c$-grouping in
Figure~\ref{fig:15M_FULLNSE}. 

In a second set of simulations with the $15$-$M_\odot$ progenitor, we
investigate the sensitivity of the collapse and postbounce evolution
to the number of nuclides included in the low-density NSE part of the
EOS. We choose the frequently used LS220 Skyrme parametrization for
the high-density SNA part and match it to a set of low-density NSE
EOSs with $23$, $82$, $206$, $837$, and $3\,335$ nuclides, using the
same matching parameters as before.  Each larger list of nuclides
includes all of the nuclides of the smaller nuclide lists and
we provide all lists at \url{https://stellarcollapse.org/SROEOS}. We
also carry out a simulation with an EOS table that uses the SNA at all
densities.

Since the low-density EOS is dominated by relativistic degenerate
electrons, differences in the number of NSE nuclei have only a mild
effect on the collapse dynamics. We find times to core bounce that
vary by less than $2\,\mathrm{ms}$. The SNA simulation reaches bounce
at $0.334\,\unit{s}$, while all simulations that include nuclides in
NSE reach bounce within a very similar time,
$t_{\mathrm{bounce}}=0.332\pm0.001\,\unit{s}$. The
  close agreement of the SNA and NSE bounce times is particular to the
  LS220 parametrization and the $15$-$M_\odot$ progenitor. For the
  same progenitor and other parameterizations, we find that SNA
  simulations reach bounce up to $20-30\,\mathrm{ms}$ later than NSE
  simulations. For other progenitor stars that have lower-density
  cores at the onset of collapse, the differences can be even larger
  (see Section~\ref{ssec:40M}, where we discuss results for a
  $40$-$M_\odot$ progenitor).

In Figure~\ref{fig:15M_shock}, we plot the postbounce evolution of the
shock radius, the mass accretion rate at a radius of $300\unit{km}$,
average nuclear mass number $\bar{A}$ at 5\unit{km} above the shock, and the
difference $\Delta\epsilon$ in specific internal energy between
$5\unit{km}$ above and 5\unit{km} below the shock. We focus on the
first $150\unit{ms}$ of postbounce evolution.

Figure~\ref{fig:15M_shock} shows that the the shock
  radius and the postbounce accretion rate are only mildly sensitive
  to differences between SNA and NSE at low densities.  Furthermore,
  the number of nuclides included in the NSE EOS also has little
  effect on the collapse properties. One notes that the SNA EOS leads
  to higher early accretion rates and a slightly earlier drop in the
  accretion rate, since the density discontinuity that is present at
  the edge of the iron core in the $15$-$M_\odot$ progenitor reaches
  small radii and the shock earlier. This is also reflected in the
  shock radius evolution, which shows a pronounced excursion when the
  density drop reaches the shock. This excursion is larger for the
  simulation with the pure SNA EOS since less energy is needed to
  break up the nuclei formed just above the shock radius (bottom panel
  of Figure~\ref{fig:15M_shock}).  We find that these qualitative
  findings are independent of the high-density part of the EOS.

From the third panel of Figure~\ref{fig:15M_shock}, showing the average 
nuclear mass $\bar{A}$ just above the shock, we note that $\bar{A}$ and 
the nuclear binding energy predicted by the LS220 SNA is very different 
from what NSE predicts. It also appears that one needs in excess of 
$\sim$\,$82$ nuclides for NSE to predict a converged $\bar{A}$, though 
this is likely sensitive to the specific set of nuclides included. The 
large differences in $\bar{A}$ and nuclear binding energy translate to 
the differences in $\Delta \epsilon$ shown in the bottom panel.  
These, in turn, explain the different shock radii plotted in the top panel
and discussed in the above.

\subsubsection{$40$-$M_\odot$ Progenitor}
\label{ssec:40M}

\begin{figure}[tb]
\centering
\includegraphics[trim = 0.in 0.25in 0.in 0.in, clip, width=\figurehalfwidth\textwidth]{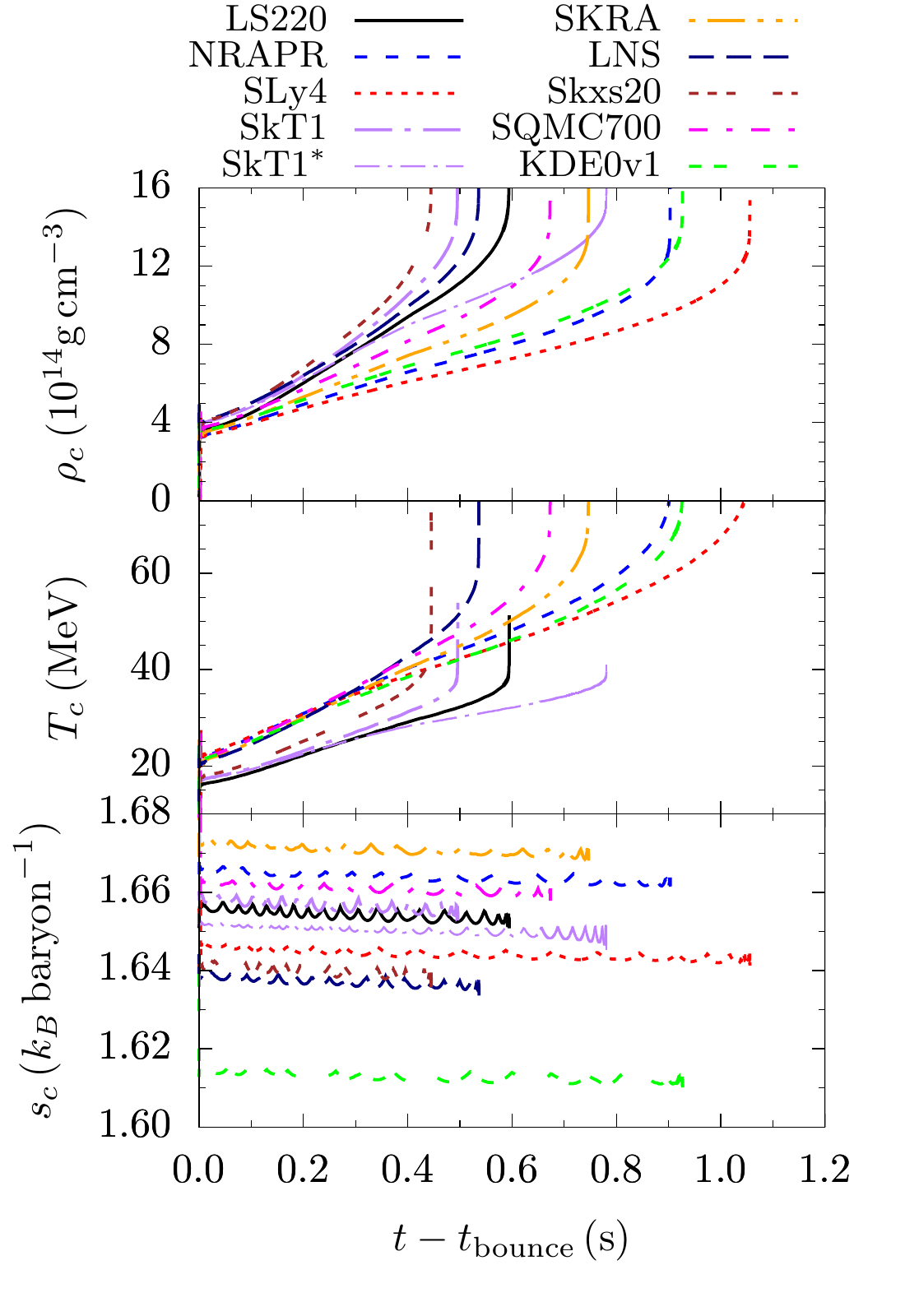}
\caption{(Color online) From top to bottom we plot the postbounce time
  evolution of the central density $\rho_c$, central temperature
  $T_c$, and central specific entropy $s_c$ for the black hole (BH)
  formation simulations with the $40M_\odot$ progenitor.  Proto-NS
  collapse and BH formation are marked by a sudden extreme steepening
  of the $\rho_c$ slope. The different graphs correspond to
  simulations with different Skyrme parametrizations. We employ a
  $3\,335$-nuclide NSE EOS at low densities
  (cf.\ Section~\ref{sec:NSE}). Note that the specific entropy stays,
  as it should, roughly constant (modulo numerical noise that can be reduced with
  higher-resolution EOS tables) throughout
  the postbounce evolution and up to BH formation. Thermal pressure
  support in the proto-NS mantle plays an important role in supported
  proto-NS masses that are $0.2-0.6\,M_\odot$ higher than the maximum
  cold NS mass. Thermal contributions are largest for those
  parametrizations that result in low effective nucleon masses and
  higher proto-NS temperatures.  }
\label{fig:40M_NSE}
\end{figure}

\begin{table*}[htbp]
\caption{\label{Tab:bounce} Bounce time $t_{\mathrm{bounce}}$, BH
  formation time $t_{\mathrm{BH}}$,
  $t_{\mathrm{BH}}-t_{\mathrm{bounce}}$, and proto-neutron star
  maximum gravitational mass $M_g$ for the $40M_\odot$ progenitor of
  W\&H. Results are for EOSs using different Skyrme parametrizations
  in the single-nucleus approximation (SNA) and in the SNA merged with
  an NSE EOS at $n_t=10^{-4}\unit{fm^{-3}}$ with $n_\delta=0.33$, see
  Section \ref{sec:NSE}.  SkT1$^*$ is the modified versions of the
  SkT1 parametrization studied in Section \ref{ssec:highNS} with
  $\delta_2=5$ and $d_2=0$.  }
\begin{ruledtabular}
\begin{tabular}{l | c c c c | c c c c}
\multicolumn{1}{c|}{EOS} &
\multicolumn{4}{c|}{SNA+NSE} &
\multicolumn{4}{c}{SNA} \\
 & $t_{\mathrm{bounce}}$ (s)
 & $t_{\mathrm{BH}}-t_{\mathrm{bounce}}$ (s)
 & $t_{\mathrm{BH}}$ (s)
 & $M_{g}$ ($M_\odot$)
 & $t_{\mathrm{bounce}}$ (s)
 & $t_{\mathrm{BH}}-t_{\mathrm{bounce}}$ (s)
 & $t_{\mathrm{BH}}$ (s)
 & $M_{g}$ ($M_\odot$) \\
\hline
LS220 \cite{lattimer:91}    & 0.490 & 0.595 & 1.085 & 2.260 & 0.513 & 0.565 & 1.078 & 2.275 \\
KDE0v1 \cite{agrawal:05}    & 0.490 & 0.927 & 1.417 & 2.425 & 0.515 & 0.914 & 1.429 & 2.441 \\
LNS \cite{cao:06}           & 0.490 & 0.537 & 1.027 & 2.234 & 0.525 & 0.515 & 1.040 & 2.255 \\
NRAPR \cite{steiner:05}     & 0.491 & 0.902 & 1.393 & 2.418 & 0.531 & 0.873 & 1.404 & 2.442 \\
SKRA \cite{rashdan:00}      & 0.491 & 0.746 & 1.237 & 2.345 & 0.532 & 0.716 & 1.248 & 2.371 \\
SkT1 \cite{tondeur:84}      & 0.489 & 0.497 & 0.986 & 2.204 & 0.530 & 0.478 & 1.008 & 2.214 \\
SkT1$^*$\cite{tondeur:84}   & 0.489 & 0.782 & 1.271 & 2.327 & 0.532 & 0.741 & 1.273 & 2.334 \\
Skxs20 \cite{brown:07}      & 0.488 & 0.448 & 0.936 & 2.182 & 0.545 & 0.406 & 0.951 & 2.182 \\
SLy4 \cite{chabanat:98}     & 0.493 & 1.053 & 1.546 & 2.488 & 0.630 & 0.936 & 1.566 & 2.525 \\
SQMC700 \cite{guichon:06}   & 0.493 & 0.671 & 1.164 & 2.310 & 0.741 & 0.488 & 1.229 & 2.368 \\
\end{tabular}
\end{ruledtabular}
\end{table*}

The $40$-$M_\odot$ progenitor is expected to result in BH
formation with no or only a very weak explosion (e.g.,
\cite{oconnor:11}). We carry out two sets of simulations with this
progenitor. In the first set, we employ ten different Skyrme
parametrizations combined with a $3\,335$-nuclide NSE EOS at low
densities using the same matching parameters as in the previous
Section~\ref{ssec:15M}. In the second set, we use the same Skyrme
parametrization, but with SNA at all densities. We summarize key
simulation results in Table~\ref{Tab:bounce} for both sets to
facilitate comparison.

In Figure~\ref{fig:40M_NSE}, we present the postbounce central density
$\rho_c$, central temperature $T_c$, and central entropy $s_c$
evolutions in the model set with an NSE treatment at low
densities. First, we note that the central entropy stays roughly
constant as it should (modulo numerical noise) throughout the
evolution to BH formation. Proto-NS collapse and BH formation is
marked by a dramatic increase in the slope of $\rho_c$, which is
mirrored by $T_c$. At this point, the \texttt{GR1D} simulations crash,
since the formulation of Einstein's equations used in \texttt{GR1D}
does not permit the evolution to continue beyond BH formation (see
\cite{oconnor:10} for details).

The time to BH formation is sensitive to the Skyrme parametrization
and set by accretion rate and the maximum proto-NS mass that can be
supported by the parametrization.  Comparing the maximum mass entries
in Table~\ref{Tab:TOV} with those in Table~\ref{Tab:bounce}, we note
that the maximum proto-NS mass is systematically $0.2 - 0.6\,M_\odot$
higher than the maximum cold NS mass. As shown by O'Connor
\& Ott \cite{oconnor:11}, this is a consequence of thermal pressure
support in the proto-NS mantle where shocked material is compressed,
reaching temperatures in excess of $100\,\mathrm{MeV}$ at late
times. As discussed in the context of the $15$-$M_\odot$ progenitor in
Section~\ref{ssec:15M}, Skyrme parametrizations that yield small
effective nucleon masses result in higher temperatures. In turn, such
parametrizations produce proto-NSs with more thermal pressure support
and see a greater increase in the maximum mass from cold NS to hot
proto-NS. For example, LS220, which has $m_t^* = m_t$ and a
maximum cold NS mass of $2.04\,M_\odot$, has a proto-NS mass of
$2.26\,M_\odot$ at BH formation ($\Delta M = 0.22\,M_\odot$).  The
SLy4 parametrization has a cold NS mass of $2.05\,M_\odot$, but its
proto-NS collapses at a mass of $2.488\,M_\odot$ ($\Delta M =
0.438\,M_\odot$). This is a direct consequence of SLy4's low effective
nucleon masses ($m_t^* = 0.695\, m_t$; cf.~Table~\ref{Tab:bulk})
and the consequently much higher temperatures reached in its proto-NS.

Like LS220, the SkT1 parametrization also has large effective nucleon
masses, resulting in lower temperatures.  Its maximum cold NS mass is
$1.846\,M_\odot$ and the proto-NS collapses at $2.204\,M_\odot$ ($\Delta M
= 0.358\,M_\odot$). Its variant SkT1$^*$ that we stiffened at high
density (see Section~\ref{ssec:highNS} and Table~\ref{Tab:TOV_SkT1})
has a maximum cold NS mass of $2.318\,M_\odot$. Interestingly, its
proto-NS collapses at a mass of only $2.327\,M_\odot$ ($\Delta M =
0.009\,M_\odot$). This at first surprising result can be understood
by considering that the SkT1$^*$ stiffening affects only the cold
high density core, but not the hot proto-NS mantle, where most of the
extra mass is located. The softness of the SkT1 parametrization
combined with the relatively modest temperatures reached in the mantle
thus explain our result for SkT1$^*$'s proto-NS mass at BH
formation.

In Table~\ref{Tab:bounce}, we compare the times to core bounce and BH
formation between simulations run with SNA at high densities and an
NSE EOS at low densities (SNA+NSE) and with SNA at all
densities. First we note that in the SNA+NSE case the time to core
bounce is insensitive to the Skyrme parametrization since the
transition to dynamical collapse is controlled by the NSE part that is
identical in all simulations. In the pure SNA simulations, this is
different and the time to core bounce can vary by hundreds of
milliseconds between some parametrizations. This is a consequence of
the metastability of the inner iron core at the onset of collapse
where small EOS differences can have substantial impact on when the
collapse becomes fully dynamical.

Finally, comparing BH formation times $t_\mathrm{BH}$ (measured from
the start of the simulation) predicted by SNA+NSE and pure SNA
simulations for a given Skyrme parametrization, we note that
$t_\mathrm{BH}$ appears insensitive to the low-density EOS
treatment. This can be understood by recalling that much of the
material that is accreted by the proto-NS to reach its maximum mass
comes from regions in the outer core and silicon and oxygen
shells. These regions are initially in hydrostatic equilibrium since
our simulations preserve the pressure stratification from the
precollapse stellar profile. Once the rarefaction wave from the core's
collapse reaches these regions, they proceed to collapse with
supersonic velocities in free fall. Hence, the collapse of the outer
regions is much less sensitive to variations in the EOS than the
collapse of the initially metastable inner core.

\section{Conclusions}\label{sec:Conclusions}

{In the twenty-six years since the seminal
  Lattimer \& Swesty (\LSn) paper \cite{lattimer:91} describing their
  finite-temperature nuclear equation of state (EOS), much progress
  has been made in both astrophysics simulation capability and in
  experimental and astrophysical constraints on the nuclear EOS. The
  \LS EOS has had tremendous impact on simulations of core-collapse
  supernovae (CCSNe) and neutron star (NS) mergers. This is due not
  least to \LS providing their EOS code as open source to the
  community.}

{In this study, we built upon the work of \LS
  and presented a generalized method for generating EOSs for CCSN and
  NS matter, using the compressible non-relativistic liquid-drop model
  with the Skyrme interaction.  With this paper, we make publicly
  available a modern, modular, and parallel Fortran 90 code for
  building EOS tables for application in CCSN and NS merger
  simulations. The code and EOS tables for the Skyrme parametrizations
  considered in this paper are available at
  \url{http://stellarcollapse.org/SROEOS}.}

Our method differs from the original \LS approach in the following
significant ways: (1) EOSs can be generated for most\footnote{Our
  method cannot presently handle Skyrme parametrizations that mix
  proton and neutron densities and kinetic energy densities in the
  nucleon effective masses (compare Equation~\ref{eq:mstar} with
  Equation 5 of Dutra \etaln~\cite{dutra:12}).} Skyrme parametrization
in the literature and for future parametrizations. This feature will
facilitate EOS parameter studies in astrophysics simulations within a
consistent EOS framework. (2) Our method includes nucleon effective
masses different from the rest masses and we obtain nuclear surface
properties self-consistently for each parametrization.  (3) Instead of
relying on Maxwell constructions that must be pre-computed for each
parametrization, we treat the transition from non-uniform to uniform
nuclear matter as a first-order phase transition that is determined as
the EOS is calculated. (4) The EOS obtained in the single-nucleus
approximation (SNA) can be smoothly merged at low densities with a
nuclear-statistical-equilibrium (NSE) EOS containing thousands of
nuclides. (5) We provide for the possibility of introducing additional
terms to Skyrme parametrizations that stiffen the EOS above saturation
density.
  (6) Our method converges reliably over a wide range of temperatures
  ($10^{-4}\,\mathrm{MeV} \lesssim T \lesssim
  10^{2.5}\,\mathrm{MeV}$), proton fractions
  ($10^{-3}\lesssim{y}\lesssim 0.7$), and densities
  ($10^{-13}\unit{fm}^{-3}\lesssim{n}\lesssim 10\unit{fm}^{-3}$). This makes
  it easy to generate EOS tables covering the space in $(n,T,y)$ required
  for simulations of CCSNe and NS mergers.

{ Using our new method, we generated EOS tables
  for nine Skyrme parametrizations: the \LS parametrization with
  $K_0=220\unit{MeV}$ (LS220) \cite{lattimer:91}, NRAPR
  \cite{steiner:05}, SLy4 \cite{chabanat:98}, SkT1 \cite{tondeur:84},
  SKRA \cite{rashdan:00}, LNS \cite{cao:06}, SQMC700
  \cite{guichon:06}, Skxs20 \cite{brown:07} and KDE0v1
  \cite{agrawal:05}. We thoroughly tested these EOSs, demonstrated
  thermodynamic consistency, and showed that our method can reproduce
  the results of the original \LS routines.  We computed cold
  beta-equilibrated NS mass-radius relationships for all EOS and
  explored the ad-hoc high-density modifications that stiffen the
  EOS. We showed that these modifications can raise the maximum NS
  mass above the astrophysical lower limit of $2\,M_\odot$ while leaving
  EOS properties at saturation density largely unaffected.
}

As a first application of our new EOS tables to astrophysics
simulations, we considered the spherically-symmetric collapse and
postbounce CCSN evolution in $15$-$M_\odot$ and $40$-$M_\odot$
progenitor star models. We tracked the $40$-$M_\odot$ models to black
hole (BH) formation. We compared SNA and NSE
  treatments of the EOS at low densities and found that subtle
  differences in the thermodynamics can affect the inner core's
  collapse time to core bounce and the postbounce accretion rate.
  Overall, as pointed out by Burrows \& Lattimer \cite{burrows:84},
  the thermodynamical properties are similar in both approaches and
  small differences translate to only minor variations in the
  postbounce evolutions.

In the case of BH formation, we find that the maximum proto-NS mass
supported by a given EOS correlates with the maximum cold NS mass of
the employed Skyrme parametrization, but is also highly sensitive to
the treatment of the nucleon effective masses. The maximum proto-NS
mass is typically substantially higher than the maximum cold NS mass
due to thermal pressure support from compression-heated accreted outer
core material. EOSs with lower effective nucleon masses lead to higher
temperatures and thus more pressure support and a higher maximum
proto-NS mass.

{Our goal with this study was to build a new
  and robust method for generating finite-temperature nuclear EOS
  tables. These can facilitate CCSN and NS merger simulations that
  explore the sensitivity of these phenomena to EOS parameters and
  predict multi-messenger (neutrino, gravitational wave,
  nucleosynthetic) signatures whose observation could help constrain
  the EOS.  We have realized this goal for the non-relativistic
  temperature-dependent liquid-drop model with Skyrme
  interaction. Much work lies ahead to generalize our method to
  include other mean-field parametrizations of nuclear interactions. A
  further important step will be to couple our new EOS tables to an
  efficient nuclear reaction network for accurately treating the
  regime in density, temperature, and composition space that is not in
  NSE.
}

\begin{acknowledgments}

We acknowledge helpful discussions with C.~J.~Horowitz, J.~Lattimer,
H.~Nagakura, A.~Ohnishi, and S.~Richers. We thank Y.~Suwa for pointing
out that it is advantageous to preserve the precollapse pressure
stratificiation when mapping stellar profiles into the core collapse
code. A.\,S.\,S. was supported in part by the Conselho Nacional de
Desenvolvimento Cient\'ifico e Tecnol\'ogico (201432/2014-5). This
research was also funded by the National Science Foundation under
award No. AST-1333520, CAREER PHY-1151197, PHY-1404569, and by the
Sherman Fairchild Foundation. C.~D.~O. thanks the Yukawa Institute for
Theoretical Physics (YITP) for support and hospitality during the
completion of this paper, which has been assigned report
no.~YITP-17-32.

\end{acknowledgments}

\appendix

\section{Leptons and photons}\label{app:ele}

We use the Timmes EOS to determine the properties of photons and
leptons \cite{timmes:99}.  The only leptons considered here are
electrons and positrons.  The photon gas is assumed to be generated by
a blackbody in local thermodynamic equilibrium.  Its pressure,
internal energy, and entropy are given by
\begin{equation}
 P_{\mathrm{rad}}=\frac{4\sigma_{SB}T^4}{3c}\,,\quad
 E_{\mathrm{rad}}=3P_{\mathrm{rad}}\,, \quad
 S_{\mathrm{rad}}=\frac{4E_{\mathrm{rad}}}{3 T}\,,
\end{equation}
where $\sigma_{SB}$ is the Stephan-Boltzmann constant, $c$ the speed of
light and $n$ the baryon number density.  The electron and positron
contributions are determined assuming charge neutrality, \ie
\begin{equation}\label{eq:neutrality}
 yn=n_{\mathrm{ele}}-n_{\mathrm{pos}}\,.
\end{equation}
Recall that $y$ is the proton fraction of the system.
Here $n_{\mathrm{ele}}$ and $n_{\mathrm{pos}}$ are, respectively, the electron and positron number densities given by
\begin{subequations}
\begin{align}
 n_{\mathrm{ele}}&=K\beta^{3/2}\left[\mathcal{F}_{1/2}(\eta,\beta)+\mathcal{F}_{3/2}(\eta,\beta)\right]\,,\\
 n_{\mathrm{pos}}&=K\beta^{3/2}\left[\mathcal{F}_{1/2}(\kappa,\beta)+\mathcal{F}_{3/2}(\kappa,\beta)\right]\,,
\end{align}
\end{subequations}
where we define the constant $K=8\pi\sqrt{2}m_e^3c^3/h^3$ with $m_e$
being the electron mass.  Furthermore, $\beta=T/(m_ec^2)$ is the
relativity parameter, $\eta=\mu/T$ is the degeneracy parameter of
electrons where $\mu$ is the electron chemical potential, and we
define $\kappa=-\eta-2/\beta$.  The function
$\mathcal{F}_k(\eta,\beta)$ is the Fermi-Dirac integral
\begin{equation}\label{eq:FermiDirac}
 \mathcal{F}_k(\eta,\beta)=\int_0^\infty \frac{u^k(1+0.5\beta u)^{1/2}}{1+\exp({u-\eta})}du\,.
\end{equation}
Note that the Fermi integral, Equation \eqref{eq:Fermi}, is a special
case of the Fermi-Dirac integral with $\beta=0$.  The degeneracy
parameter $\eta$ is found from the solution of Equation
\eqref{eq:neutrality} and can be used to obtain the thermodynamic
variables of the electron and positron gas.  Their pressures and energies per 
volume are given by 
\begin{subequations}
\begin{align}
 P_{\mathrm{ele}}&=\frac{2K}{3}m_ec^2\beta^{5/2}\left[\mathcal{F}_{3/2}(\eta,\beta)+\frac{\beta}{2}\mathcal{F}_{5/2}(\eta,\beta)\right],\\
 P_{\mathrm{pos}}&=\frac{2K}{3}m_ec^2\beta^{5/2}\left[\mathcal{F}_{3/2}(\kappa,\beta)+\frac{\beta}{2}\mathcal{F}_{5/2}(\kappa,\beta)\right],\\
 E_{\mathrm{ele}}&=K m_ec^2\beta^{5/2}\left[\mathcal{F}_{3/2}(\eta,\beta)+\beta \mathcal{F}_{5/2}(\eta,\beta)\right],\\
 E_{\mathrm{pos}}&=K m_ec^2\beta^{5/2}\left[\mathcal{F}_{3/2}(\kappa,\beta)+\beta \mathcal{F}_{5/2}(\kappa,\beta)\right]\nonumber\\
               &\qquad+2n_{\mathrm{pos}}m_ec^2\,,
\end{align}
where the subscripts ele and pos refer, respectively, to electrons and
positrons.  Meanwhile, their entropy densities are
\begin{align}
 S_{\mathrm{ele}}&=\frac{P_{\mathrm{ele}}+E_{\mathrm{ele}}}{T}-n_{\mathrm{ele}}\eta\,,\\
 S_{\mathrm{pos}}&=\frac{P_{\mathrm{pos}}+E_{\mathrm{pos}}}{T}-n_{\mathrm{pos}}\kappa\,.
\end{align}

\end{subequations}
For details on how these calculations are performed see
\cite{timmes:99}.

\section{Nuclear Statistical Equilibrium}\label{app:NSE}

In NSE, the chemical potential of nuclear species $i$ is given by
\begin{eqnarray}
\mu_i &=& m_i + E_{c,i} + T \log \left[ \frac{n_i}{g_i(T)} \left(\frac{2\pi}{m_i T}\right)^{3/2} \right], \nonumber\\
&=& Z_i \mu_p + (A_i - Z_i) \mu_n\,,
\end{eqnarray}
where $m_i$ is the mass, $A_i$ is the nucleon number, $Z_i$ is the
proton number, $n_i$ is the number density, and $g_i(T)$ is the internal
partition function of species $i$.  We use the partition functions of
Rauscher \& Thielemann \cite{rauscher:00} and nuclear masses
from the JINA REACLIB database. See Cyburt \etal \cite{cyburt:10} and
references therein. The partition function tables and nuclear mass
tables are available at \url{http://stellarcollapse.org/SROEOS}.  The
Coulomb correction in the Wigner-Seitz approximation is
\begin{equation}
E_{c,i} = \frac{3 \alpha_C Z_i^2}{5 r_i}\left(\frac{1}{2} u_i - \frac{3}{2}u_i^{1/3} \right),
\end{equation}
where the nuclear radius $r_i = (3 A_i / 4 \pi n_0)^{1/3}$, $u_i = y n/n_0 A_i/Z_i$, and $\alpha_C$ is the fine structure constant.
Imposing mass and charge conservation, this system of equations can be solved for the composition.
When calculating NSE, we assume that the neutrons and protons are arbitrarily degenerate, non-relativistic particles.
We neglect Coulomb corrections for the protons.

The pressure, energy density, and entropy density 
  of the nuclei ensemble in NSE
is given by
\begin{eqnarray}
P_n &=& \sum_i n_i \left\{T + \frac{\partial E_{c,i}}{\partial \ln n} \right \}\,, \\
E_n &=& \sum_i n_i \left\{ \frac{3}{2} T + E_{c,i} - \mathrm{B_i} + T \frac{d \ln g_i}{d
\ln T} \right\}\,, \\
S_n &=& \sum_i n_i \left\{\frac{5}{2} + \ln\left[\frac{g_i}{n_i} \left(
\frac{m_i T_i}{2 \pi} \right)^{3/2} \right]  + \frac{d \ln g_i}{d\ln T}
\right\}\,,
\end{eqnarray}
where $\mathrm{B_i}$ is the binding energy of species $i$ relative to
$A_i$ neutrons.  The contribution of the nucleons is given by the
expressions in Section~\ref{sec:bulk} with the Skyrme parameters set
to zero.  The free energy density of the nuclei ensemble is set by
\begin{equation}\label{eq:FNSE}
 F_{\mathrm{NSE}}=E_n-TS_n.
\end{equation}

\section{Critical temperature coefficients}
\label{app:Tcrit}

In Section \ref{ssec:surface} we present a method for determining the
critical temperature $T_c$ below which nuclear matter may phase
separate into two phases of different densities, $n_i$ and $n_o$, and
proton fractions, $y_i$ and $y_o$.  In Table \ref{Tab:Tcrit}, we
present the coefficients calculated for the critical temperature
approximation $T_c\equiv T_c(y_i)$, Equation \eqref{eq:Tcrit}.  Since
we do not obtain the surface properties for the LS220 parametrization,
we set the coefficients $T_c(y_i)$ to match those of \LSn.  Note that
for all other parametrizations we have $a_c\simeq1.00$ and
$b_c\simeq-1$.  In fact, the EOS we calculate are not significantly
altered by enforcing $a_c=1$ and $b_c=-1$.

\begin{table}[htbp]
\caption{\label{Tab:Tcrit} Coefficients for the fit of the proton-fraction 
  dependence of the critical temperature $T_c(y)$ given by Equation~\eqref{eq:Tcrit}. 
  These  coefficients depend on the Skyrme parametrization and we provide
  them here for completeness. $T_c$ is in MeV while $a_c$, $b_c$, $c_c$, and
  $d_c$ are dimensionless. }
\begin{ruledtabular}
\begin{tabular}{l  D{.}{.}{2.2} D{.}{.}{2.4} D{.}{.}{2.4} D{.}{.}{2.4} D{.}{.}{2.4}}
\multicolumn{1}{c}{Parametrization}&
\multicolumn{1}{c}{$T_c$}&
\multicolumn{1}{c}{$a_c$}&
\multicolumn{1}{c}{$b_c$}&
\multicolumn{1}{c}{$c_c$}&
\multicolumn{1}{c}{$d_c$}\\
\hline
LS220 \cite{lattimer:91}        & 16.80 & 1.0000 & -1.0000 & 0.0000 & -0.0000 \\
KDE0v1 \cite{agrawal:05}        & 14.85 & 1.0035 & -1.1600 & 0.7797 & -1.6822 \\
LNS \cite{cao:06}               & 14.92 & 1.0017 & -1.2052 & 0.2432 & -0.6667 \\
NRAPR \cite{steiner:05}         & 14.39 & 1.0029 & -1.0029 & 0.4679 & -0.9929 \\
SKRA \cite{rashdan:00}          & 14.35 & 1.0031 & -1.1227 & 0.4336 & -0.9523 \\
SkT1 \cite{tondeur:84}          & 17.05 & 1.0022 & -1.1921 & 0.4371 & -0.7393 \\
Skxs20 \cite{brown:07}          & 15.37 & 1.0017 & -1.3778 & 0.4015 & -0.6087 \\
SLy4 \cite{chabanat:98}         & 14.52 & 1.0038 & -1.0127 & 0.7771 & -1.6520 \\
SQMC700 \cite{guichon:06}       & 14.72 & 1.0022 & -1.1794 & 0.3284 & -0.8968 \\
\end{tabular}
\end{ruledtabular}
\end{table}

\bibliography{SROEOS}

\end{document}